\documentclass[fleqn,usenatbib]{mnras}

\usepackage{newtxtext,newtxmath}

\usepackage[T1]{fontenc}
\usepackage{xcolor}

\DeclareRobustCommand{\VAN}[3]{#2}
\let\VANthebibliography\thebibliography
\def\thebibliography{\DeclareRobustCommand{\VAN}[3]{##3}\VANthebibliography}

\usepackage{graphicx}	
\usepackage{amsmath}	
\usepackage{bm}
\usepackage{physics}
\usepackage{cancel}

\usepackage{yhmath}
\usepackage{array}
\usepackage{makecell}

\newcommand{\bn}{\bm\nabla}
\newcommand{\Pm}{\mathrm{Pm}}
\newcommand{\Ek}{\mathrm{Ek}}
\newcommand{\Em}{\mathrm{Em}}
\newcommand{\Le}{\mathrm{Le}}
\newcommand{\Lep}{\mathrm{Le_p}}
\newcommand{\Let}{\mathrm{Le_t}}
\newcommand{\Pt}{P_\mathrm{t}}
\newcommand{\ct}{C_\mathrm{t}}
\newcommand{\edr}{E_\mathrm{dr}}
\newcommand{\Mp}{M_\mathrm{p}}
\newcommand{\Mt}{M_\mathrm{t}}
\newcommand{\Rm}{\mathrm{Rm}}
\newcommand{\tap}{t_\mathrm{ap}}
\newcommand{\Bp}{B_\mathrm{p}}
\newcommand{\ls}{l_\Omega}
\newcommand{\Lec}{\mathrm{Le_c}}

\renewcommand{\Re}{\mathrm{Re}}
\newcommand{\Lo}{\mathrm{L_o}}
\newcommand{\Lz}{\omega_\mathrm{A}}
\newcommand{\Lp}{\omega_\mathrm{A_\varphi}}
\newcommand{\Ekk}{\omega_\nu}
\newcommand{\Emk}{\omega_\eta}
\newcommand{\Som}{S_\Omega}
\newcommand{\vaz}{v_{\mathrm{A}_z}}

\definecolor{db}{HTML}{191586}
\definecolor{vi}{HTML}{5E1111}

\graphicspath{{figures/}}

\title[Tidal flows and magnetic fields]{Interplay between tidal flows and magnetic fields in nonlinear simulations of stellar and planetary convective envelopes}

\author[A.~Astoul  \& A.J.~Barker]{
Aur\'{e}lie Astoul$^{1}$\thanks{E-mail: a.a.v.astoul@leeds.ac.uk (AA)}
and Adrian J. Barker$^{1}$\thanks{E-mail: A.J.Barker@leeds.ac.uk (AJB)}
\\
$^{1}$School of Mathematics, University of Leeds, Leeds LS2 9JT, UK\\
}

\date{Accepted Nov 6 666. Received Nov 6 666; in original form Nov 6 666}

\pubyear{\the\year{}}

\begin{document}
\label{firstpage}
\pagerange{\pageref{firstpage}--\pageref{lastpage}}\maketitle

\begin{abstract}
Stars and planets in close systems are magnetised but the influence of magnetic fields on their tidal responses (and vice versa) and dissipation rates has not been well explored. We present exploratory nonlinear magnetohydrodynamical (MHD) simulations of tidally-excited inertial waves in convective envelopes. These waves probably provide the dominant contribution to tidal dissipation in several astrophysical settings, including tidal circularisation of solar-type binary stars and hot Jupiters, and orbital migration of the moons of Jupiter and Saturn. We model convective envelopes as incompressible magnetised fluids in spherical shells harbouring an initially (rotationally-aligned) dipolar magnetic field. We find that depending on its strength (quantified by its Lehnert number $\Le$) and the magnetic Prandtl number $\Pm$, the magnetic field can either deeply modify the tidal response or be substantially altered by tidal flows. Simulations with small $\Le$ exhibit strong tidally-generated differential rotation (zonal flows) for sufficiently large tidal amplitudes, such that both the amplitude and topology of the initial magnetic field are tidally impacted. In contrast, strong magnetic fields can inhibit these zonal flows through large-scale magnetic torques, and by Maxwell stresses arising from magneto-rotational instability, which we identify and characterise in our simulations, along with the role of torsional Alfv\'{e}n waves. Without tidally-driven zonal flows, the resulting tidal dissipation is close to the linear predictions. We quantify the transition $\Le$ as a function of $\Pm$, finding it to be comparable to realistic values found in solar-like stars, such that we predict complex interactions between tidal flows and magnetic fields.
\end{abstract} 




\section{Introduction}
Tidal interactions are a key driver of orbital and rotational evolution in compact stellar and exoplanetary systems \citep[e.g.][]{O2014,Mathis2019}. Solar-like (low-mass) stars and giant gaseous planets feature convective envelopes in which inertial waves restored by Coriolis forces -- and in magnetised stars, magneto-inertial waves restored by both Coriolis and Lorentz forces -- can be tidally excited. Their dissipation is believed to contribute significantly to angular momentum exchanges and spin-orbit evolution in these systems. For example, simplified calculations indicate that these waves may largely explain the observed orbital evolution of Jupiter's and Saturn's moons \citep[e.g.][]{OL2004,L2023,D2023,DBMDAR2024,PBH2024}, the circularisation periods of solar-type binary stars \citep[e.g.][]{B2022}, and the eccentricity distributions of hot and warm Jupiters \citep[e.g.][]{LBdVA2024}. However, many aspects of these waves in realistic astrophysical environments, like in differentially-rotating and/or magnetised envelopes with convective motions and density stratification, and their resulting contributions to tidal dissipation are still poorly understood.

Inertial waves are thought to be particularly important for tidal evolution in systems with fast rotators, such as young stars or Jupiter-like planets. One reason is that inertial waves are only (linearly) excited for tidal frequencies satisfying $|\omega|\leq 2|\Omega_0|$, where $\omega$ is the Doppler-shifted frequency in the frame rotating with the stellar angular velocity $\Omega_0$. This condition is more easily satisfied for faster rotation (larger values of $\Omega_0$). Secondly, the (frequency-averaged) tidal dissipation rate scales approximately with the square of the rotational frequency in linear theory \citep[e.g.][]{O2013}, so faster rotating stars or planets tend to be more dissipative than more slowly rotating ones (all else being equal), as been shown in \citet{M2015,BM2016,GB2017} and \citet{B2020} using 1D stellar evolution models.

The solar convection zone is known to harbour a predominantly dipolar magnetic field with an overall strength of approximately 20 Gauss at the surface, but this varies substantially across the surface and inside the convective envelope \citep[possibly up to several Tesla for sunspot-forming toroidal flux ropes,][]{C2013,C2014}.
Convective fluid motions are thought to play a key role in the solar dynamo, even if the extent to which this is the case is debated \citep[e.g.][]{BB2017}, as well as the location of the dynamo, either deep down in the convective envelope \citep[and top of the radiative layer][]{P1993}, in the bulk \citep[e.g.][]{SB2017}, or in the near surface shear layers \citep[for instance,][]{VL2024}. Observations also indicate that magnetic fields are ubiquitous in low-mass stars, as revealed by spectropolarimetry \citep[e.g.][]{DL2009,R2012}, which probes the large-scale magnetic fields at their surfaces, and as predicted by 3D MHD simulations of convective dynamos \citep[for a review see][and references therein]{Kapyla2023}.

While Alfvén waves restored by the Lorentz force are the only low frequency (i.e. ignoring surface gravity and acoustic) waves in a magnetised envelope, in a rotating and magnetised medium, mixed types of wave arise due to the combined action of the Coriolis acceleration and magnetic tension. These include slow and fast magneto-inertial waves (also called magneto-Coriolis or MC waves), which arise depending on whether the two restoring forces cancel each other or sum up in the dispersion relation \citep[][for a review]{L1954,M1967,F2008}. In the latter case, the wave frequency exceeds the cut-off frequency $2\Omega_0$ of inertial waves, while in the former case the frequency is lower than the rotation frequency and the waves are often called magnetostrophic waves, for fast rotators. Magneto-inertial waves have been largely studied in the context of the geodynamo, as they are expected in the liquid outer core of the Earth, and they have been studied experimentally, for example via rotating spherical Couette flow in liquid metal or sodium experiments \citep[][for a review]{SA2008,SC2013,LB2022}, or through numerical simulations \citep[e.g.][]{S2010,A2018}. One type of wave that arises in this context is torsional Alfvén waves, which are (cylindrically) radially propagating waves in differentially rotating fluids that are invariant along the rotation axis. They may reflect and form standing modes, the so-called torsional oscillations \citep[e.g.][for a review]{B1970,SJ2017,HN2023}.

In stellar envelopes, the effect of a magnetic field has been studied on some specific torsional inertial modes \cite[the r-modes, as in][for neutron stars]{AR2012,LJ2010}, and the propagation of shear Alfvén waves (with dissipative processes) has been investigated by \cite{RR2003} and \cite{RR2004} in incompressible (though non-rotating) shells with a dipolar magnetic field. 
Very few global non-linear studies have been performed to explore the tidal response and its dissipation in convective envelopes of rotating stars and planets \citep[e.g.][]{T2007,FB2014,AB2022,AB2023}, and none with a magnetic field \citep[though][simulated the elliptical instability in a full sphere in the presence of a magnetic field to study tidal dynamos]{CH2014}. 
Previous theoretical studies of tidal inertial waves have primarily involved linear calculations or explored nonlinear tidal waves in the absence of magnetic fields. 
With rotation, an important prior study is \citet[see also \citealt{W2016} in a local box]{LO2018}, who performed linear calculations of tidal magneto-inertial waves in convective envelopes with an imposed rotationally-aligned dipolar magnetic field \citep[or a uniform field aligned with the rotation axis, also in][for rapid rotators]{W2018}. They found that the (low-frequency) frequency-averaged tidal dissipation when inertial waves are excited is unmodified by a magnetic field. However, the dissipation at a given frequency, as well as the nature of the waves and the mechanisms of their dissipation -- whether this is due to viscosity or Ohmic diffusion, and whether it is due to turbulent or microscopic processes -- can be very different when considering a magnetic field. Considering stellar tides in hot Jupiter systems\footnote{These systems are composed of a Jupiter-like planet orbiting within a few days (typically) around a low-mass star.}, it has been shown that magnetic effects should not be neglected when computing tidal dissipation at a given frequency \citep{AM2019}. 

This motivates us to study here the interplay between tidal flows and magnetism using three-dimensional nonlinear simulations of rotating stellar or planetary convection zones, building upon our prior hydrodynamical studies in \citet[][hereafter AB22]{AB2022} and \citet[][hereafter AB23]{AB2023}. 

In AB22 and AB23, we found that non-linear self-interactions of tidally forced inertial waves induce cylindrical-like differential rotation \citep[also called zonal flows, and also found in][using different kinds of boundary forcing instead of the effective body force in our studies]{ML2010,FB2014,CV2021}. This differential rotation is particularly strong for thin convective envelopes, high tidal amplitudes (e.g.~relevant for tides inside the closest hot Jupiters and stellar binaries) and low viscosities (relevant for the microscopic values in stars and planets), where nonlinear effects (including wave-wave and wave-zonal flow interactions and instabilities) were observed to play an important role, as shown in AB2023. In such cases, we have found that nonlinear simulations exhibit important deviations from linear predictions for tidal dissipation. In the following, we are particularly interested in exploring the effects of magnetism on the generation of differential rotation and how it modifies tidal dissipation rates in simulations of tidally-driven inertial waves in convective envelopes. To do so, we impose an initial (axially aligned) dipolar magnetic field, with a strength that we vary, along with varying the value of the Ohmic diffusivity.

We structure this paper as follows. In \S~\ref{sec:model}, we describe our magnetohydrodynamical (MHD) model of tidal flows in convective envelopes, including how we drive tidal waves and initialise the magnetic field. We also derive the energetic balances in our model. In \S~\ref{sec:res}, we describe and analyse the results of our simulations varying the Lehnert number, including exploring the evolution of differential rotation and magnetic field, angular momentum fluxes, and identifying the presence of torsional waves. In \S~\ref{sec:Pm} we vary the magnetic Prandtl number and also examine the occurrence of magnetic instabilities. We present our conclusions, discuss the astrophysical implications of our results and fruitful avenues for further work in \S~\ref{sec:con}.

\section{Nonlinear MHD tidal model with an initial imposed dipolar magnetic field} 
\label{sec:model}

We build on the hydrodynamical and nonlinear tidal model described in detail in AB22, to which we add an initial dipolar magnetic field. We turn to solving here the MHD equations for tidally excited magneto-inertial waves in an incompressible and adiabatic (neutrally stratified) convective envelope of a low-mass star or giant planet. The size of the inner core (normalised to the total radius $R$) is fixed to $\alpha=0.5$, which represents a slightly thicker-than-solar envelope. This value is relevant for modelling lower-mass M or K stars throughout certain stages of their evolution, or to a giant planet with an extended dilute core that is sufficiently stably stratified such that it acts like a rigid boundary for low frequency waves in the convection zone \citep[e.g.][]{MF2021,PBH2024}. This restriction is made in this initial study of magnetic effects because envelopes with $\alpha=0.5$ have been the most-widely studied in prior linear and nonlinear studies to-date \citep{O2009,FB2014,LO2018,AB2022,AB2023}. The shell rotates at a frequency $\Omega_0$ along the vertical unit vector $\bm e_z$, and we assume that the envelope consists of fluid with a constant density $\rho$ (which is set to 1, without loss of generality). The momentum, induction, and continuity (plus solenoidal constraint on the magnetic field) equations for the tidally-excited magneto-inertial waves are given by:
\begin{subequations}
    \begin{align}
        \partial_t\bm u+2\bm e_z\wedge\bm u+(\bm u\cdot\bn)\bm u &=-\bn p+\Le^2(\bn\wedge\bm B)\wedge\bm B +\Ek\,\Delta\bm u+\bm f_\mathrm{t},\\
        \partial_t\bm B &= \bn\wedge(\bm u\wedge\bm B)+\Em\,\Delta\bm B, \label{eq:ind}\\
        \bn\cdot\bm u&=0, \label{eq:solu}\\
        \bn\cdot\bm B&=0, \label{eq:solb}
    \end{align}
\label{eq:sys}
\end{subequations}
\hspace{-0.23 cm}
with $\bm u$, $\bm B$ and $p$ the dimensionless velocity, magnetic field and pressure, and $\Delta=\bm\nabla^2$ the Laplacian operator. We adopt $R$, $\Omega_0^{-1}$ and $B_0$ as units of length, time, and magnetic field, respectively, where the latter is a typical strength of the magnetic field. We have introduced several dimensionless parameters, including the Lehnert number $\Le=B_0/(\sqrt{\mu\rho}R\Omega_0)$ (with $\mu$ the vacuum magnetic permeability) which is a measure of the magnetic field strength in rotational units (it is the ratio of the rotational timescale to the Alfv\'{e}n propagation timescale over the distance $R$), and the viscous and magnetic Ekman numbers $\Ek=\nu/(R^2\Omega_0)$ and $\Em=\eta/(R^2\Omega_0)$. In the latter two, $\nu$ and $\eta$ are the (assumed) constant kinematic viscosity and Ohmic diffusivity of the fluid, which can be considered to represent turbulent values (e.g.~from mixing-length theories). We also define the magnetic Prandtl number $\Pm=\nu/\eta=\Ek/\Em$, which we vary in our simulations from $10^{-1}$ to $5$, while keeping the Ekman number constant and set to $\Ek=10^{-5}$. This choice is motivated by mixing-length values for the solar convective envelope \citep[assuming such a turbulent viscosity damps tidal waves, e.g.][for inertial waves in the Sun]{OL2007,BC2022}, by our previous hydrodynamical simulations that have explored this value extensively (AB22,AB23), and finally by computational limitations that prevent much smaller values of $\Ek$ (such as the tiny microscopic ones that are of order $10^{-15}$ in the solar envelope) from being simulated. 

We decompose the tidal flow into an equilibrium/non-wavelike tide and a dynamical/wavelike tide as in AB22 and AB23. The former is assumed to be perfectly maintained on the timescale of our simulations -- which are designed to probe a brief snapshot in the evolution of a system -- and to be described within linear theory\footnote{It is by definition the quasi-hydrostatic adjustment of a body, and its associated flow, due to the tidal and self-gravitational potentials from the perturber and the perturbed body, respectively \citep[e.g.][]{O2014}.}, but it satisfies the correct tidally-perturbed free surface boundary condition at $r=R$ along with the conditions for a rigid core at $r=\alpha R$. This is most directly applicable to modelling the equilibrium tide in a giant planet with a solid core, but it approximately describes the flow in the convective envelope atop a radiative core. The wavelike tide is forced by the effective tidal forcing:
\begin{equation}
    \bm f_\mathrm{t}=\mathrm{Re}\left\{ -\bm e_z\wedge\bn\left[\left(r^2+\frac{2}{3}\alpha^5 r^{-3}\right)Y^2_2(\theta,\varphi)\right]\frac{\omega \, C_\mathrm{t}}{1-\alpha^5}\mathrm{e}^{-\mathrm{i}\omega t}\right\},
    \label{eq:forcing}
\end{equation}
written here in dimensionless units, with $(r,\theta,\varphi)$ and $t$ being spherical polar coordinates and time, $Y^2_2$ a quadrupolar spherical harmonic, $\omega$ the tidal forcing frequency, and $C_\mathrm{t}$ the dimensionless tidal amplitude \citep[see also][]{O2013,LO2018}. We treat $C_\mathrm{t}$ as an input parameter in our study, and this is related to the dimensionless tidal amplitude $\epsilon=(M_2/M_1)(R/a)^3$ by $C_\mathrm{t}=(1+k_2)\epsilon$, where $M_2$ and $M_1$ are the masses of the perturber and perturbed body, respectively, $a$ is the orbital semi-major axis, and $k_2$ is the real part of the quadrupolar Love number (typically approximated by its hydrostatic value). 

Note that Eq.~(\ref{eq:forcing}) is purely hydrodynamical, namely, it only takes into account the non-inertial term (i.e. the Coriolis pseudo-force) acting on the equilibrium tide that is omitted from its definition and which is taken to drive tidal waves here. \cite{AM2019} have studied how magnetism can modify the equilibrium tide, and thus the excitation of magneto-inertial waves.  
However, the impact of this magnetic contribution to the tidal forcing is negligible (in the linear regime) compared with Eq.~(\ref{eq:forcing}), when considering either the amplitude of large-scale magnetic fields of low-mass stars hosting hot Jupiters (for the tide in the star) or the magnetic field inside a hot Jupiter itself (for the tide in the planet); therefore, we neglect magnetic effects on equilibrium tides in this study.

We follow AB22 and retain the nonlinear terms involving tidal waves only, but not those involving the equilibrium tidal flow. As explained in AB22, this is justified if the wavelengths of the waves are much shorter than the radius of the body (and the tidal velocity magnitudes of the waves are larger), which is typically satisfied in our calculations and is usually also expected in reality. Incorporating nonlinear interactions with the equilibrium tide in our spherical shell geometry has also been found to lead to unphysical angular momentum evolution \citep[][AB22]{FB2014}. The same arguments are expected to hold for the nonlinear Lorentz force and magnetic induction terms involving the equilibrium tidal flow with the tidal waves and their magnetic field perturbations. Future work should explore the additional contributions of nonlinear interactions with the equilibrium tidal flow in realistic tidally-deformed geometries, but performing such studies will be a formidable task.

We adopt stress-free and impenetrable boundary conditions for the velocity of the tidal waves, and current-free (i.e.~insulating, $\bm e_r\cdot(\bn\wedge\bm B)=\bm 0$) boundary conditions for the magnetic field at both the inner and outer shells, where the field also continuously matches onto a potential field in the core and exterior of the body. Note that impenetrable conditions are not applied to the equilibrium tidal flow at the surface, which satisfies the correct free surface condition.

We use the spherical pseudo-spectral code MagIC to perform MHD simulations of nonlinear tidal waves in the presence of magnetic fields (see Sect.~\ref{sec:magic}). In MagIC, the velocity and magnetic field are described using a poloidal/toroidal decomposition since they are both solenoidal (Eqs.~(\ref{eq:solu}) and (\ref{eq:solb})). In other words, they can be decomposed as (here for the magnetic field):
\begin{equation}
\bm B=\bn\wedge(\bn\wedge g\bm e_r)+\bn\wedge h \bm e_r,
\end{equation}
where $g$ and $h$ are poloidal and toroidal scalar potentials, respectively. In the following, we will refer to the poloidal and toroidal magnetic fields as $\bm B_\mathrm{p}=\bn\wedge(\bn\wedge g\bm e_r)$ and $\bm B_\mathrm{t}=\bn\wedge h \bm e_r$, respectively, with corresponding poloidal and toroidal energies $\Mp=\Le^2\langle |\bm B_\mathrm{p}|^2/2\rangle$ and $\Mt=\Le^2\langle |\bm B_\mathrm{t}|^2/2\rangle $, respectively (with $\langle\cdot\rangle$ denoting a volume integral over the whole shell).

Our model and governing equations (Eqs.~(\ref{eq:sys})) are similar to those in \cite{LO2018}, except that we solve the fully nonlinear system for magnetic tidal waves whereas they performed linear calculations. We also adopt an initial dipolar magnetic field \citep[rather than a ``background field" as in][but of the same form, with opposite sign]{LO2018}: 
\begin{equation}
    \bm B(t=0)=\bm B_0=-\left(\frac{\alpha}{r}\right)^3\left[ \cos\theta\,\bm e_r+\frac{\sin\theta}{2} \bm e_\theta \right],
    \label{eq:B0}
\end{equation}
which has dimensional amplitude $B_0$. Thus, initially, the magnetic energy in the whole shell of volume $V$ is:
\begin{equation}
M_0=\Le^2\int_V \frac{|\bm B_0|^2}{2} r^2 \sin\theta\,\mathrm{d}r\,\mathrm{d}\theta\,\mathrm{d}\varphi= \alpha^3 \Le^2\frac{\pi}{3}\left(1-\alpha^3\right).
\end{equation}
The magnetic field $\bm B_0$ is allowed to evolve, but it is not self-sustained by convective motions, so it will decay Ohmically. We define a new Lehnert number which follows the evolution of the poloidal magnetic field $M_\mathrm{p}(t)$:
\begin{equation}
    \Lep(t)=\sqrt{\frac{3\Mp}{\alpha^3\pi\left(1-\alpha^3\right)}},
    \label{eq:lep}
\end{equation}
such that $\Lep=\Le$ at $t=0$, where $\Mp$ is the poloidal magnetic energy (with $\Mp=M_0$ at $t=0$). We also define a corresponding toroidal Lehnert number, which takes a similar form:
$\Let(t)=\sqrt{3M_t/\left[\alpha^3\pi\left(1-\alpha^3\right)\right]}$, where $\Let(t=0)=0$ initially.

The sum of the integrated magnetic ($M=\Le^2\,\langle|\bm B|^2\rangle/2=\Mp+\Mt$) and kinetic ($K=\langle |\bm u|^2\rangle/2$) energies satisfies the energetic balance described by:
\begin{equation}
    \partial_t(M+K)=P_\mathrm{t}+\mathcal{F}_\mathrm{P}-D_\nu-D_\eta.
    \label{eq:bal}
\end{equation}
This can be derived using Eqs.~(\ref{eq:sys}) and the boundary conditions. When our simulations reach an overall steady state, the rate of energy injected by the tide, i.e.~the tidal power, defined by $P_\mathrm{t}=\langle\bm u\cdot\bm f_\mathrm{t} \rangle$, is mostly balanced by the sum of viscous and Ohmic dissipation, $D_\nu=-\langle\Ek\,\bm u\cdot\Delta\bm u \rangle$ and $D_\eta=-\left\langle \mathrm{Le}^2\,\Em\,\bm B\cdot\Delta\bm B\right\rangle$. The Poynting flux 
$ \mathcal{F}_\mathrm{P}=\Le^2\int_{\delta V} \bm u\cdot\bm B\, B_r\,\mathrm{d}S$, quantifies the transfer of magnetic energy through the inner and outer boundaries, but this is almost always found to be negligibly small in our simulations \citep[see also][]{AJ2005} so can be ignored for our purposes\footnote{It is strictly non-zero, taking the value $\mathcal{F}_\mathrm{P}\approx-2.18\times 10^{-6}$ in linear theory for a fixed field with strength $\Le=10^{-2}$ with our default parameters. This is much smaller than the corresponding $D_\nu\approx 3.5\times 10^{-3}$ (and $D_\eta\approx 9.31\times 10^{-4}$).}. Furthermore, the total angular momentum is conserved in the simulations given our choice of boundary conditions for the magnetic field and velocity, following Appendix A of \cite{JB2011}, combined with the fact that the effective wavelike tidal torque $\boldsymbol{r}\wedge\boldsymbol{f}_t\propto e^{2i\phi}$ also vanishes when integrating over the whole shell (see AB22 for further details of the latter). We have verified that both energetic and angular momentum balances are accurately satisfied in our simulations, as we will discuss further below.

As in \citet{FB2014}, AB22, and AB23, we also define the mean rotation rate of the fluid, relative to the reference frame rotating at the rate $\Omega_0$, by $\delta\Omega=\langle u_\varphi/(r\sin\theta)\rangle/V$, where $V$ is the fluid volume, and the energy in the differential rotation $\edr=\langle\left(\langle u_\varphi\rangle_\varphi-\delta\Omega\, r\sin\theta\right)^2\rangle/2$,  where $\langle\cdot\rangle_\varphi$ denotes a $\varphi$-average. Note that $\delta \Omega$ can be non-zero despite global angular momentum conservation, as a result of differential rotation produced within the fluid volume, though it is typically quite small. The energy in differential rotation $\edr$ is helpful in quantifying the generation of zonal flows by nonlinear self-interactions of tidal waves.

\begin{figure*}
    \centering
    \includegraphics[trim=0cm 0cm 2cm 0.87cm,clip,clip,width=0.285\linewidth]{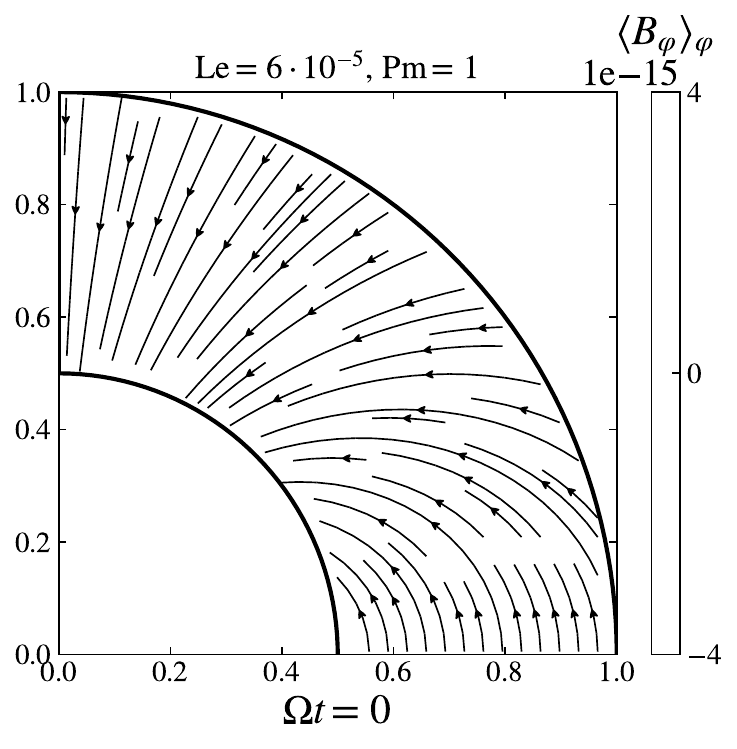}
    \includegraphics[width=0.345\linewidth]{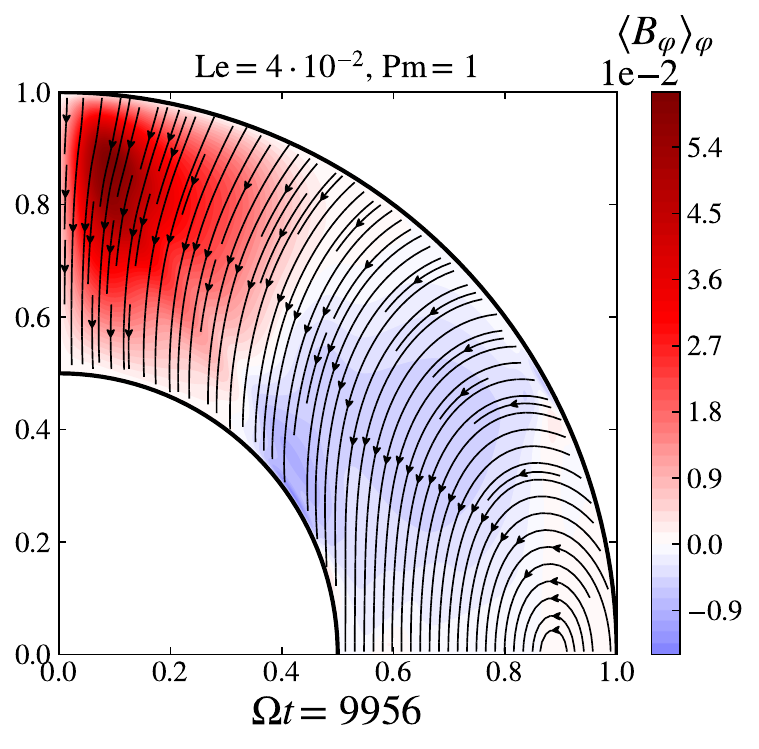}
    \includegraphics[width=0.35\linewidth]{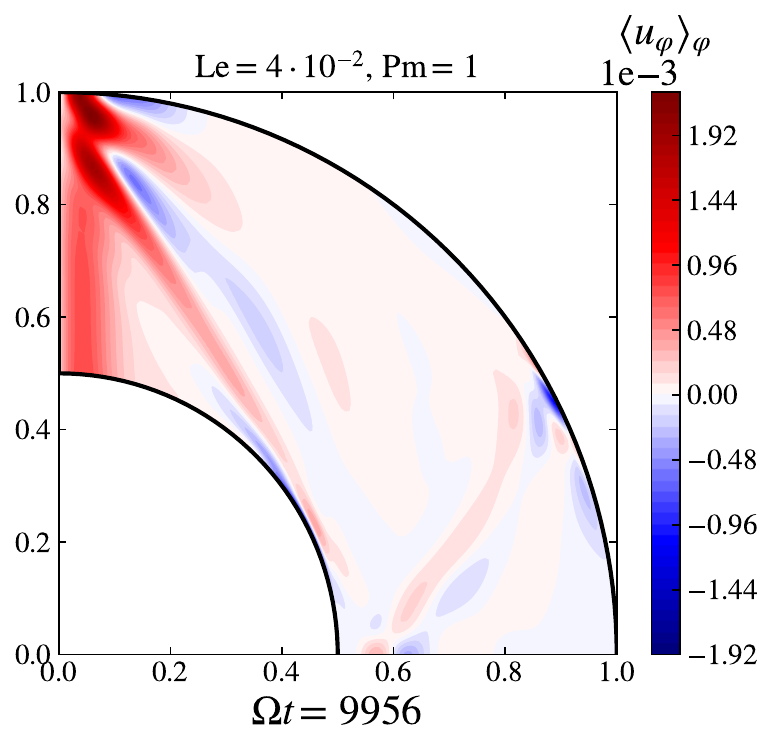}
    \includegraphics[width=0.33\linewidth]{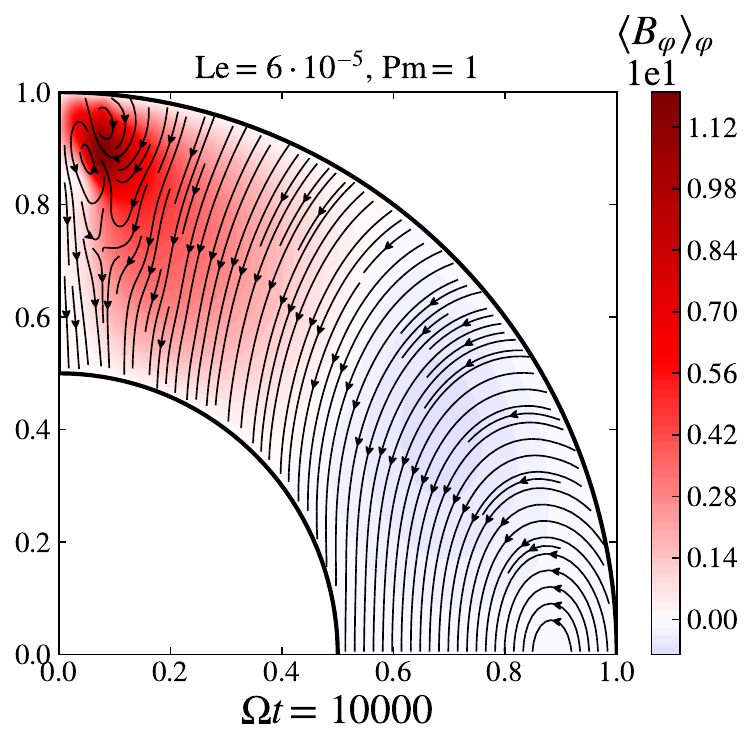}
    \includegraphics[width=0.33\linewidth]{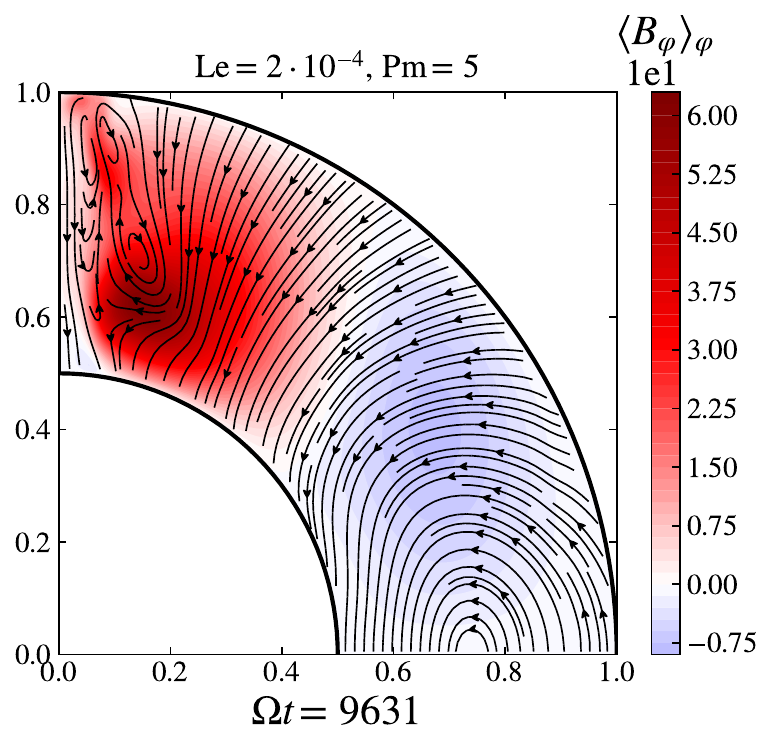}
    \includegraphics[width=0.33\linewidth]{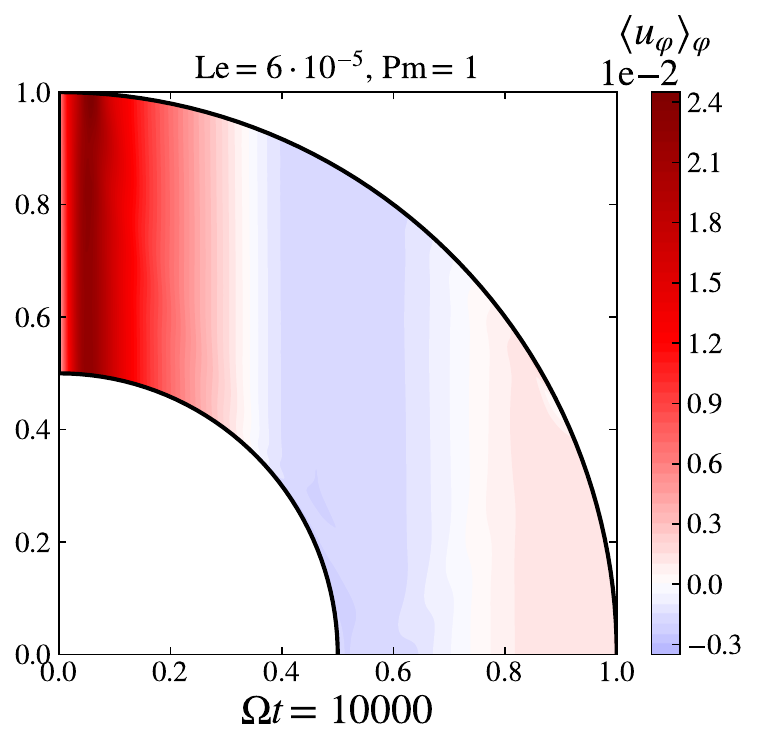}
    \caption{\textit{First and second columns:} Meridional, in the $(r,\theta)$ plane, magnetic field lines (in black, with arrows), along with the azimuthally-averaged azimuthal magnetic field, $\langle B_\varphi\rangle_\varphi$, at $t=0$ (top left) and at the times indicated in three different simulations. $\langle B_\varphi\rangle_\varphi$ is predominantly antisymmetric about the equator so we have only plotted one quadrant. \textit{Upper Left:} Initial dipolar magnetic field lines in all simulations. \textit{Right column:} Azimuthally averaged azimuthal velocity (symmetric about the equator) $\langle u_\varphi\rangle_\varphi$ in two simulations at the times indicated. The zonal flow for $\Pm=5$, $\Le=2\cdot10^{-4}$ cannot be distinguished from the flow shown here for $\Le=6\cdot10^{-5}$, $\Pm=1$ (that is why the former is not shown here).
    }
    \label{fig:phiavg}
\end{figure*}
\subsection{MagIC code and ranges of parameter values}
\label{sec:magic}
We solve the system of equations (\ref{eq:sys}) with the 3D pseudo-spectral MHD code MagIC\footnote{\url{https://magic-sph.github.io/}} (version 6.2). We set the Ekman number to $\Ek=10^{-5}$, the radial aspect ratio to $\alpha=0.5$, the tidal frequency to $\omega=1.1$ (motivated by former studies as discussed for the aspect ratio in the beginning of Sect. \ref{sec:model}), and finally the tidal forcing amplitude $\ct=10^{-2}$, unless otherwise stated. This choice allows us to explore the impact of magnetism on tidal flows in the convective shell for a tidal frequency relevant for inertial wave excitation, and we explore variations in the Lehnert number $\Le$ in the range $[10^{-5},1]$, and the magnetic Prandtl number $\Pm$ in the range $[10^{-1},5]$. The range of $\Le$ covers weakly magnetised cases with $\Le=10^{-5}$ and cases that are strongly magnetised with $\Le\gtrsim 0.1$. It can be compared with the range $\Le\in [10^{-4},10^{-1}]$ computed in solar-type convective envelopes from the pre-main sequence (PMS) to the end of the main sequence (MS) by \citet{AM2019} using a 1D stellar evolution code and scaling laws to estimate the overall amplitude of the magnetic field, as reported in Table \ref{tab:Le} for $M=1M_\odot$ \footnote{Values of $\Le$ are found to be of the same order of magnitude for other masses with $M=0.6,~0.8,~1.2~M_\odot$.}.
$\Le$ increases as the star gets older, mainly because the star slows down its rotation.
Smaller values with $\Le\lesssim 10^{-4}$ are expected in giant planets like Jupiter \citep[using values, and the magnetostrophic scaling law that is likely to provide an upper bound, in e.g.][]{WH2017,AM2019}, and $\Le\lesssim 10^{-3}$ in spin-synchronised hot Jupiters \citep[also using values from Fig.~11 of][]{dVBH2023}. Realistic values of $\Pm$ in stars are typically $\mathcal{O}(10^{-2})$ using microscopic diffusion coefficients, and can be even smaller in giant planets \citep[even $\mathcal{O}(10^{-5})$, e.g.][where $\Em$ can be estimated between $10^{-13}$ and $10^{-11}$ in the convective region of a Jupiter-like model]{FB2012}. However, if $\Ek$ is considered to represent a turbulent viscosity in mixing-length theory, we might expect $\eta$ to also represent a turbulent value, in which case $\Pm$ may be $O(1)$ \citep[e.g.][]{K2020}. Note that $\ct=10^{-2}$ is the approximate value for the tide in a hot Jupiter orbiting a solar-type star in one day, or for the tide in a solar-type binary star also with an orbital period of one day.

\begin{table}\centering
    \begin{tabular}{c|cc|cc}
         zone & \multicolumn{2}{c|}{BCZ} & \multicolumn{2}{c}{TCZ} \\
 age $\backslash$ ini. rot. & slow & fast & slow & fast \\
        \hline\hline 
        PMS ($t\sim10^7$ yr) & $10^{-2}$ & $4\cdot10^{-3}$ & $2\cdot10^{-3}$ & $8\cdot10^{-4}$ \\
        beg. MS ($t\sim10^9$ yr) & \multicolumn{2}{c|}{$2\cdot10^{-2}$} & \multicolumn{2}{c}{$4\cdot10^{-3}$} \\
        end MS ($t\sim10^{10}$ yr) & \multicolumn{2}{c|}{$4\cdot10^{-2}$} & \multicolumn{2}{c}{$5\cdot10^{-3}$}
    \end{tabular}
\caption{Values of the Lehnert number $\Le$ at different ages in a $1M_\odot$ star at the base (BCZ) and the top (TCZ) of the convective envelope for slow and fast initial rotation from \citet{AM2019}. These values adopt the magnetostrophic regime to estimate a mean magnetic field \citep[this is likely to provide an upper limit on $\Le$ and agrees best with observations for low-mass stars over other scaling laws discussed in][]{AM2019}.}
\label{tab:Le}
\end{table}

Our simulations are usually run for times $t\gtrsim 10\, 000$, corresponding to more than $0.1$ global viscous times, which is usually sufficient to reach a time-averaged steady state (for the tidal power and dissipation, although the magnetic field continues to slowly decay). We use a CNAB2 scheme for time integration with an adaptive timestep ($\mathrm{d}t$) satisfying a Courant–Friedrichs–Lewy (CFL) condition, which is no larger than $\mathrm{d}t=10^{-2}$ to guarantee adequate time resolution. We adopt a Chebyshev collocation method in the radial direction and spherical harmonics in the horizontal directions in MagIC. The spatial resolution used in the simulations varies from case to case, from $n_{r,\mathrm{max}}=97$ (number of radial grid points), $l_\mathrm{max}=213$ (maximum spherical harmonic degree), and $m_\mathrm{max}=10$ (maximum spherical harmonic order) for the less demanding simulations at low Lehnert numbers $\Le$, up to $n_{r,\mathrm{max}}=193$, $l_\mathrm{max}=341$, and $m_\mathrm{max}=100$ for the more demanding simulations at high $\Le$ and/or high magnetic Prandtl numbers $\Pm$.  
To guarantee adequate (horizontal and radial) spatial resolution, we used an empirical ``rule of thumb" that ensures that there is at least 3 orders of magnitude of difference between the peak of the energy spectrum and the energy in the highest resolvable wavenumbers, both for horizontal (spherical harmonic degree $\ell$ and azimuthal order $m$) and radial directions (in terms of Chebyshev spectral coefficients) when the simulation reaches an approximate steady state.

\section{Numerical results varying $\Le$ for $\Pm=1$}\label{results}
\label{sec:res}
Our simulations begin with an aligned dipole field (Eq.~\ref{eq:B0}) in the absence of flow, with the tidal forcing being switched on at $t=0$. Tidal waves are excited and subsequently interact with the magnetic field, which gradually decays on a long timescale because of Ohmic diffusion. The resulting dynamics is due to a competition between tidal forcing, hydrodynamic nonlinearity and viscous damping, as well as the interaction of tidal flows with the magnetic field through Lorentz forces, and the modification of the field by both the flows and Ohmic diffusion. We are particularly interested in the effects of the field on tidal flows (both tidal waves and zonal flows) -- and the modifications of the resulting tidal dissipation with magnetic fields -- but also in studying the modification of the field by the tidal flows. 

\subsection{Evolution of the magnetic field and differential rotation for weaker fields}
\begin{figure*}
    \centering
    \includegraphics[width=0.49\textwidth]{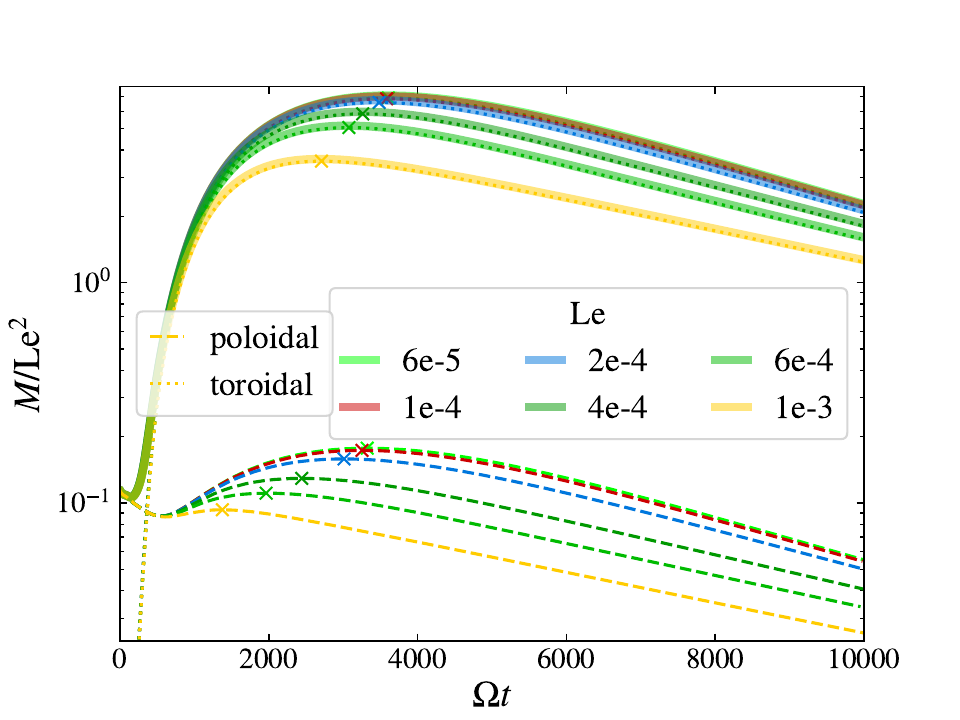}
    \includegraphics[width=0.49\textwidth]{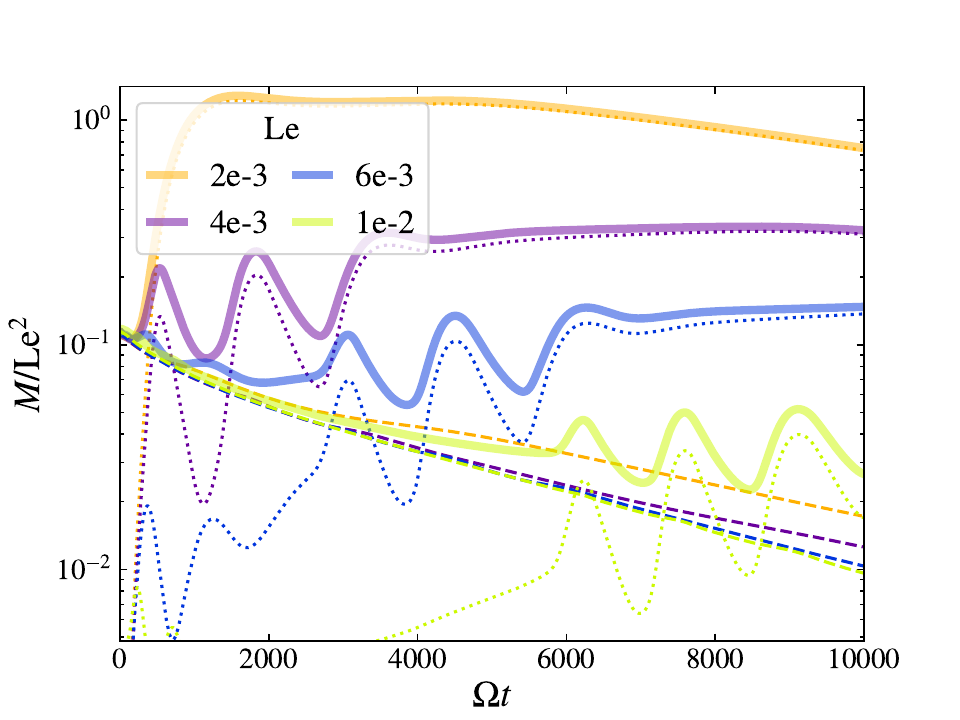}
    \caption{Total, poloidal, and toroidal magnetic energy normalised by the Lehnert number squared ($M/\Le^2$) for $\Pm=1$ in solid, dashed, and dotted lines, respectively. Each cross indicates a local maximum of either the poloidal or toroidal magnetic energy for each simulation. 
    \textit{Left:} For low initial Lehnert numbers for which the tidally-driven zonal flow is not destroyed. \textit{Right:} For higher initial Lehnert numbers for which the tidally-driven zonal flow is inhibited even shortly after the start of the simulation.
    }
    \label{fig:M_t_Pm1}
\end{figure*}
\begin{figure*}
     \centering
     \includegraphics[width=0.49\linewidth]{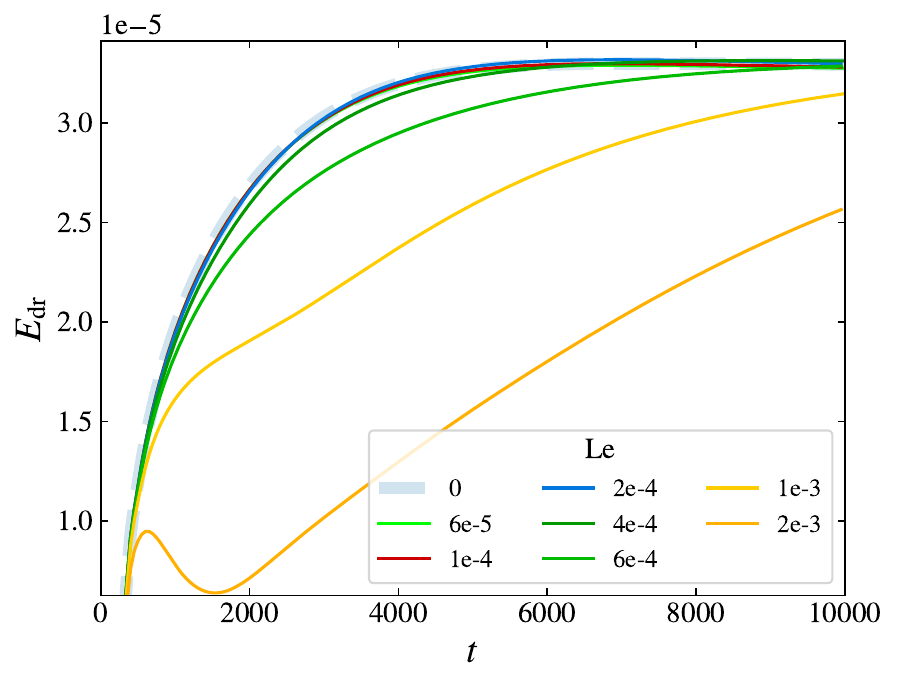}
     \includegraphics[width=0.49\linewidth]{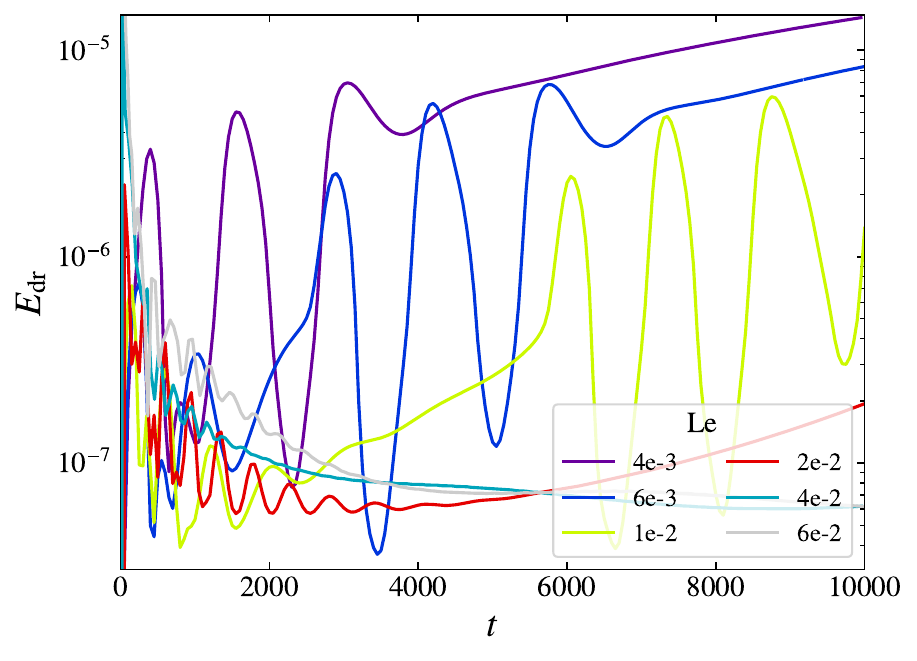}
     \caption{Energy in the differential rotation $\edr$ against time for low (\textit{left}) to high (\textit{right}) initial Lehnert numbers $\Le$ (in different colours). The hydrodynamical evolution of $\edr$ (i.e.~for $\Le=0$) is indicated in the pale dashed line in the left plot for reference.}
     \label{fig:edr_t}
 \end{figure*}

We first explore the impact of a weak dipolar magnetic field on tidal wavelike flows, setting the initial Lehnert number $\Le\in[10^{-5},10^{-3}]$ and $\Pm=1$. The initial dipolar magnetic field lines in the meridional plane are shown in the top left panel of Fig.~\ref{fig:phiavg}, with the other panels in this figure showing the meridional (equivalently axisymmetric poloidal) magnetic field lines and mean azimuthal components at the specified times in simulations with different $\Le$, together with examples of the zonal flows in the right panels (a weak field case in the bottom right panel and a strong field case in the top right). In Fig.~\ref{fig:M_t_Pm1}, we display the evolution of the poloidal and toroidal components of the magnetic energy, normalised by the initial squared Lehnert number. For early times $t\lesssim 500$, the magnetic field is primarily poloidal (as imposed by the initial conditions), before a toroidal component grows and becomes dominant. The latter is produced by the cylindrical differential rotation (zonal flows) --  created by non-linear self-interactions of inertial waves \citep[see][AB22]{T2007,FB2014} -- stretching poloidal field lines, i.e., by the so-called $\Omega$-effect \citep[e.g.][]{M1978,S1999} which we explore in detail below. The differential rotation is found to be stronger for weaker magnetic fields, as we observe in the left panel of Fig.~\ref{fig:edr_t}. This figure shows the energy in the differential rotation $\edr$ and its tendency to evolve towards the hydrodynamical value ($\Le=0$) for smaller $\Le$.
For simulations with $\Le\leq2\cdot10^{-3}$, these zonal flows efficiently stretch the poloidal magnetic field lines to produce a toroidal magnetic field, as is shown (in colour) in the bottom right (zonal flow) and bottom left (toroidal field) panels of Fig.~\ref{fig:phiavg}. 

Since the tidally-driven zonal flow (shown in the bottom right panel of Fig.~\ref{fig:phiavg}) is axisymmetric and mainly independent of the vertical coordinate $z$ (along the rotation axis), we can write it as $\bm U(s)=\langle u_\varphi\rangle_{z,\varphi}\bm e_\varphi=s\delta\Omega(s)\bm e_\varphi$ with $s=r\sin\theta$ the cylindrical radius, and where the azimuthal velocity is vertically and azimuthally averaged. By ignoring Ohmic diffusion in Eq. (\ref{eq:ind}), the interaction of the initial magnetic field $\bm B_0$ with the zonal flow induces an axisymmetric toroidal magnetic field $\bm B_\Omega$ satisfying: 
\begin{equation}
\begin{aligned}
    \partial_t\bm B_\Omega&=\bn\wedge(\bm U\wedge\bm B_0)\\
    &=\bm e_\varphi\,s(\bm B_0\cdot\bm\nabla)\delta\Omega(s)=\frac{3}{2} B_0^r s\sin\theta\,\bm e_\varphi\partial_s\delta\Omega(s),
    \label{eq:omeffect}
\end{aligned}
\end{equation}
using $B_0^r=-\left(\frac{\alpha}{r}\right)^3\cos\theta=2B_0^\theta\cot{\theta}$ from Eq. (\ref{eq:B0}).
It is not straightforward to verify this quantitatively in simulations in which the tidally-driven zonal flow first develops before winding up the initial magnetic field. Thus, we have tested this mechanism by also restarting a hydrodynamical simulation (described in AB22) with a steady zonal flow with $\omega=0.2$ (for which we had suitable hydrodynamical data, but the mechanism is the same for $\omega=1.1$ except that the zonal flows take a different form) and $\ct=5\cdot10^{-2}$, 
by injecting a dipolar magnetic field (Eq. (\ref{eq:B0})) and solving Eq. (\ref{eq:sys}) from an initial time (relative to the hydrodynamic initial state) $t=10000$.
\begin{figure*}
    \centering
    \includegraphics[width=0.48\linewidth]{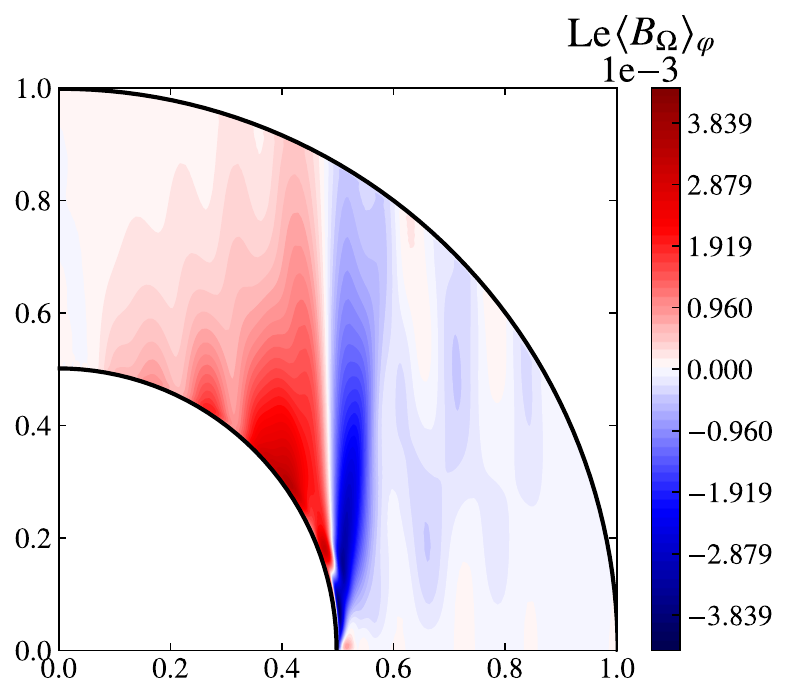}
    \includegraphics[width=0.48\linewidth]{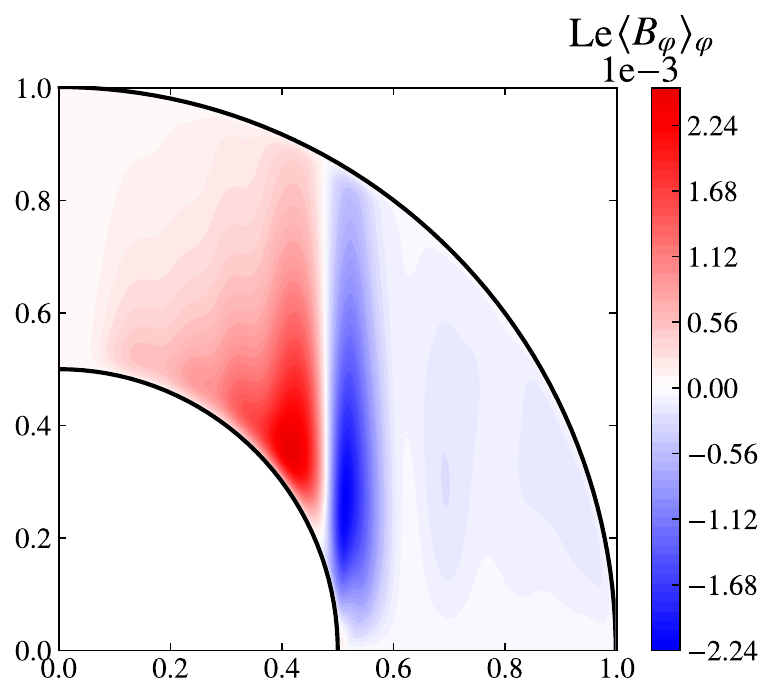}
    \caption{Azimuthal average of the toroidal magnetic field produced by predictions of the $\Omega$-effect $\langle B_\Omega\rangle_\varphi$ (\textit{left}) and azimuthal magnetic field $\langle B_\varphi\rangle_\varphi$ (\textit{right})
    at $t=1050$ from the outputs of a hydrodynamical simulation (from AB22) restarted at $t=10^4$ with an initial dipolar magnetic field with $\Le=10^{-3}$, $\Pm=2$, and $\omega=0.2$. \textit{Left:} $\langle B_\varphi\rangle_\varphi$ computed using the right hand side of Eq. (\ref{eq:omeffect}), $\times50$ for time integration.
    \textit{Right:} the colour range shown is the same as in the left panel (though the values taken slightly differ).
    }
    \label{fig:omeffect}
\end{figure*}
From the $\varphi$-averaged snapshots shown in Fig.~\ref{fig:omeffect}, it is clear that the $\Omega$-effect explains the amplitude and the structure of the azimuthal magnetic field early in the simulation. The snapshot after $50$ rotation periods (right panel) matches quite well with the prediction $\bm B_\Omega$ derived from Eq. (\ref{eq:omeffect}). 

\begin{figure}
    \centering
    \includegraphics[width=\linewidth]{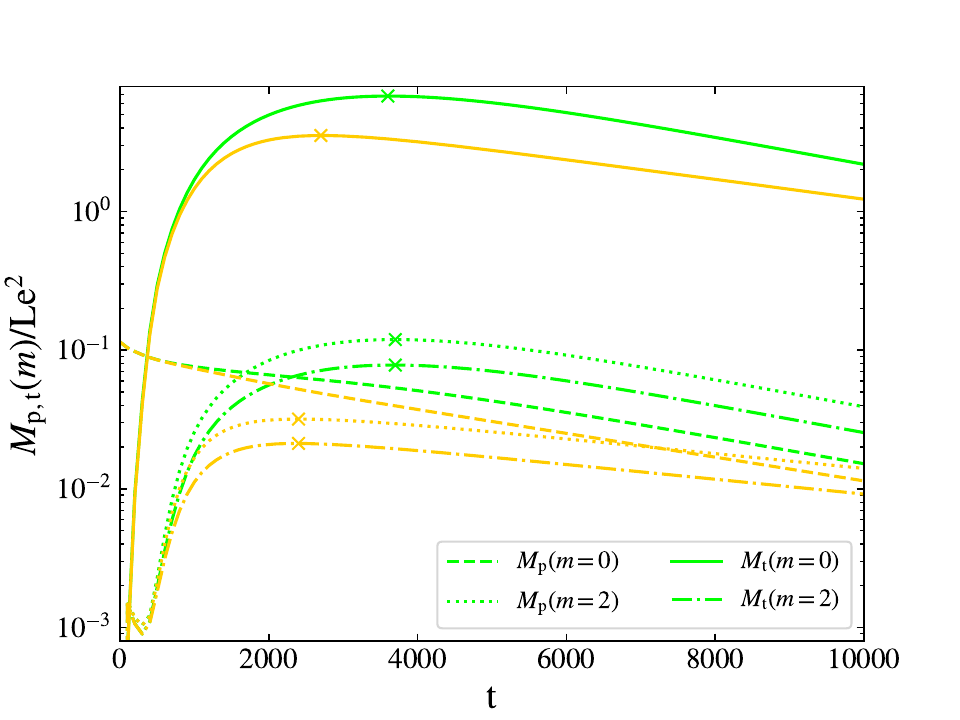}
    \caption{Evolution of the poloidal and toroidal  magnetic energies $M_\mathrm{p}$ and $M_\mathrm{t}$ (normalised by $\Le^2$) for various azimuthal wavenumber components, the $m=0$ (axisymmetric) and $m=2$ (non-axisymmetric) components, for two simulations with $\Le=6\cdot10^{-5}$ (in green) and $\Le=10^{-3}$ (in yellow). Crosses indicate maxima.}
    \label{fig:M_t_m}
\end{figure}

The time taken for differential rotation to build up a toroidal magnetic field of the same magnitude as the poloidal magnetic field can be estimated from Eq. (\ref{eq:omeffect}). We refer to it as the winding-up time, similarly as in \citet{AW2007} and \citet{JG2015}:
\begin{equation}
t_\Omega=\left(\frac{\partial\delta\Omega}{\partial\ln s}\right)^{-1}.
\label{eq:tOm}
\end{equation}
Note that, since the differential rotation becomes strong close to the poles in our simulations, $\sin\theta$ (in Eq. (\ref{eq:omeffect})) differs substantially from one, reducing the right-hand side by almost an order of magnitude. 
Taking into account this reduction, we estimate this timescale (early in the simulation) to be $t_\Omega=\mathcal{O}(10)$ in the restarted hydrodynamic case for $\Le=6\cdot10^{-5}$ and $t_\Omega=\mathcal{O}(100)$ in other simulations for $\Le\lesssim2\cdot10^{-2}$. These are consistent with the times where $\Mp=\Mt$ (see e.g.~Fig.~\ref{fig:M_t_Pm1}). 
After that (for $t\gtrsim500$), the poloidal magnetic energy increases, possibly due to the $m=2$ wavelike flow stretching the newly created axisymmetric toroidal magnetic field, as suggested in Appendix \ref{sec:ind} and by Fig. \ref{fig:phiavg} (bottom left panel), where poloidal magnetic field lines are significantly modified close to the poles where the toroidal magnetic field is strong.  
The resulting poloidal magnetic field should then have a quadrupolar component, which is strongly corroborated by Fig.~\ref{fig:M_t_m} that depicts evolution of the dominant magnetic energy components\footnote{The next strongest component $m=4$ \citep[due to super-harmonics at $2\omega$,][]{AB2022} is at least one order of magnitude lower than $m=0$ or $m=2$ throughout these simulations.}. Indeed, we observe a strong increase in $\Mp(m=2)$ and $\Mt(m=2)$ shortly after $t\sim500$. Similarly, the rise of the latter may result from the action of the tidal flow on $\bm B_\mathrm{p}(m=0)$.
From $t=1700$, the amplitude of $\Mp(m=2)$ for $\Le=6\cdot10^{-5}$ exceeds that of $\Mp(m=0)$ (which is decaying), and peaks at $t=3700$ (as does $M_\mathrm{t}(m=2)$), 
later than $\Mt(m=0)$ which peaks around $t=3600$, supporting our inference that the wavelike flow $\bm u_\mathrm{w}(m=2)$ acts on $\bm B_\Omega$
to create $\bm B_\mathrm{p}(m=2)$.
As a result, after a few thousand rotation units, the poloidal magnetic energy is dominated by its quadrupolar component. This observation, coupled with the fast decay of $\Mp(m=0)$, explains why the total poloidal magnetic energy $\Mp\approx\Mp(m=0)+\Mp(m=2)$ increases and reaches a maximum earlier, around $t=3300$ in Fig.~\ref{fig:M_t_Pm1}, while $\Mt$, which is dominated by its axisymmetric component from the beginning, peaks later, still around $t=3600$. 

It is difficult to predict the maximum amplitude reached by the axisymmetric toroidal magnetic field, especially in simulations where both magnetism and differential rotation are evolving on similar timescales\footnote{For comparison, we observe that in the restarted hydrodynamical simulation with $\omega=0.2$, the maximum toroidal magnetic energy is twice the maximum of $\Mt$ taken in the non-restarted simulation for $\Le=6\cdot10^{-5}$.}. The saturation of the toroidal magnetic field amplitude could result from a complex interplay between Ohmic diffusion, magnetic tension from the Lorentz force (this is very weak for $\Le=6\cdot10^{-5}$ but is appreciable for larger $\Le$) and the production of axisymmetric and quadrupolar poloidal and toroidal components of the magnetic fields. 
We estimate the shear lengthscale (i.e., the lengthscale over which the rotation varies substantially) close to the poles, where the zonal flow is produced, to be 
\begin{align}
\ls=(\partial\ln\Omega/\partial s)^{-1},
\end{align}
which takes the value $\ls=\mathcal{O}(10^{-1})$ for $\Le\lesssim10^{-3}$, where $\Omega=1+\delta\Omega$. For this lengthscale, the Ohmic diffusion timescale is $t_\eta=\ls^2/\Em=\mathcal{O}(10^3)$, which is on the same order as the timescale for the saturation of the magnetic field. However, the root mean square (RMS) amplitude of the Ohmic diffusion term is one order of magnitude lower than the RMS induction terms in Eq.~(\ref{eq:ind}), so Ohmic diffusion is unlikely to be solely responsible for saturating $\Mt$.

On the other hand, the Alfvén timescale, which quantifies the time taken for magnetic perturbations to propagate energy out of the shear region (defined later), varies from $\tap\sim10^5$ ($\tap\sim10^4$) for $\Le=10^{-5}$, to $\tap\sim10^3$ ($\tap\gtrsim10^2$) for $\Le=10^{-3}$, down to $\tap\gtrsim10$ ($\tap\gtrsim10$) for $\Le=6\cdot10^{-2}$ when estimated at $t\approx100$ (at $t\approx500$). It will be shown that $\tap$ is relevant to explain the transition between regimes with strong or magnetically inhibited zonal flows, but it is too long to explain the saturation of $\Mt$ occurring at a few thousand rotation units in simulations with low Lehnert numbers. Moreover, we have to account for decay of the poloidal magnetic field due to Ohmic diffusion, and for the newly created toroidal magnetic field $\bm B_\Omega$ being converted into $\Bp(m=2)$ by the stretching effects of the quadrupolar tidal flow (as explained before and in App. \ref{sec:ind}).
These additional effects may reduce overall production of toroidal magnetic energy. Finally, the evolution of the strength of the zonal flow, with a peak of $\edr$ 
around $t=7000$ for $\Le=6\cdot10^{-5}$ (see Fig. \ref{fig:edr_t}), is also likely to play a role. In the restarted hydrodynamical simulation, where the zonal flow is nearly steady, both poloidal and toroidal magnetic energies saturate a bit sooner (after $2000$ rotation units) than in the non-restarted simulations (after nearly $4000$ rotation units).

\begin{figure}
    \centering
    \includegraphics[width=\linewidth]{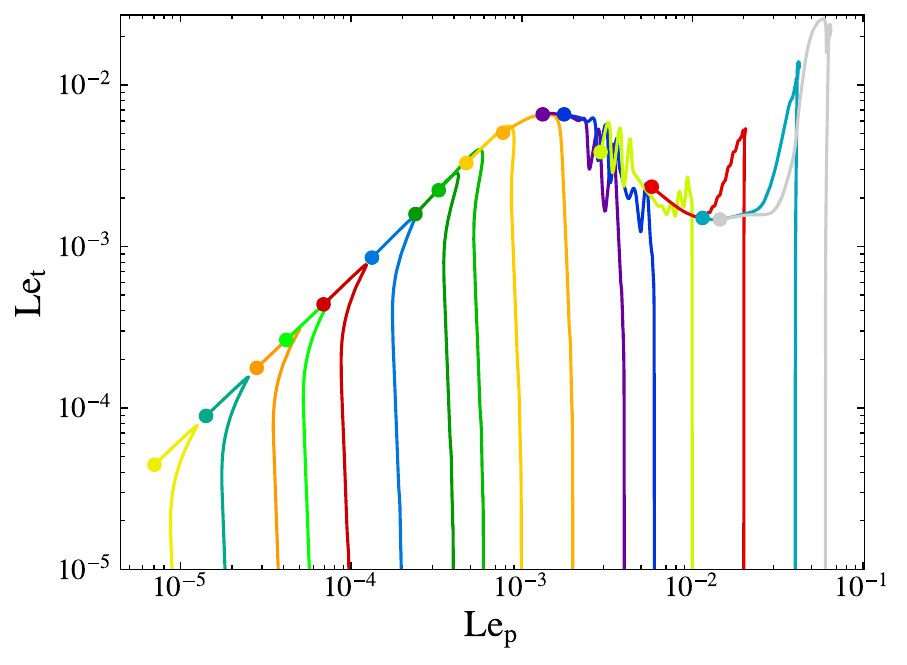}
    \caption{$\Let(t)$ versus $\Lep(t)$ for simulations with $\Pm=1$ having different initial Lehnert numbers $\Le$ (with $\Lep(t=0)=\Le$) in different colours. Bullet points indicate the values reached at the end of each simulation. The left of the figure where $\Let\propto \Lep$ indicates a kinematic regime where Lorentz forces are weak. 
    }
    \label{fig:lep_let}
\end{figure}

As we decrease the initial Lehnert number, magnetic energies converge towards an asymptotic limit that seems to be reached for $\Le\approx6\cdot10^{-5}$, with magnetic energy evolutions for lower $\Le$ being identical when rescaled by $\Le^2$ (omitted from the figure). This implies a kinematic regime for low enough Lehnert numbers, in which the role of Lorentz forces becomes negligible and the tidal flow is not affected by the magnetic field any more. In this regime, the induction equation is linear in $\bm B$. This kinematic effect of the tidal flows on the magnetic field is further supported by Fig.~\ref{fig:lep_let}, which shows the relationship between the RMS toroidal and poloidal magnetic fields (Eq. (\ref{eq:lep})) as they evolve in time in each simulation. The simulations for low Lehnert number $\Le\leq10^{-3}$ are first dominated by the $\Omega$-effect due to stretching of the poloidal field by differential rotation (vertical increase of $\Let$ in Fig.~\ref{fig:lep_let}) and by advection of the newly created azimuthal field by the tidal flow to a lesser extent (slight horizontal bend toward higher $\Lep$ until a maximum is reached). The maximum toroidal magnetic field is approximately 6 times larger than the maximum poloidal magnetic field, given by the ratio $\Let/\Lep$. Once this value is reached and the differential rotation attains a steady state, $\Let$ just decays linearly due to Ohmic diffusion, i.e., $\Let\propto\Lep$ for $\Le\leq10^{-3}$.
\begin{figure*}
    \centering
    \includegraphics[width=\linewidth]{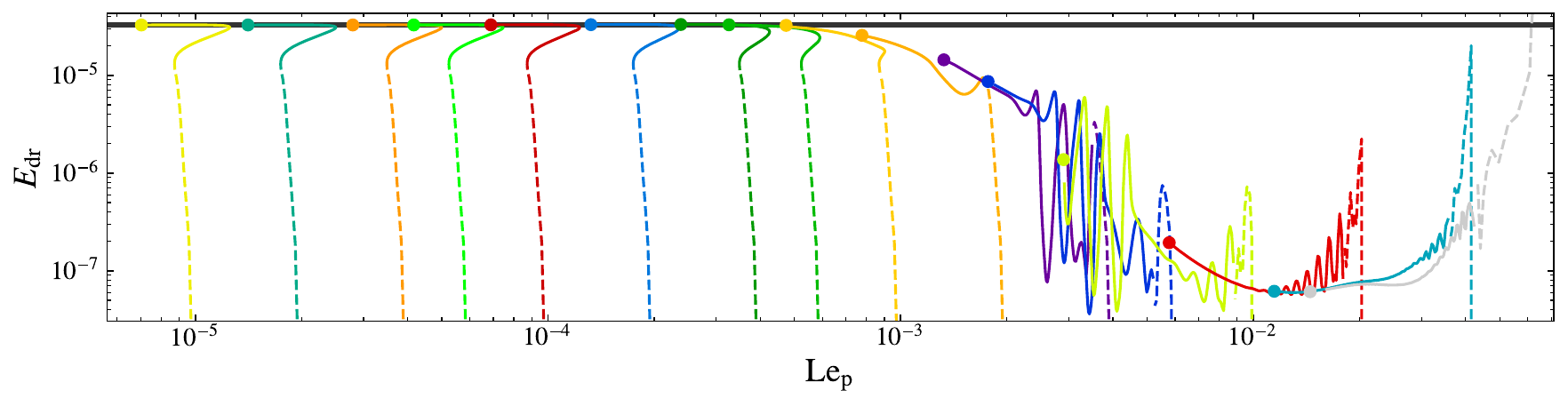}
    \includegraphics[width=\linewidth]{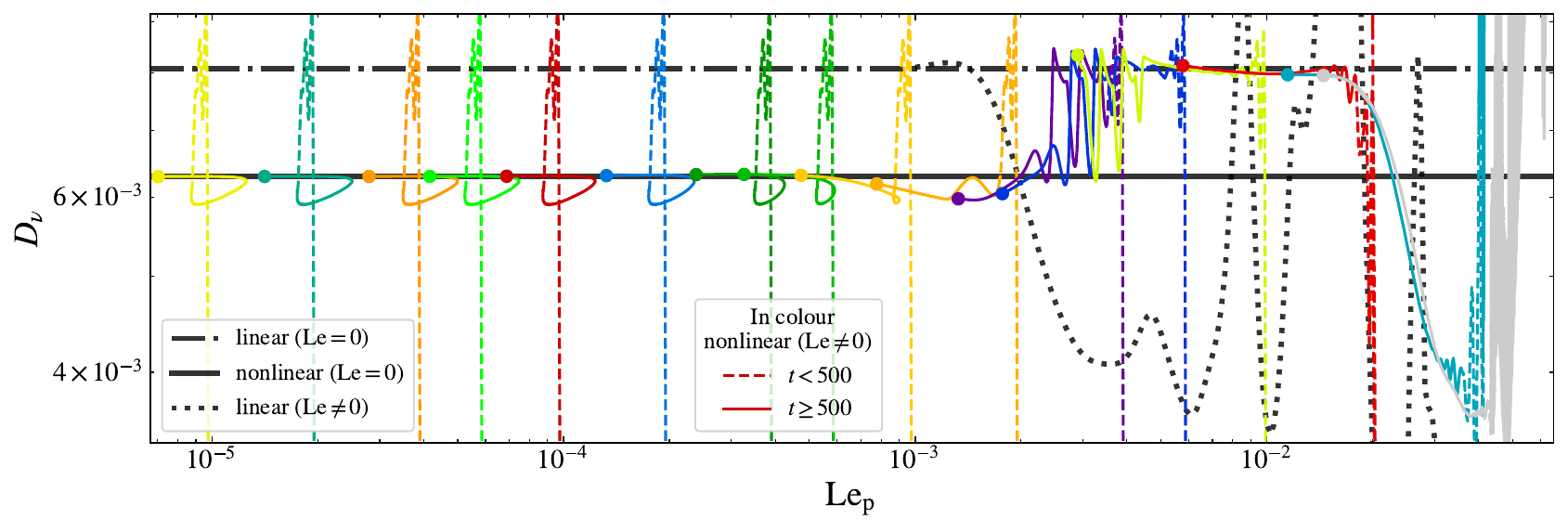}
    \caption{Energy in differential rotation $\edr$ (\textit{top}) and viscous dissipation $D_\nu$ (\textit{bottom}) against the poloidal Lehnert number $\Lep$ for $\Pm=1$ and $\Ek=10^{-5}$. Non-linear MHD simulations are shown in colour (non black), each colour referring to a simulation with a different initial Lehnert number $\Le$ (with $\Lep(t=0)=\Le$). 
    Dots indicate the values reached at the end of the simulation around $t=10^4$. Dashed lines become solid for $t\geq 500$ to indicate the evolution in time. For comparison, linear (magneto-)hydrodynamical predictions (black dotted and black dashed dotted lines) and non-linear hydrodynamical predictions (black solid lines) have been added.    
    }
    \label{fig:varLe_Pm1}
\end{figure*}

In our simulations, which do not model turbulent convective motions and the resulting dynamos, the magnetic field is not self-sustained. Hence, in the absence of fluid motions substantially maintaining (or amplifying) the field, magnetic energy is expected to decay due to Ohmic diffusion, approximately according to
\begin{equation}
    \partial_t M=-\Le^2\Em \langle (\bm\nabla\wedge\bm B)^2\rangle=-D_\eta,
    \label{eq:diff}
\end{equation}
when the induction term $\nabla\wedge\bm (\bm u\wedge\bm B$) is neglected. Using a poloidal/toroidal decomposition of the magnetic field, we can solve the induction equation to find freely decaying modes \citep[as in Appendix \ref{sec:appfm}, following][]{M1978}. For the single largest radial wavelength $l=1$ mode, we predict a decay rate for the poloidal magnetic energy (i.e. for $-\mathrm{d}\ln\Mp/\mathrm{d}t$) of approximately 
$1.79\cdot10^{-4}$.
This is quite close to the observed decay rate for the same component of the magnetic energy, which we measure to be $2.1\cdot10^{-4}$ in the last thousand time units for\footnote{We measure the same decay for a simulation with $\Le=10^{-3}$ running until $40000$.} $\Le\in[10^{-5},6\cdot10^{-2}]$. Since we observe the same decay for a comparison simulation with a lower tidal forcing amplitude $\ct=10^{-4}$, we can rule out tidal flows acting to enhance Ohmic decay, which could in principle explain the small difference in values by producing smaller magnetic length-scales. Similar values, but one order of magnitude higher, are found when $\Pm=0.1$. For simulations with $\Pm=2$ and $\Pm=5$, the predicted decay rates are, respectively, $8.96\cdot10^{-5}$ and $3.59\cdot10^{-5}$, while the measured decay rates are again slightly higher, $\sim 1.1\cdot10^{-4}$ and $\sim 6\cdot10^{-5}$, respectively. The small discrepancies between predicted and observed values could come from the initial dipole not being a single free decay eigenmode. It can be represented as a sum of free decay eigenmodes (with different $k_\alpha$'s, using the notation of Appendix \ref{sec:appfm}), since the set of these forms an orthogonal basis for a single $l,m$. Thus, we would not expect a single free decay eigenmode to exactly explain the observed decay rate because the field in our simulations is a superposition of these with different decay rates. Nevertheless, the approximate agreement between predicted and observed values indicates that in most of our simulations the field is decaying Ohmically, and is not being sustained by, or subject to enhanced (turbulent) diffusion by, the tidal flows.

\subsection{Evolution for stronger fields, angular momentum fluxes and torsional waves}

\begin{figure*}
    \centering
    \includegraphics[width=0.49\linewidth]{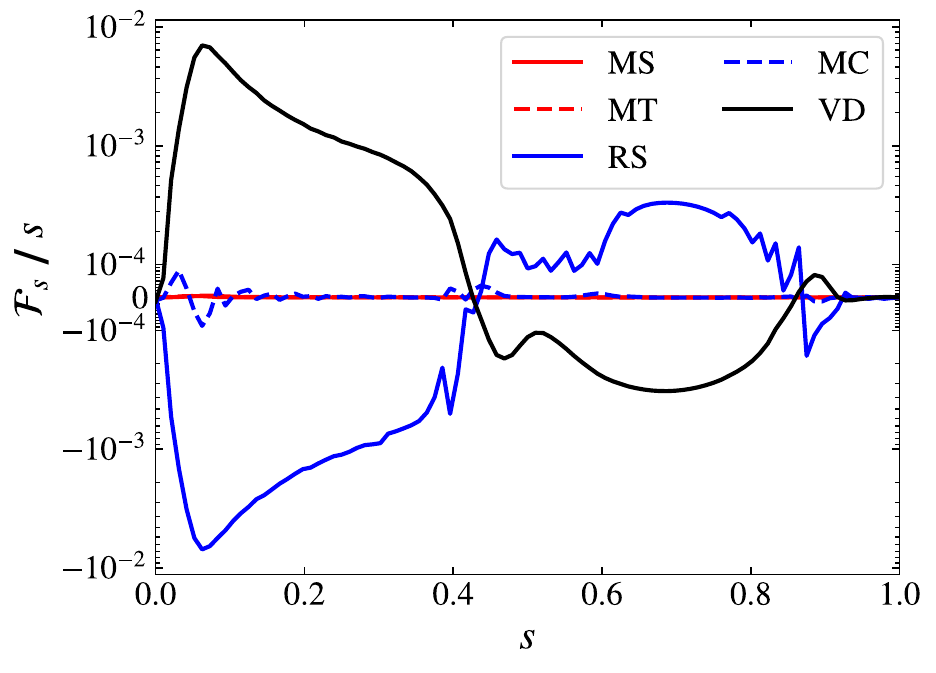}
    \includegraphics[width=0.49\linewidth]{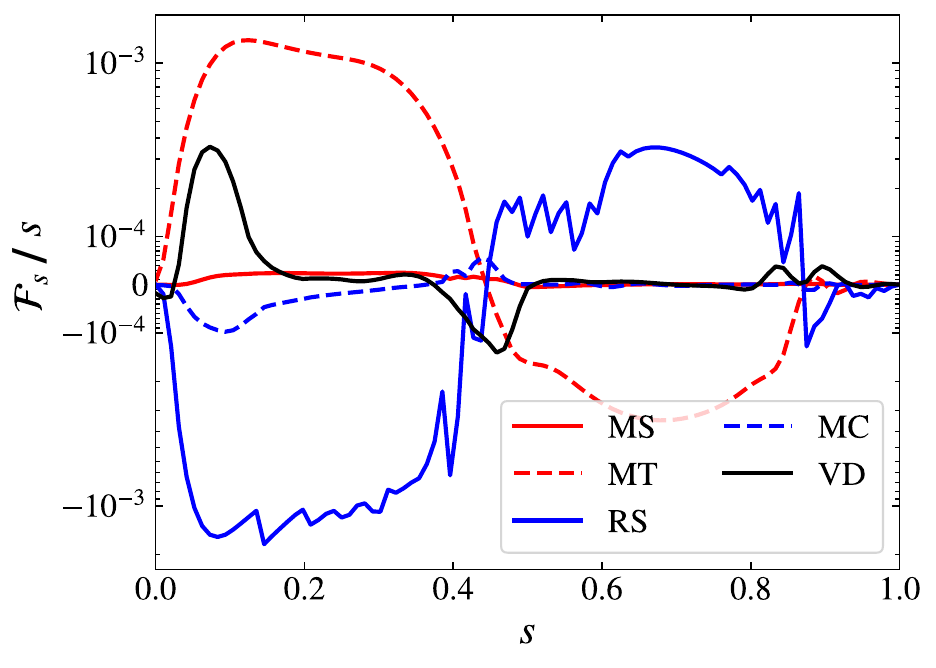}
    \includegraphics[width=0.49\linewidth]{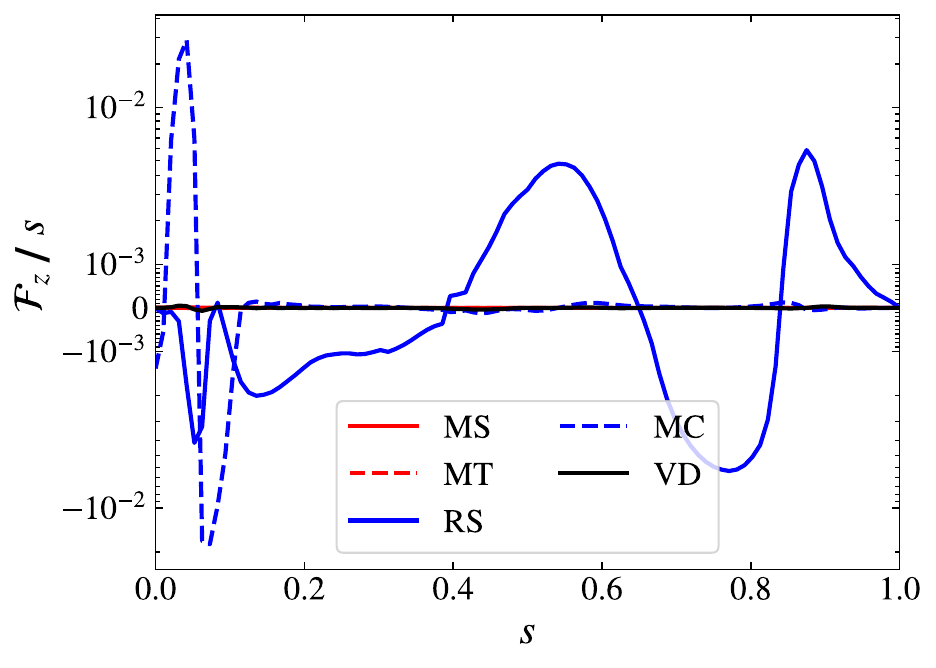}
    \includegraphics[width=0.49\linewidth]{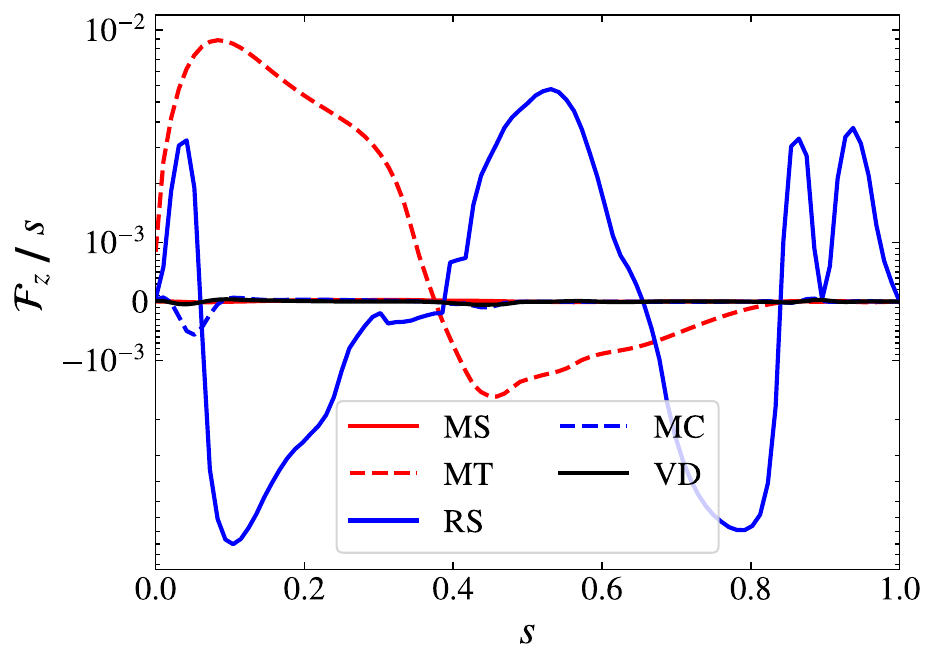}
    \caption{Angular momentum flux contributions within the cylindrical radial $\mathcal{F}_s/s$ (\textit{top}) and vertical $\mathcal{F}_z/s$ (\textit{bottom}) components, which are time integrated over the last $3000$ time units and integrated along $z$ over the northern hemisphere only (because $\mathcal{F}_z$ is anti-symmetric about the equator, while $\mathcal{F}_s$ is symmetric). We show Maxwell stresses (MS), magnetic torques (MT), Reynolds stresses (RS), meridional circulations (MC), and viscous diffusion contributions (VD). \textit{Left:} $\Le=6\cdot10^{-5}$ and $\Pm=1$. \textit{Right:} $\Le=4\cdot10^{-2}$ and $\Pm=1$.}
    \label{fig:fluxes}
\end{figure*}
For higher initial Lehnert numbers $\Le\gtrsim\Lec$, where $\Lec=3\cdot10^{-3}$ is a critical value indicating a change of regime, the differential rotation becomes substantially inhibited by the stronger initial magnetic fields within the first thousand time units. This difference between stronger field cases and those with $\Le\lesssim\Lec$ is demonstrated particularly clearly by comparing the bottom (small $\Le$) and top (large $\Le$) right panels of Fig.~\ref{fig:phiavg}. In the latter, polar zonal flows are substantially weaker and the net azimuthal flows are strongest in the (magneto-)inertial wave shear layers instead. The evolution of the differential rotation is also shown in the right panel of Fig.~\ref{fig:edr_t} (and in the upper panel of Fig.~\ref{fig:varLe_Pm1}) where $\edr$ is observed to be smaller by more than two orders of magnitude for $\Le\sim10^{-2}$ compared to $\Le\lesssim 10^{-3}$. As a result of the correspondingly weaker $\Omega$-effect, the toroidal magnetic energy is also reduced, as is evidenced by the 2D snapshots in Fig.~\ref{fig:phiavg} (when comparing the overall amplitude of $\langle B_\varphi\rangle_\varphi$ in the top middle panel with the bottom left panels), and also in volume-integrated energies in Figs.~\ref{fig:M_t_Pm1} (right panel) and \ref{fig:lep_let}. For the largest $\Le$, the poloidal magnetic energy becomes dominant over the toroidal one throughout these simulations. 

To better understand what causes the transition between regimes in which tidally driven zonal flows develop strongly or not, we derive an equation governing evolution of angular momentum. We take the azimuthal component of the momentum equation in cylindrical coordinates, multiplied by the cylindrical radius $s$, to obtain:
\begin{equation}
    \partial_t(s \widehat{u_\varphi})+\frac{1}{s}\partial_s(s\mathcal{F}_s)+\partial_z\mathcal{F}_z=0,
    \label{eq:angmom}
\end{equation}
where the $\widehat\cdot$ symbol denotes the combination of taking a $\varphi$-average, $z$-integration, and time integration (over an arbitrary time $\tau$), such that for a variable $A$, $\widehat{A}=\int^\tau\int_{z_\mathrm{i}}^{z_\mathrm{o}}\langle A\rangle_\varphi\,\mathrm{d}z\,\mathrm{d}t$, with $z_\mathrm{i}$ and $z_\mathrm{o}$ the inner and outer vertical heights of the spherical boundaries which depend on $s$. In Eq. \ref{eq:angmom}, we have introduced the cylindrical radial angular momentum flux
\begin{equation}
    \mathcal{F}_s=s\left[-\Ek\, s\partial_s(\widehat{u_\varphi}/s)+\widehat{u_s}\widehat{u_\varphi}+\widehat{u'_su'_\varphi}-\Le^2\widehat{B_s}\widehat{B_\varphi}-\Le^2\widehat{B'_sB'_\varphi}\right],
    \label{eq:Fs}
\end{equation}
and the vertical angular momentum flux
\begin{equation}
    \mathcal{F}_z=s\left[-\Ek\, s\partial_z(\widehat{u_\varphi}/s)+\widehat{u_z}\widehat{u_\varphi}+\widehat{u'_zu'_\varphi}-\Le^2\widehat{B_z}\widehat{B_\varphi}-\Le^2\widehat{B'_zB'_\varphi}\right].
\end{equation}
In these, we define the prime $'$ symbol to denote non-axisymmetric fluctuations to distinguish them from axisymmetric ones, such that $u_i=\langle u_i\rangle_\varphi+u'_i$ (since $\langle u'_i\rangle_\varphi=0$).  In both fluxes, from left to right, we have the contributions from viscous diffusion, meridional circulations, Reynolds stresses, magnetic torques, and Maxwell stresses \citep[similarly as in][in spherical coordinates]{B2004,B2008}.
Here, the magnetic torque (meridional circulation) represents the $m=0$ component of the magnetic (hydrodynamic) contributions to the angular momentum fluxes, whereas the Maxwell (Reynolds) stresses are defined to result from the non-axisymmetric components.

The different flux terms within $\mathcal{F}_s$ and $\mathcal{F}_z$ are each displayed in Fig.~\ref{fig:fluxes} for two examples, one with a low and one with a high initial dipolar magnetic field strength. We show results after $z-$integration and time integration over the last three thousand rotation units, when the simulation is approximately in a steady state but with a slowly decaying magnetic field (though at high Lehnert numbers the decay of the magnetic field influences the differential rotation strength substantially). When $\Le$ is low (left panels), strong Reynolds stresses in $\mathcal{F}_s$ (especially near the poles) generate differential rotation until they can be nearly perfectly compensated by viscous diffusion (with opposite signs), with negligible contributions from meridional circulations, Maxwell stresses and magnetic torques. This is similar to what we expect from purely hydrodynamical simulations when the zonal flow reaches a steady state \citep[see also Appendix A in][]{AB2023}, where Reynolds stresses from tidal waves balance viscous diffusion. The meridional circulation $\widehat{u_z}\widehat{u_\varphi}$, which represents the large-scale correlations of vertical and azimuthal axisymmetric components, plays a more important role near the poles for $\mathcal{F}_z$ than it does for $\mathcal{F}_s$. This suggests large-scale recirculations in the polar columnar flow, both up and down (because of the opposite signs), from the reflection of the shear layers at the rotation axis. Reynolds stresses are also important for $\mathcal{F}_z$ and peak in the outer tangent cylinder at locations where the shear layers reflect at the inner/outer boundaries or at the equator. Note that $\mathcal{F}_z$ is anti-symmetric about the equator (while $\mathcal{F}_s$ is symmetric), so performing $z$-integration over the two hemispheres would almost cancel out this component. 

\begin{figure*}
    \centering
    \includegraphics[width=0.33\linewidth]{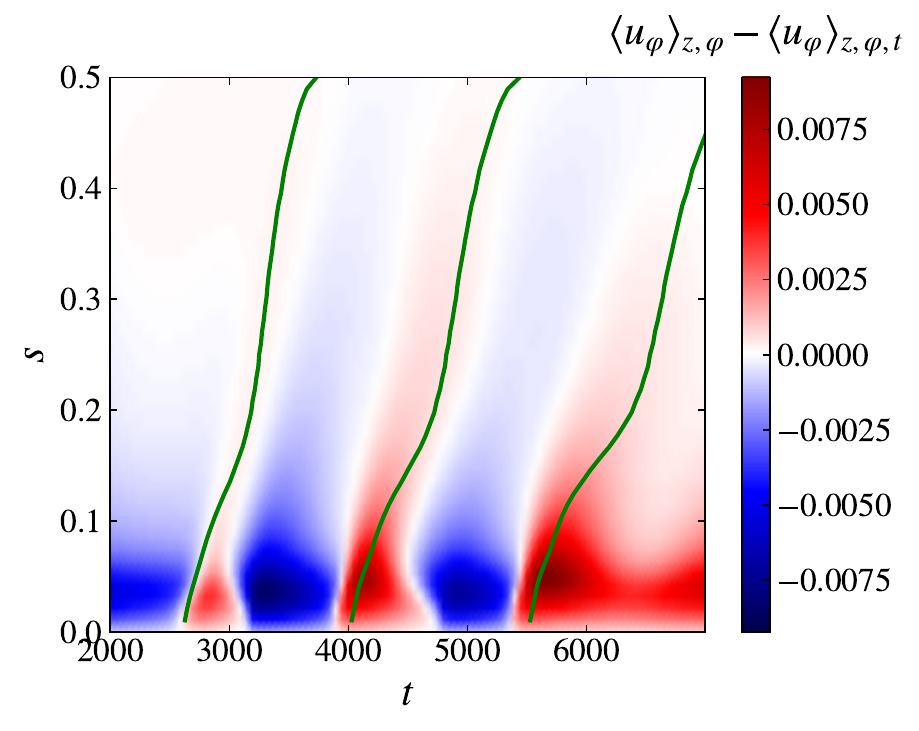}
    \includegraphics[width=0.33\linewidth]{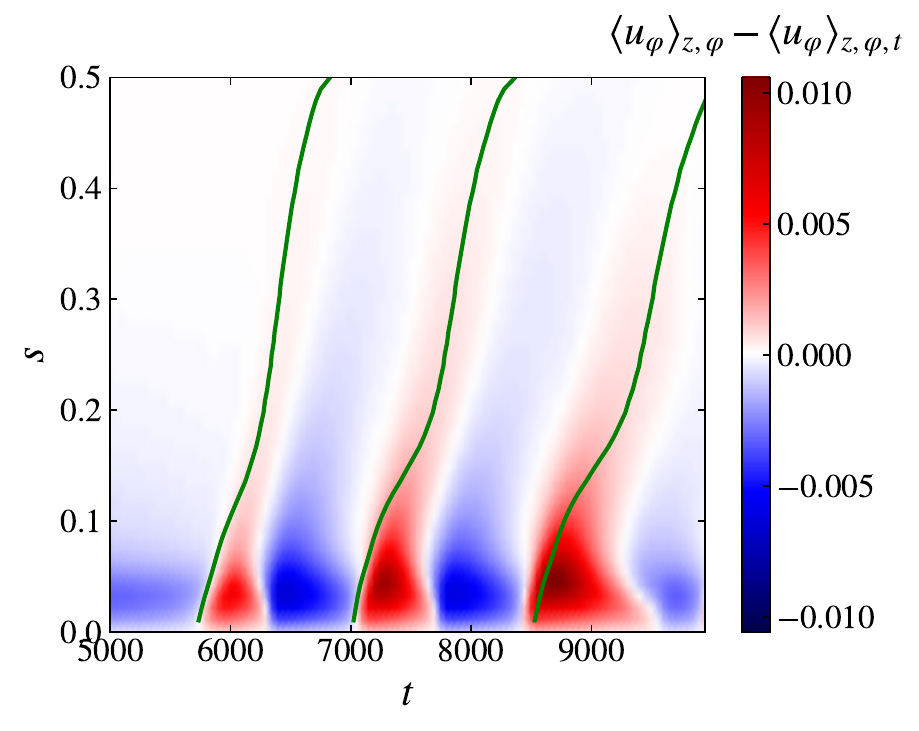}
    \includegraphics[width=0.33\linewidth]{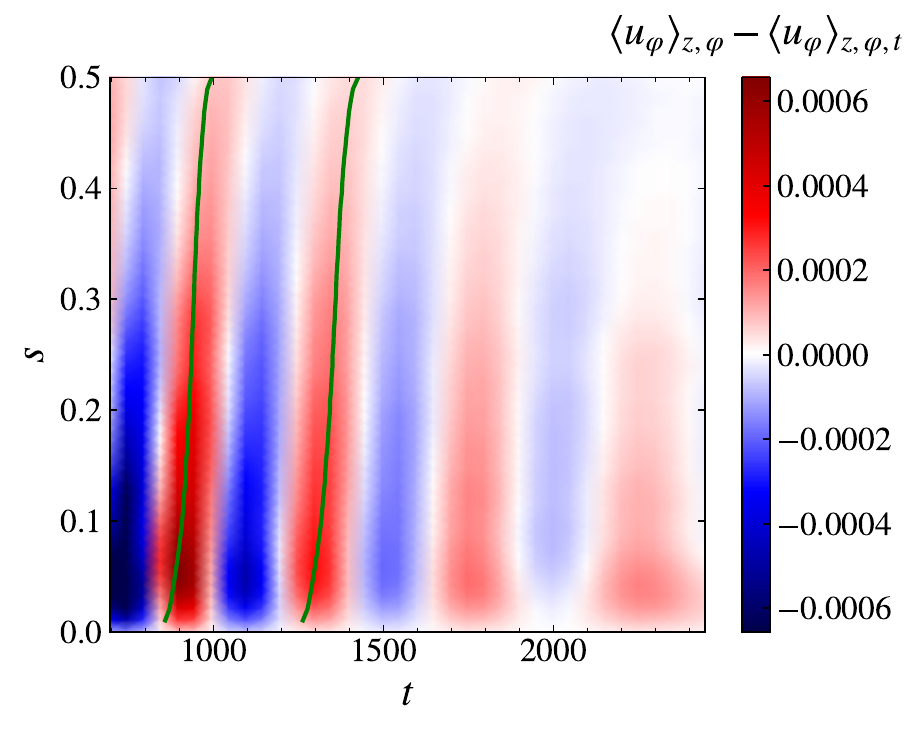}
    \caption{Amplitude of the fluctuating $z$ and $\varphi$ averaged zonal flow $\langle u_\varphi\rangle_{z,\varphi}-\langle u_\varphi\rangle_{z,\varphi,t}$ versus time $t$ and cylindrical radius $s$ for three simulations possibly exhibiting propagating torsional Alfv\'{e}n waves. The time average for the zonal flow $\langle u_\varphi \rangle_{z,\varphi,t}$ is performed over the whole time range shown in each plot where the oscillations are observed. The green curve shows the Alfvén timescale $t_\mathrm{A}$ (averaged over an approximate cycle around which $t_\mathrm{A}$ is drawn) vs $s$.
    \textit{Left:} $\Le=6\cdot10^{-3}$. \textit{Middle:} $\Le=10^{-2}$. \textit{Right:} $\Le=2\cdot10^{-2}$.}
    \label{fig:toros}
\end{figure*}

When the initial Lehnert number is high (right panels), viscous diffusion and meridional circulations are much less important than in the previous case, and Reynolds stresses are now balanced by magnetic torques. These inhibit development of zonal flows and result in much weaker differential rotation. Interestingly, Maxwell stresses (involving correlations of non-axisymmetric components) are small compared to magnetic torques here, which is the opposite of what has been found in \citet{B2004} and \citet{B2008} in their convective dynamo simulations where Maxwell stresses seem to cancel out differential rotation in some regimes.
Stronger magnetic torques may partly explain why, for $\Le\gtrsim\Lec$, tidally-driven zonal flows have more difficulty developing or are completely inhibited from the start for the highest initial Lehnert numbers. Concomitantly, we also observe periodic oscillations on long periods $\sim1500$ (or somewhat shorter) in $\edr$, $\Mt$ and $\Mp(m=2)$, early in the simulations (see the right panels of Figs.~\ref{fig:M_t_Pm1}, \ref{fig:edr_t} and \ref{fig:varLe_Pm1}). These oscillations may correspond with torsional Alfvén waves, restored by the magnetic tension component of the Lorentz force \citep[e.g. the review of][]{HN2023}. To illustrate the emergence of these waves, we compute the vertically and azimuthally averaged azimuthal velocity inside the shell, such that:
\begin{equation}
    \langle u_\varphi \rangle_{\varphi,z}=\frac{1}{h}\int_{z_\mathrm{i}}^{z_\mathrm{o}}\langle u_\varphi\rangle_\varphi\,\mathrm{d}z,
\end{equation}
from which we remove the mean background state $\langle u_\varphi\rangle_{z,\varphi,t}=1/\tau\int^\tau \langle u_\varphi\rangle_{z,\varphi}\,\mathrm{d}t$ over an arbitrary time $\tau$ (chosen to define a representative mean state) in the manner of \citet{TJ2014}. We define $h=(z_\mathrm{o}-z_\mathrm{i})/2$, $z_\mathrm{o}=\sqrt{1-s^2}$, and $z_\mathrm{i}=\sqrt{\alpha^2-s^2}$ in the inner tangent cylinder (ITC), and $z_\mathrm{i}=0$ in the outer tangent cylinder (OTC). The Alfvén time\footnote{Since $B_s$ is anti-symmetric compared to the equator, the square of it has been taken before performing the $z$ average.}, which describes the timescale for radial propagation of these waves over a distance $s$, is defined by:
\begin{equation}
    t_\mathrm{A}=\frac{s}{\Le\sqrt{\langle B_s^2\rangle_{\varphi,z,t}}}, 
    \label{eq:tA}
\end{equation}
with $B_s=B_r\sin\theta+B_\theta\cos\theta$ the cylindrical component of the magnetic field. The temporally fluctuating mean zonal flows are shown versus $s$ in Fig.~\ref{fig:toros} for three simulations with increasing initial Lehnert numbers that possibly exhibit torsional waves. The fluctuations of the zonal flows as a function of $s$ and $t$ are nicely explained by the variation of the Alfvén timescale inside the ITC (Eq. (\ref{eq:tA})), bending towards higher $s$ and $t$, when averaging over one cycle around different initial times. This suggests that the oscillatory and wave-like nature of the zonal flows in these simulations is likely to result from the propagation of torsional Alfv\'{e}n waves. These appear to be excited near the polar regions and to subsequently propagate outwards where they are primarily dissipated, rather than being reflected to form torsional (standing-mode) oscillations.

We observe that the zonal flow oscillation cycle is longer for lower initial Lehnert numbers, for example when comparing the timescale of the first oscillation for $\Le=6\cdot10^{-3}$ (about $1000$ rotation units) and $\Le=2\cdot10^{-2}$ (about $400$). It also increases with time in each panel since the poloidal magnetic field, therefore $B_s$, decreases due to Ohmic diffusion (for instance, it is about $1500$ for the $2^\mathrm{nd}$ cycle for $\Le=6\cdot10^{-3}$). For $\Le=10^{-2}$, similar timescales are found as for $\Le=6\cdot10^{-3}$ since torsional oscillations are triggered later in the simulations (see the right panel in Fig.~\ref{fig:edr_t}), so $\Lep$ (and so $B_s$) are of the same amplitude as can be seen in Fig.~\ref{fig:varLe_Pm1} (top panel).
It is interesting to see that for $\Le=10^{-2}$ (and for $\Le=6\cdot10^{-3}$), the amplitude of the magnetic torques, viscous diffusion and meridional circulations vary with the sign of the fluctuating zonal flows $\langle u_\varphi\rangle_{z,\varphi}-\langle u_\varphi\rangle_{z,\varphi,t}$, which we illustrate in Fig.~\ref{fig:flux_osc}: when $\langle u_\varphi\rangle_{z,\varphi}<\langle u_\varphi\rangle_{z,\varphi,t}$ (left panels) magnetic torques dominate, while when $\langle u_\varphi\rangle_{z,\varphi}>\langle u_\varphi\rangle_{z,\varphi,t}$ (right panels) viscous diffusion in $\mathcal{F}_s$ and meridional circulation in $\mathcal{F}_z$ take over close to the pole.
It is not clear whether the fast oscillations for $\Le=2\cdot10^{-2}$ are of the same nature since this cyclic trend is not observed and magnetic torques dominate for all times like in Fig.~\ref{fig:fluxes} (right panels).  
\begin{figure*}
    \centering
    \includegraphics[width=0.49\linewidth]{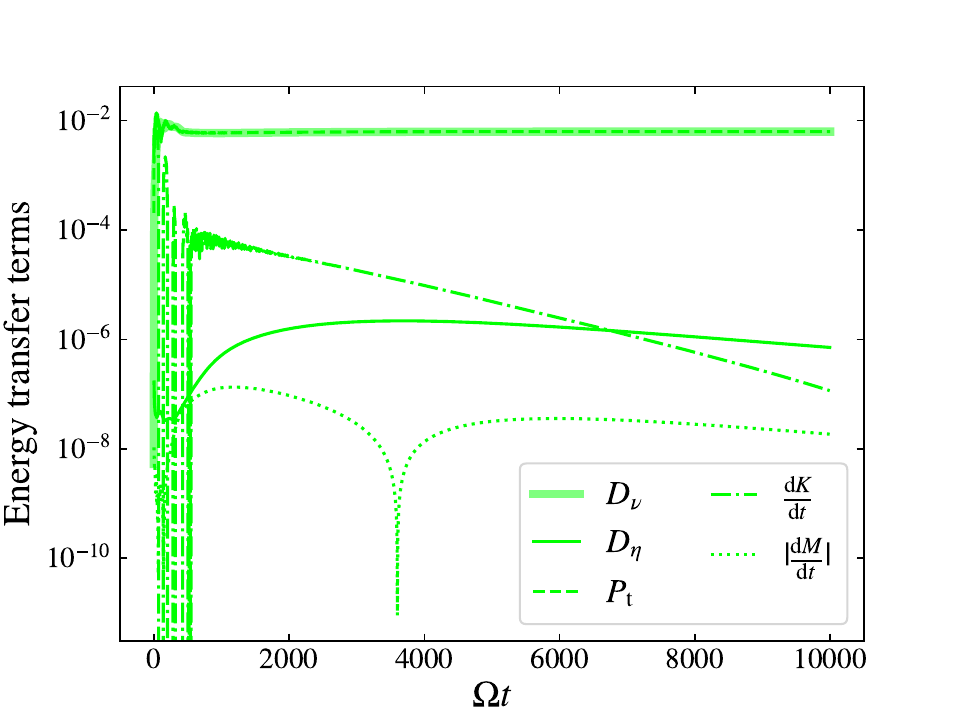}
    \includegraphics[width=0.49\linewidth]{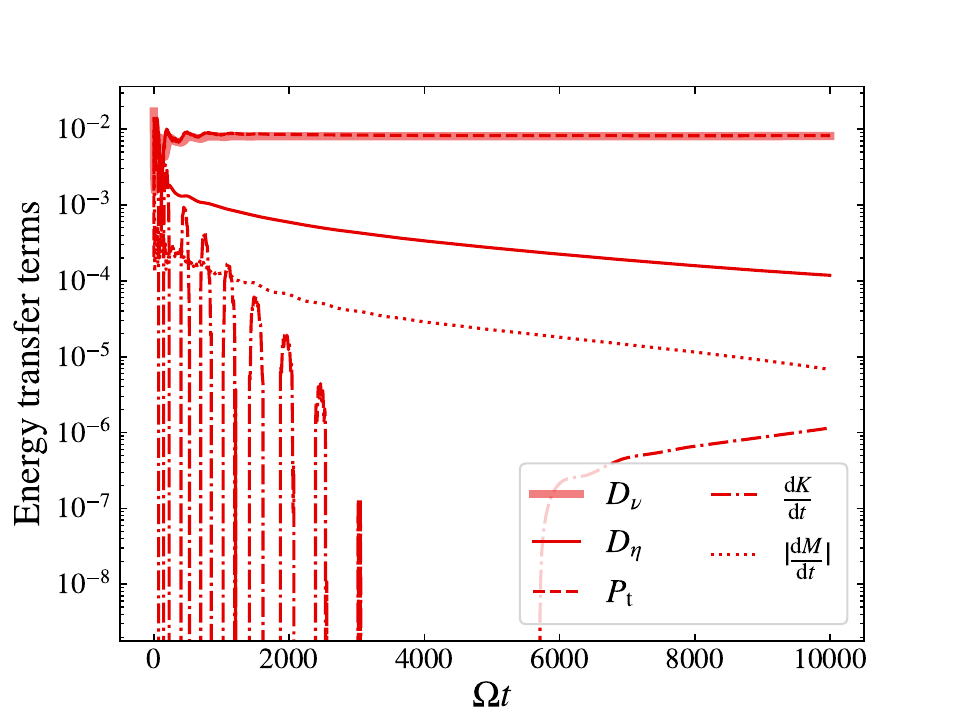}
    \caption{Time evolution of contributions to evolution of the total energy in Eq. (\ref{eq:bal}). We show, in order, viscous dissipation, Ohmic dissipation, tidal power, and time derivative of the kinetic and magnetic energies ($\Pm=1$). We omit the Poynting flux since it is found to be negligibly small in both panels. The energy equation is thus accurately satisfied in our simulations. \textit{Left:} $\Le=6\cdot10^{-5}$ (weak field). \textit{Right:} $\Le=2\cdot10^{-2}$ (strong field).}
    \label{fig:bal}
\end{figure*}
The transition between regimes where zonal flows are strong, like in hydrodynamical cases, or are substantially quenched by magnetic torques (which also corresponds with when slow torsional oscillations are observed) can be further interpreted by introducing the back-reaction timescale $t_\mathrm{ap}$ of the magnetic tension on differential rotation. We define this in a similar way (but modified) as \citet{AW2007} and \citet{JG2015}, as $t_\mathrm{ap}=\ls/v_\mathrm{ap}$, with $v_\mathrm{ap}=\sqrt{\langle B_s^2\rangle_{\varphi,z}}$ the torsional Alfvén velocity of the magnetic field in the cylindrical direction, and $\ls$ is once again the length-scale of variation of the differential rotation. 

For $\Le<10^{-3}$, the back-reaction timescale is long compared to both the winding-up timescale and the Ohmic damping timescale of Alfvén waves $t_\eta$, namely $\tap\gg t_\Omega$ and $\tap\gg t_\eta$, respectively. 
This means that differential rotation has time to stretch poloidal magnetic field lines to create a strong toroidal component, while Alfvén waves have insufficient time to propagate before being damped by Ohmic diffusion. When $\Le\approx \Lec$, we measure (at $t=100$) $\tap\simeq t_\Omega\simeq t_\eta$ (all taking values around $500$). From Fig.~\ref{fig:edr_t} (left panel), we indeed note the slight perturbation of $\edr$ to set in at early times for this transitional Lehnert number. For higher initial Lehnert numbers $\Le> \Lec$, $\tap$ is smaller at a fixed time while $t_\Omega$ and $t_\eta$ both stay the same, so that $\tap\ll t_\Omega$ and $\tap\ll t_\eta$. This means that Alfvén waves have time to propagate before being damped and their large-scale axisymmetric correlations (magnetic torque) can act on differential rotation to quench it.

\subsection{Tidal dissipation rates as a function of $\Le$}

The variation in the strength of differential rotation (with $\Le$ and time) has a substantial impact on tidal viscous dissipation rates $D_\nu$, as we show in the bottom panel of Fig.~\ref{fig:varLe_Pm1}. 
For low poloidal Lehnert numbers $\Lep$, since $\edr$ is very close to the hydrodynamical prediction (in the upper panel), $D_\nu$ also matches the prediction computed with hydrodynamic ($\Le=0$) non-linear simulations (presented in AB22) when the simulation reaches a time-averaged steady state. 

On the other hand, for much higher initial Lehnert numbers $\Le>\Lec$, $D_\nu$ ends up much closer to the linear hydrodynamical prediction, since differential rotation is too weak to impact viscous dissipation in these simulations. Thus, for our set of parameters, the main ingredient controlling the magnitude of viscous dissipation is the strengths of the zonal flows, with the magnetic fields themselves playing only an indirect role on $D_\nu$.
For $\Pm=1$, the Ohmic dissipation $D_\eta$ is quite low in all simulations compared to the tidal power and viscous dissipation, which mainly balance each other, as  is shown in Fig.~\ref{fig:bal} for two different initial Lehnert numbers. In these simulations, the energy balance is well satisfied with an error at the end of the simulation of one tenth of a percent for $\Le=2\cdot10^{-2}$ (and much lower for $6\cdot10^{-5}$), which could be reduced even further by increasing the spatial (and temporal) resolution. 

For large Lehnert numbers, the values of the dissipation rates $D_\eta$ and $D_\nu$ are quite different from what is predicted by linear MHD models with an imposed dipolar magnetic field only \citep[dotted black lines in Fig.~\ref{fig:varLe_Pm1}, and][]{LO2018}. Indeed, linear viscous dissipation is predicted to be strongly affected by the presence of a magnetic field (with clear peaks and troughs) for $\Le\gtrsim\Lec$, while linear Ohmic dissipation is expected to be dominant. 
However, these linear predictions for $D_\nu$ and $D_\eta$ do not account for toroidal and non-axisymmetric components of the magnetic field, the modest differential rotation that is present, and the fact the initial dipole decays over time, which explains why $D_\eta$ drops over time in Fig.~\ref{fig:bal}, leaving $D_\nu$ as the only dominant sink of energy injected by tidal power. Early time values of $D_\nu$ and $D_\eta$ in nonlinear simulations may be closer to linear MHD predictions, but it is difficult to be definite because the time to reach steady state is almost always comparable with (or longer than) the Ohmic decay time for the imposed field.
%
\section{Dependence on the magnetic Prandtl number}
\label{sec:Pm}
\begin{figure*}
    \includegraphics[width=0.48\textwidth]{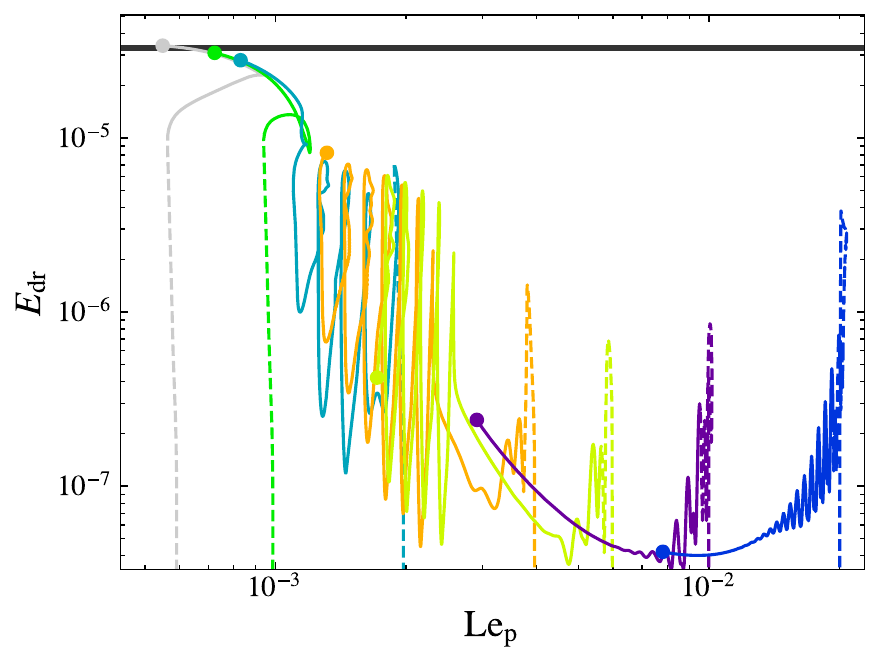}
    \includegraphics[width=0.5\textwidth]{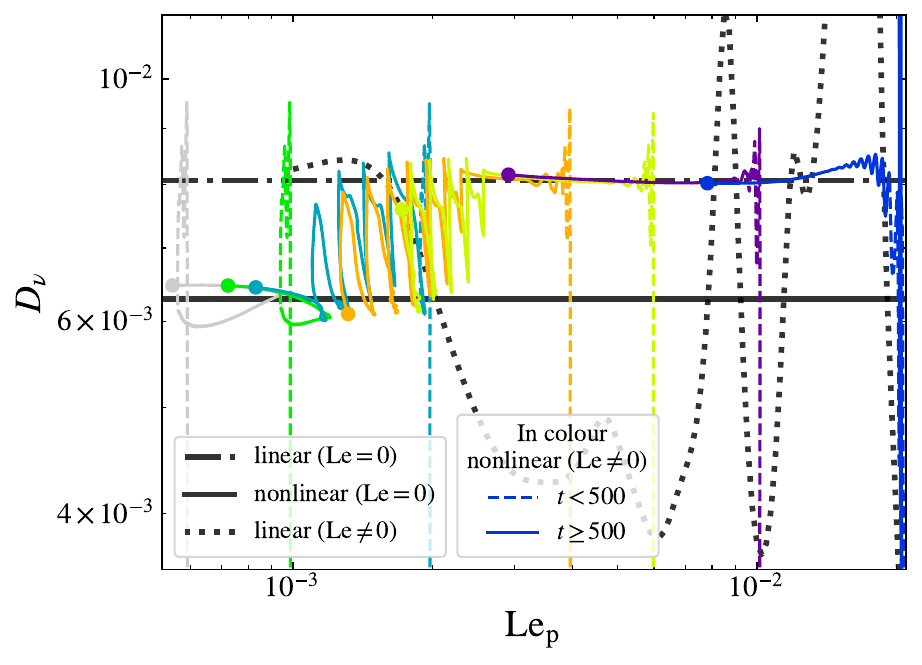}
    \includegraphics[width=0.48\textwidth]{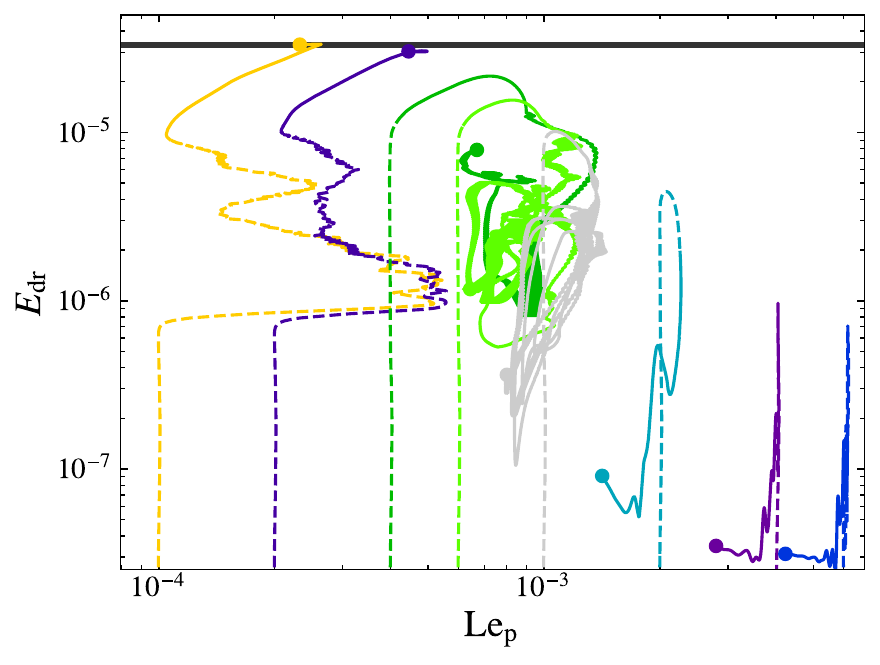}
    \includegraphics[width=0.5\textwidth]{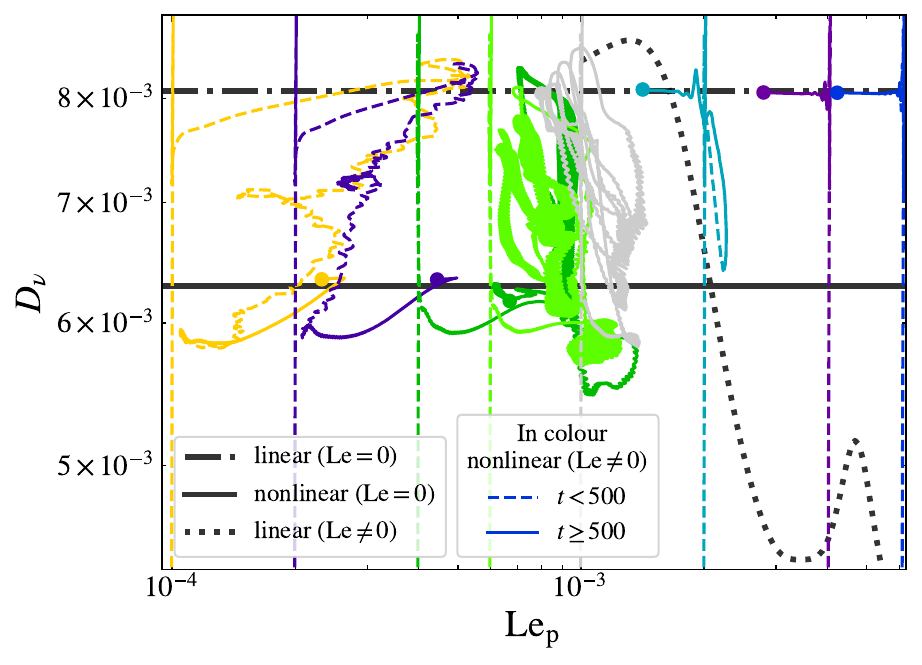}
    \includegraphics[width=0.48\textwidth]{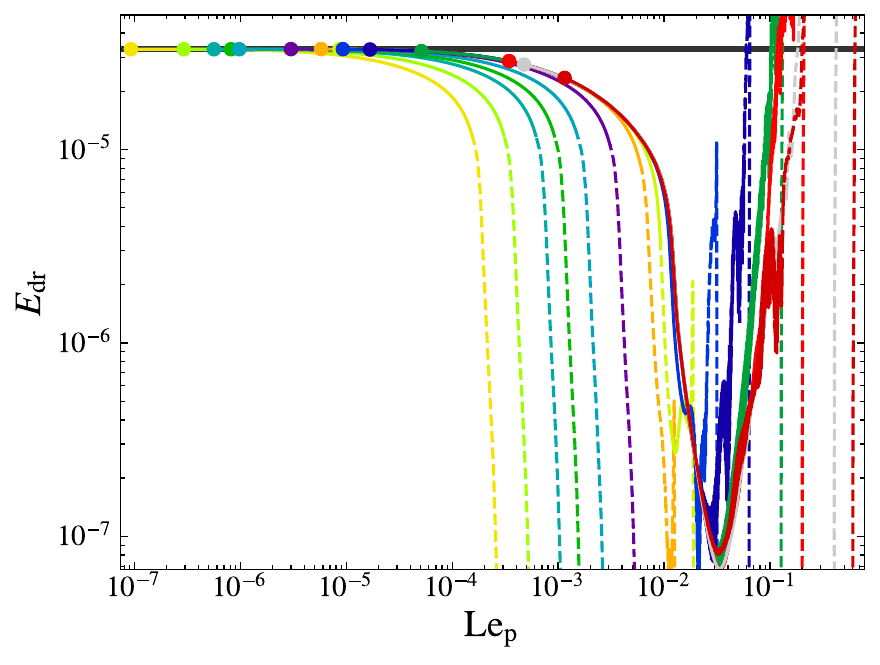}
    \includegraphics[width=0.5\textwidth]{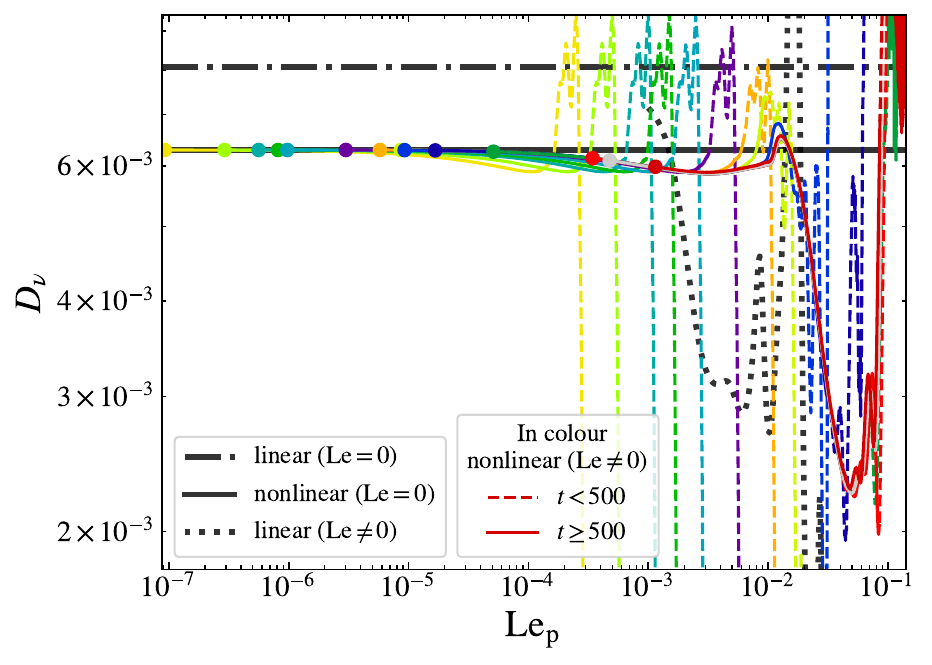}
    \caption{Same as Fig. \ref{fig:varLe_Pm1} but for $\Pm=2$ (\textit{top}, simulations ran until $t=20000$ here), $\Pm=5$ (\textit{middle}), and $\Pm=0.1$ (\textit{bottom}). 
    }
    \label{fig:varLe_Pm2}
\end{figure*}
\begin{figure}
    \centering
    \includegraphics[width=\linewidth]{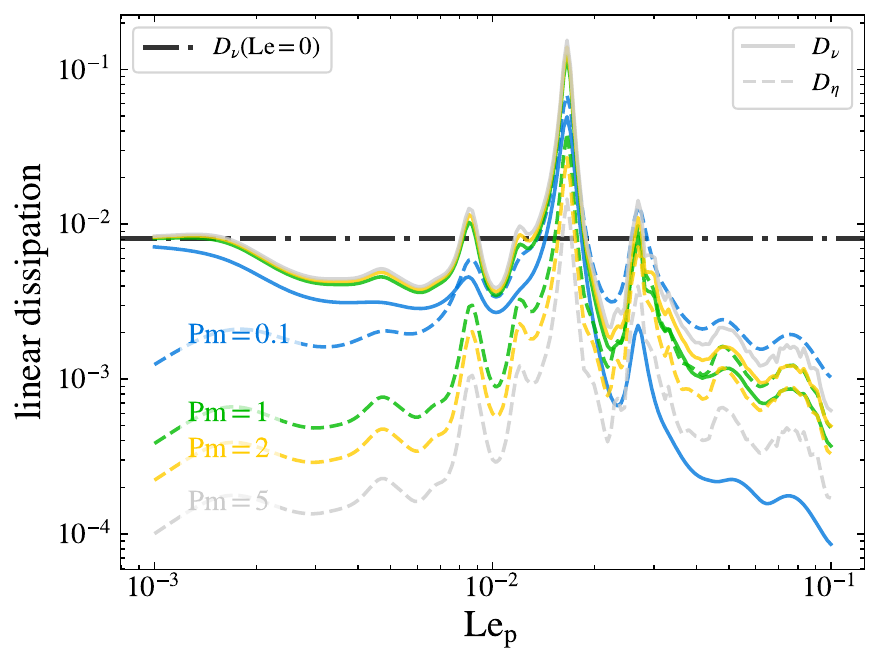}
    \caption{Viscous and Ohmic dissipation $D_\eta$ and $D_\nu$ computed in linear MHD calculations with a fixed background dipolar magnetic field as in \citet{LO2018}, versus dipolar Lehnert number $\Lep$ for various magnetic Prandtl numbers $\Pm$. Viscous dissipation from a linear hydrodynamical calculation has been put for reference in black.}
    \label{fig:MHDlinear}
\end{figure}
\begin{figure*}
    \centering
    \includegraphics[width=0.49\linewidth]{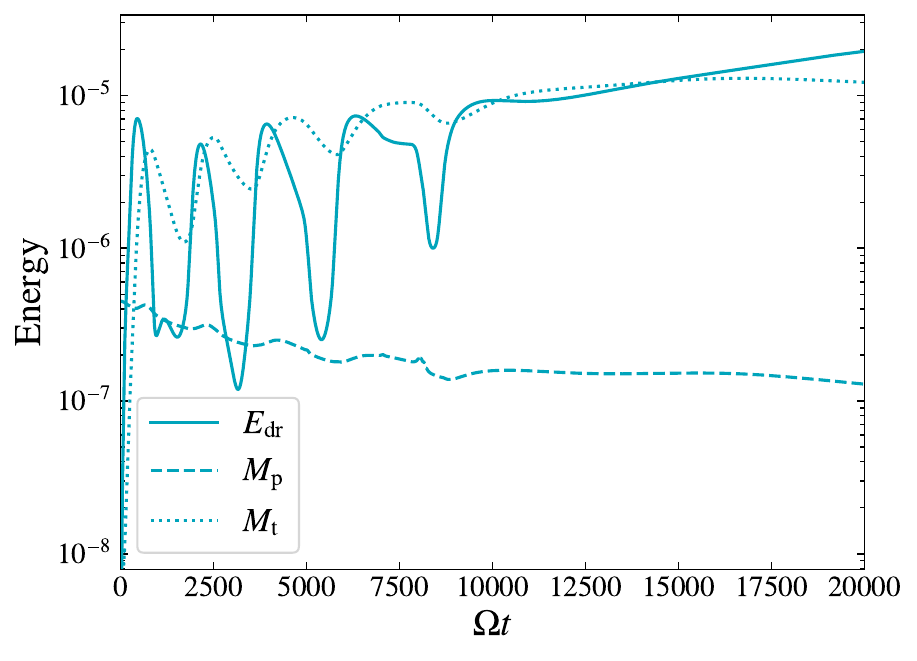}
    \includegraphics[width=0.47\linewidth]{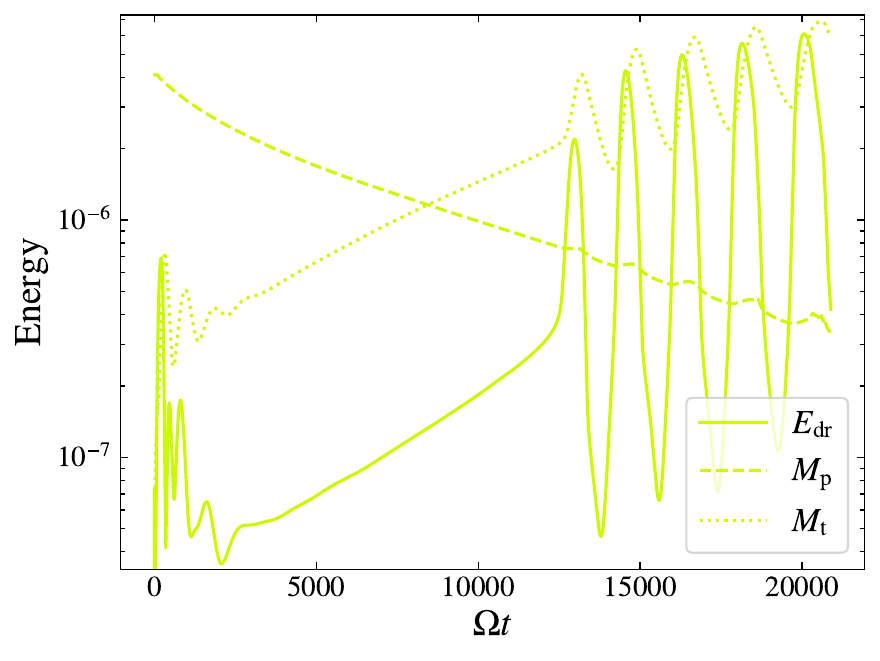}
    \caption{Poloidal $\Mp$ and toroidal $\Mt$ magnetic energies along with energy in differential rotation $\edr$ versus time for 2 simulations with $\Pm=2$ and $\Le=2\cdot10^{-3}$ (\textit{left}) and $\Le=6\cdot10^{-3}$ (\textit{right}).
    }
    \label{fig:compEn}
\end{figure*}
\begin{figure*}
    \centering
    \includegraphics[width=0.49\linewidth]{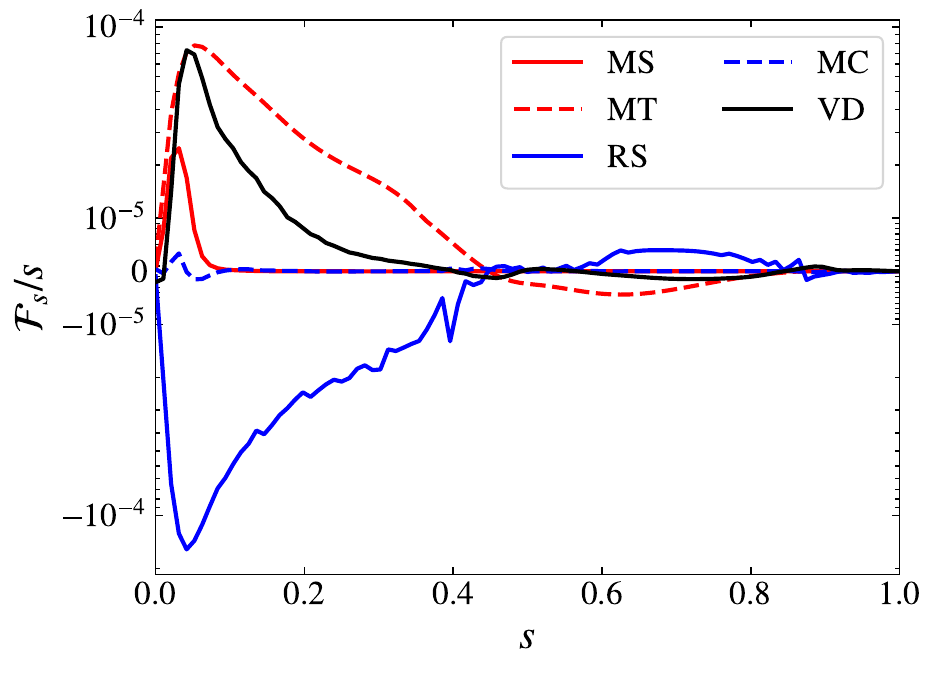}
    \includegraphics[width=0.49\linewidth]{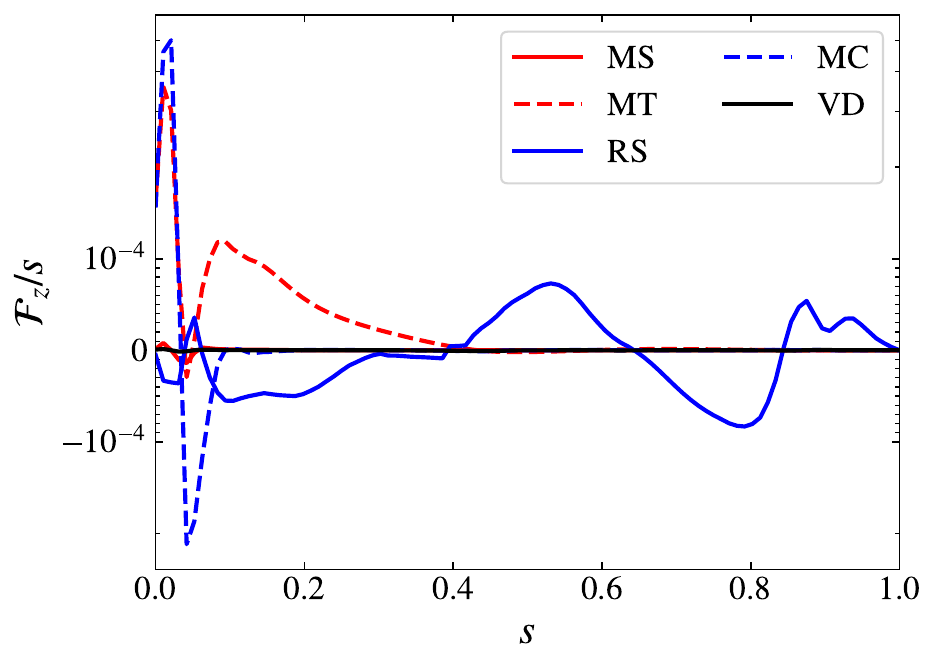}
    \caption{Flux terms within the cylindrical $\mathcal{F}_s/s$ (\textit{left}) and vertical $\mathcal{F}_z/s$ (\textit{right}) angular momentum fluxes, integrated along $z$ over the northern hemisphere only, for $\Le=2\cdot10^{-3}$ and $\Pm=2$ at $t=650$.}
    \label{fig:flux_osc_Pm2}
\end{figure*}
\begin{figure*}
    \centering
    \includegraphics[trim=0cm 1cm 0cm 0cm,clip,width=0.48\linewidth]{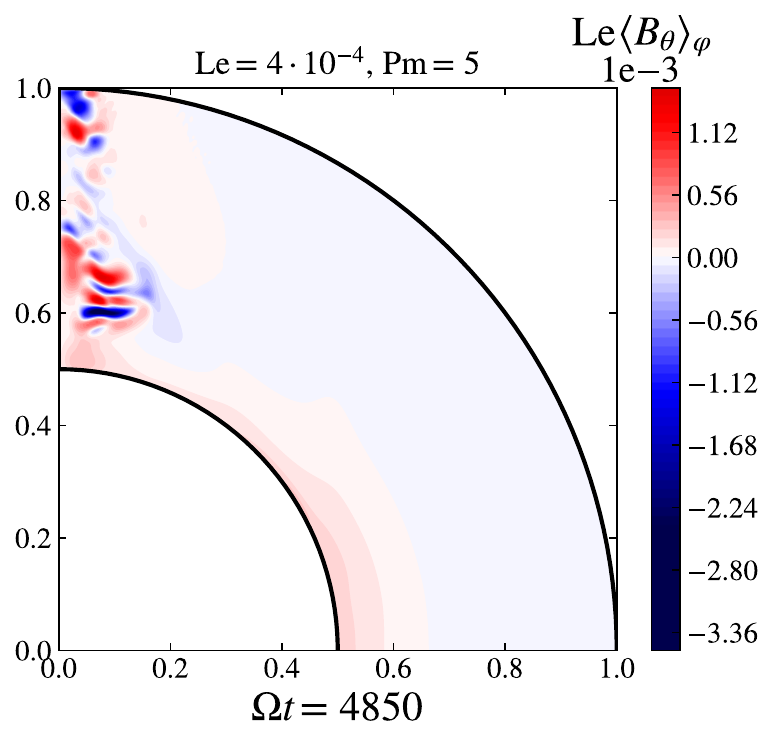}
    \includegraphics[width=0.48\linewidth]{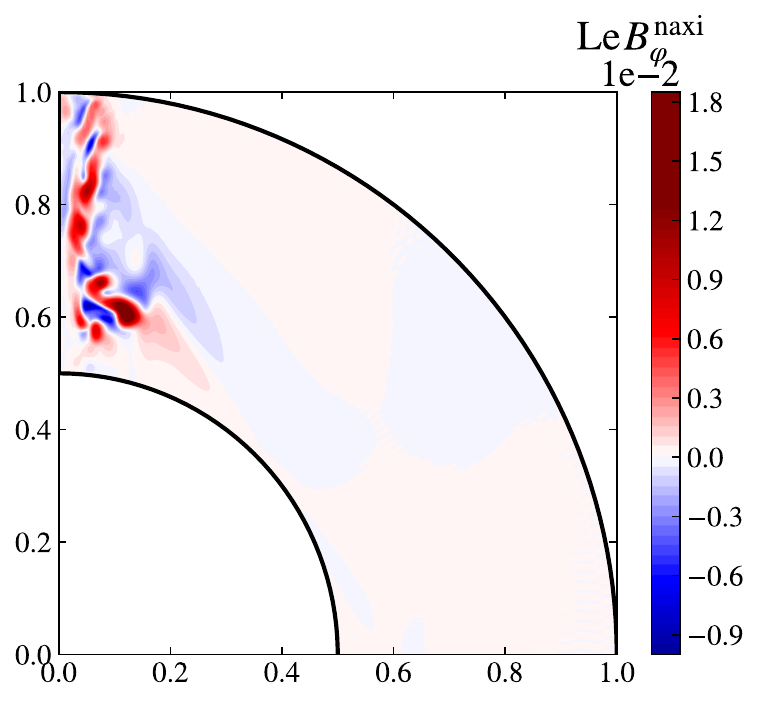}
    \includegraphics[trim=0cm 1cm 0cm 0cm,clip,width=0.48\linewidth]{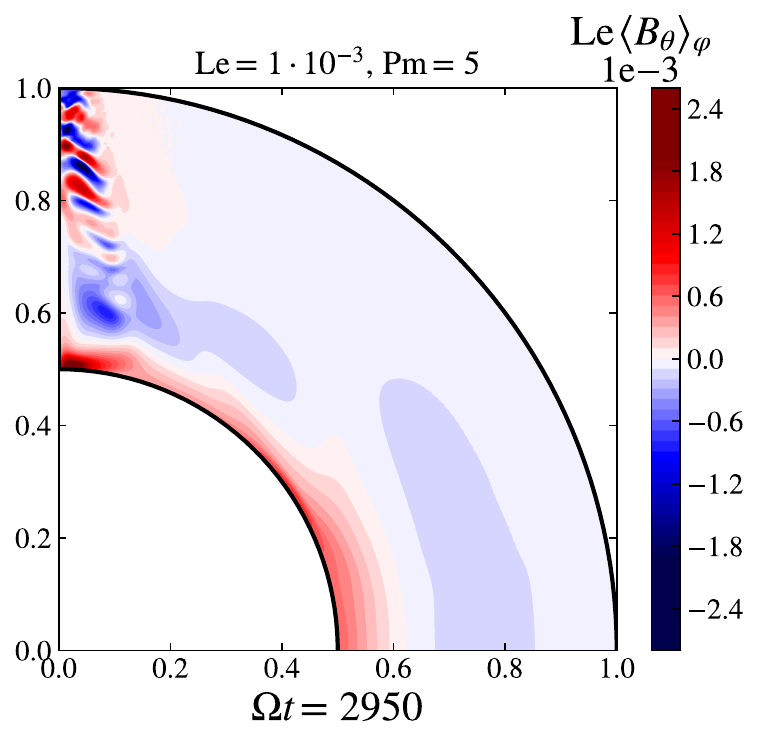}
    \includegraphics[width=0.48\linewidth]{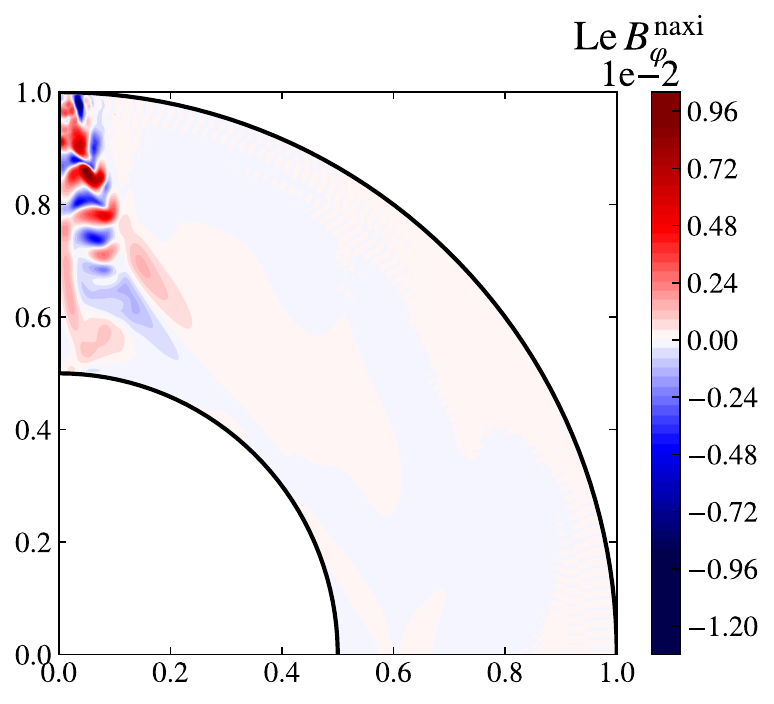}
    \caption{Axisymmetric (\textit{left}) and non-axisymmetric (\textit{right}) latitudinal and azimuthal magnetic fields ($\Le\langle B_\theta\rangle_\varphi$ and $\Le B_\varphi^\mathrm{naxi}$, respectively) for two simulations exhibiting instabilities near the poles. These cases have $\Le=4\cdot10^{-4}$ at $t=4850$ \textit{(top)} and $\Le=10^{-3}$ at $t=2950$ (\textit{bottom}), both with $\Pm=5$.}
    \label{fig:mri_map}
\end{figure*}
\subsection{Trends varying the magnetic Prandtl number}
We have also performed simulations varying the initial Lehnert number with different magnetic Prandtl numbers $\Pm\in[10^{-1},2,5]$ (for the same $\Ek$) to explore the impact of variations in the Ohmic diffusivity (relative to the viscosity). The energy in the differential rotation $\edr$ and the viscous dissipation $D_\nu$ versus $\Lep$ are displayed in Fig.~\ref{fig:varLe_Pm2}, with linear and nonlinear hydrodynamical  predictions (in black dashed dotted and solid lines, respectively) as in Fig. \ref{fig:varLe_Pm1} for $\Pm=1$. Similarly, 
we also plot linear MHD predictions (in black dotted lines) with an imposed dipolar magnetic field with a strength given by the initial $\Lep$ \citep[following][]{LO2018}. As for $\Pm=1$ (Fig. \ref{fig:varLe_Pm1}), this exhibits a complicated dependence on $\Lep$ for each $\Pm$. These predictions do not clearly match our simulations for larger $\Le$ (where they differ from hydrodynamical predictions), probably because the field in them has substantial toroidal and non-axisymmetric components, and more importantly because the initial dipole decays, unlike what is assumed in the linear calculations used to compute the black dotted lines.
In Fig. \ref{fig:MHDlinear} we compute linear MHD values (for a fixed background dipolar field) of the viscous and Ohmic dissipations for the Lehnert and magnetic Prandtl numbers investigated here. For high magnetic Prandtl numbers $\mathrm{Pm=2}$ and $\mathrm{Pm}=5$, it is interesting to see that viscous dissipation always dominates Ohmic dissipation over $\Le$. It differs from what has been found in \citet[Fig. 4]{LO2018} for $\Pm\ll1$, where $D_\eta$ takes over $D_\nu$ for $\Le\gtrsim10^{-4}$. For $\mathrm{Pm}=1$, $D_\eta$ starts to be dominant for $\Le\gtrsim2\times10^{-2}$ (in Fig. \ref{fig:MHDlinear}), which may explain why the nonlinear viscous dissipation in Fig. \ref{fig:varLe_Pm1} starts to be seriously mitigated at this high Lehnert number. Lastly, for $\Pm=0.1$, $D_\eta\gtrsim D_\nu$ from $\mathrm{Le}\gtrsim8\times10^{-3}$ (in Fig. \ref{fig:MHDlinear}), and the surge (and drops) in nonlinear $D_\nu$ in Fig. \ref{fig:varLe_Pm2} (right bottom panel) around $\Le=10^{-2}$, may correspond to the highest peak (and drops) found in Fig. \ref{fig:MHDlinear}. However, although Ohmic dissipation dominates for lower $\Pm$, Ohmic decay is even faster for this low magnetic Prandtl number, making a direct comparison between linear and nonlinear MHD predictions even more difficult.

For higher magnetic Prandtl numbers, two regimes, either having strong (for small $\Lep$) or weak (for larger $\Lep$) differential rotation, are observed in a similar way as for $\Pm=1$.
Unlike cases with $\Pm=1$, oscillations at the transition for $\Pm=2$ (e.g., when $\Le=2\cdot10^{-3}$ or $\Le=6\cdot10^{-3}$) are associated with repeated brief surges of the total poloidal magnetic energy $\Mp$, concomitantly with an increase in $\Mt$, but after $\edr$ peaks. This is shown in Fig.~\ref{fig:compEn}. In these simulations, the quadrupolar component in the poloidal magnetic field (created by stretching of the axisymmetric toroidal magnetic field by tidal waves) may play a role in the cyclic mitigation of differential rotation, as evidenced in Fig.~\ref{fig:flux_osc_Pm2}.
Indeed, unlike in the previous cases when $\Pm=1$, we notice the importance of Maxwell stresses (at the first `trough' in $\edr$) which reflect here the strength of non-axisymmetric magnetic correlations.
For low $\Le$, we also note the important role of the zonal flows, which more strongly shape magnetic fields for high magnetic Prandtl numbers, and in particular poloidal magnetic field lines, as we can see in Fig.~\ref{fig:phiavg} (bottom middle panel).
 
For an even larger magnetic Prandtl number $\Pm=5$, we also observe the transition between the two regimes, but it is shifted further towards a lower Lehnert number $\Lec\approx10^{-3}$, while it was $\Lec\approx2\cdot10^{-3}$ for $\Pm=2$, and $\Lec\approx3\cdot10^{-3}$ for $\Pm=1$. With increasing $\Pm$, the magnetic Ekman number ($\Em=\Ek/\Pm$) decreases, and so does $D_\eta$, with the latter always being negligible compared to $\Pt$ and $D_\nu$. We do not observe a clear transition for a low magnetic Prandtl number $\Pm=0.1$, though it may arise around $\Le\approx10^{-2}$. In these simulations, the Ohmic dissipation $D_\eta$ is non-negligible at early times, while $D_\nu$ is smaller. However, since Ohmic diffusivity $\Em=10^{-4}>\Ek$ is important there, magnetic fields decay relatively rapidly, and so does $D_\eta$. The non-linear hydrodynamical regime is then reached more rapidly for $\Pm=0.1$ as $\Lep$ more rapidly decays with time in this case. 

\begin{figure*}
    \centering
    \includegraphics[width=0.49\linewidth]{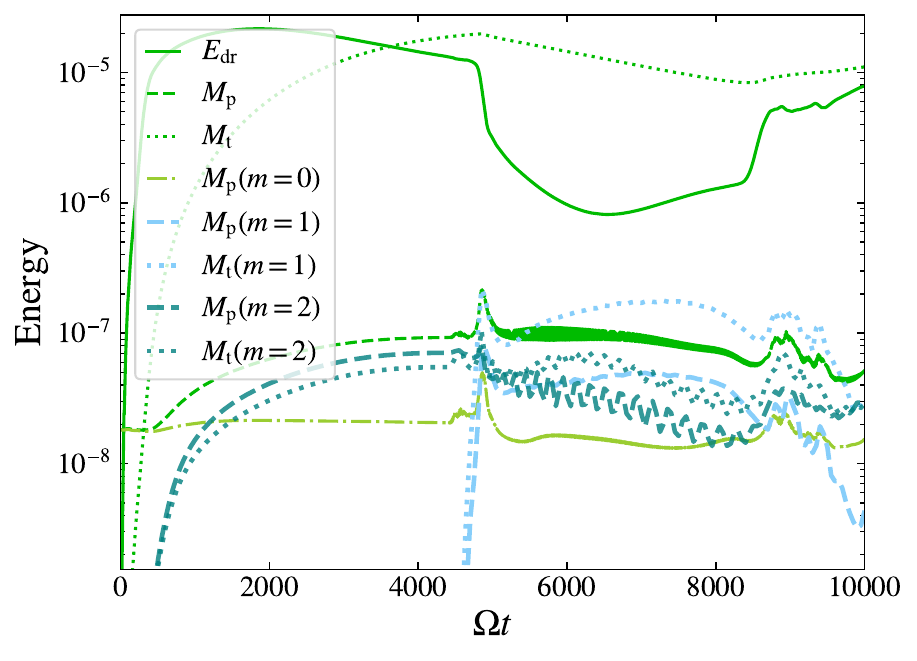}
    \includegraphics[width=0.49\linewidth]{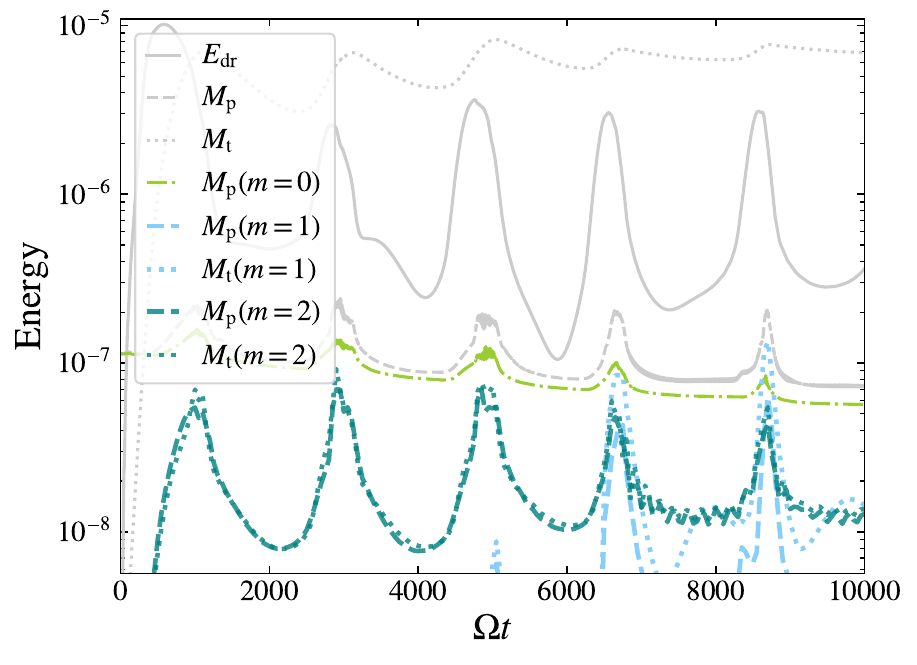}
    \caption{Evolution of various key energy terms (see legends) in two simulations when $\Pm=5$. \textit{Left:} $\Le=4\cdot10^{-4}$. \textit{Right:} $\Le=10^{-3}$. 
    }
    \label{fig:compE_pm5}
\end{figure*}
The heuristic scaling law derived in \citet[][hereafter LO18]{LO2018} to distinguish between linear regimes dominated by (approximately hydrodynamic) inertial or (inherently magnetic) Alfvén waves may be applied and modified for our study. When $\Pm\ll 1$, they found that inertial waves prevail as long as $\Le<\mathcal{O}(\Em^{2/3})$, with a prefactor depending on the direction of the magnetic field compared to that of the (wave) shear layer of width $l$ (and of the zonal flow here, which modifies the direction of wave propagation). The inertial wave propagation time and the timescale for setting up the zonal flow are constrained by either Ohmic diffusivity or viscosity, according to the value of $\Pm$. For $\Pm\gg1$, the viscous damping timescale $l^2/\Ek$ should be preferred compared to the Ohmic diffusion one. In the intermediate regime when $\Pm\sim1$, a geometric mean diffusion time could be used instead, such that the diffusion timescale across a wave beam becomes $\tau_\mathrm{i}=l^2/\sqrt{\Ek\Em}$ instead. Using the same heuristic arguments as in LO18 with the above timescale, LO18's scaling law becomes $\Le=\mathcal{O}(\Ek^{2/3}/\Pm^{1/3})=\mathcal{O}(\Em^{2/3}\Pm^{1/3})$. 
It is reported in Table \ref{tab:scal} for $\Ek$ and $\Pm$ used in our simulations, along with the values of $\Lec$ roughly estimated from Figs.~\ref{fig:varLe_Pm1} and \ref{fig:varLe_Pm2}. For $\Pm\geq1$, the scaling law may be relevant to predict the transition with an overall proportionality constant of approximately $5$, though it is less clear that it holds for $\Pm=0.1$. Such a proportionality constant could be related to the fact that the inertial wave shear layers are inclined with respect to the magnetic field lines (close to the pole it is about $\theta=\arcsin(\omega/m)$ from the rotation axis because the poloidal field is nearly vertical).
Further simulations exploring a wider range of parameters, particularly varying $\Ek$ as well as $\Pm$, would be useful to explore the validity of this trend before we can confidently extrapolate it to stars and planets. Nevertheless, the values here for $\Lec$ in our simulations are close to (or within) the ranges of values expected in stars and hot Jupiters, as discussed in \S~\ref{sec:magic}. Thus, we may expect complex interactions between tidal flows and magnetic fields in many stars and planets.

\begin{table}
    \centering
    \setlength{\tabcolsep}{4pt}
    \begin{tabular}{c||c|c|c|c}\newline
        $\Lec\,\backslash\,\Pm$ & $0.1$ & $1$ & $2$ & $5$ \\
        \hline
        $\mathcal{O}(\Em^{2/3}\Pm^{1/3})$ & $\mathcal{O}(10^{-3})$ & $\mathcal{O}(5\cdot10^{-4})$ & $\mathcal{O}(4\cdot10^{-4})$ & $\mathcal{O}(3\cdot10^{-4})$ \\
        Inferred & $\gtrsim10^{-2}$ & $3\cdot10^{-3}$ & $2\cdot10^{-3}$ & $10^{-3}$ \\
    \end{tabular}
    \caption{Transitional Lehnert numbers $\Lec$ for various magnetic Prandtl numbers $\Pm$, estimated from the scaling law or approximatively inferred from Figs.~\ref{fig:varLe_Pm1} and \ref{fig:varLe_Pm2}.}
    \label{tab:scal}
\end{table}

\subsection{Identification and analysis of magnetohydrodynamic instabilities}
In our simulations with $\Pm=5$ and $\Le\sim\Lec$ (which have both moderately strong differential rotation and magnetic field), we have observed magnetohydrodynamic instabilities to operate, particularly for\footnote{In these simulations, we increased the spatial resolution to guarantee that the instabilities observed are well resolved, with a maximum number of Chebyshev radial nodes of $n_{r,\,\mathrm{max}}=161$, and maximum degree and order of the Legendre polynomials $l_\mathrm{max}=341$ and $m_\mathrm{max}=50$. 
} $\Le\in[4\cdot10^{-4},10^{-3}]$. In Fig.~\ref{fig:mri_map}, we show snapshots of the meridional plane in two examples exhibiting possible MHD instabilities. These show growth of localised spatially oscillatory patterns in the magnetic field close to the poles where the differential rotation is strongest, as seen in both the axisymmetric ($\langle B_\theta\rangle_\varphi$) and non-axisymmetric ($B_\varphi^\mathrm{naxi}=B_\varphi-\langle B_\varphi\rangle_\varphi$) components (each chosen to most clearly visualise the instability). These snapshots have been taken at times which correspond to a maximum of the poloidal magnetic energy, as we can see in Fig.~\ref{fig:compE_pm5}. This time is also correlated with abrupt changes in the toroidal magnetic energy and differential rotation. For $\Le=4\cdot10^{-4}$ (left panel) the rapid, approximately exponential, growth of the poloidal magnetic energy is associated with a strong decrease of $\edr$ (and to a lesser extent $\Mt$) around $t=4800$, while for $\Le=10^{-3}$ (right panel) we note periodic quasi-simultaneous bursty behaviours in each of $\Mp$, $\Mt$ and $\edr$ (and $\edr$ slightly precedes the two others). These periodic bursts are associated with sinks of the viscous dissipation $D_\nu$ and tidal power $\Pt$, which drop by approximately $20-30\%$ during these periods. The periods of exponential growth in magnetic energy, and the fall in $\edr$, indicate the likely onset of MHD instabilities driven by the differential rotation. We have measured growth rates $\gamma$ during the exponential growth phases of different magnetic components, which we present in Table \ref{tab:gamma}.

\begin{figure*}
    \centering
    \includegraphics[width=0.49\linewidth]{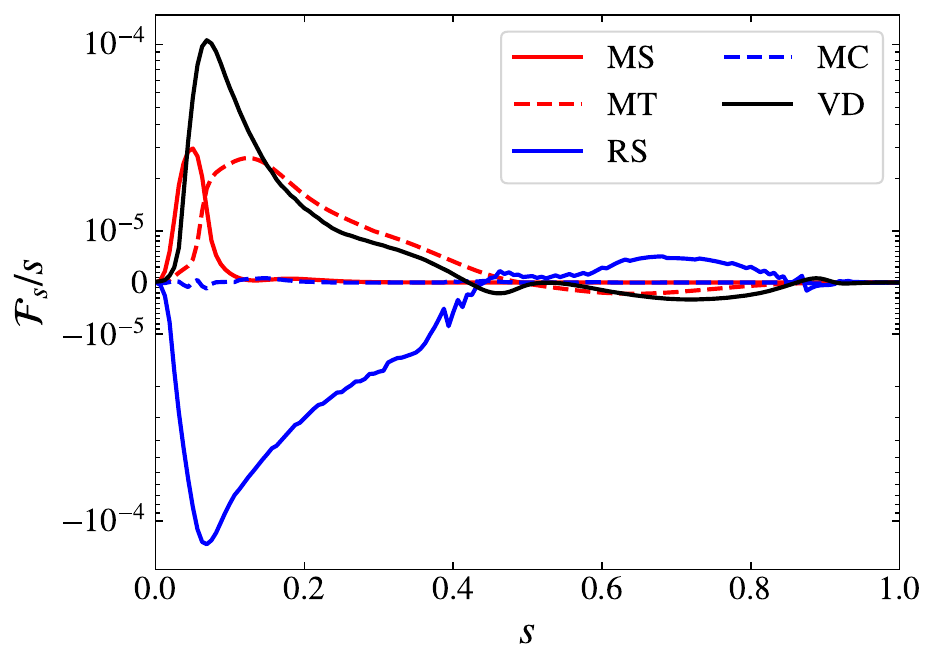}
    \includegraphics[width=0.49\linewidth]{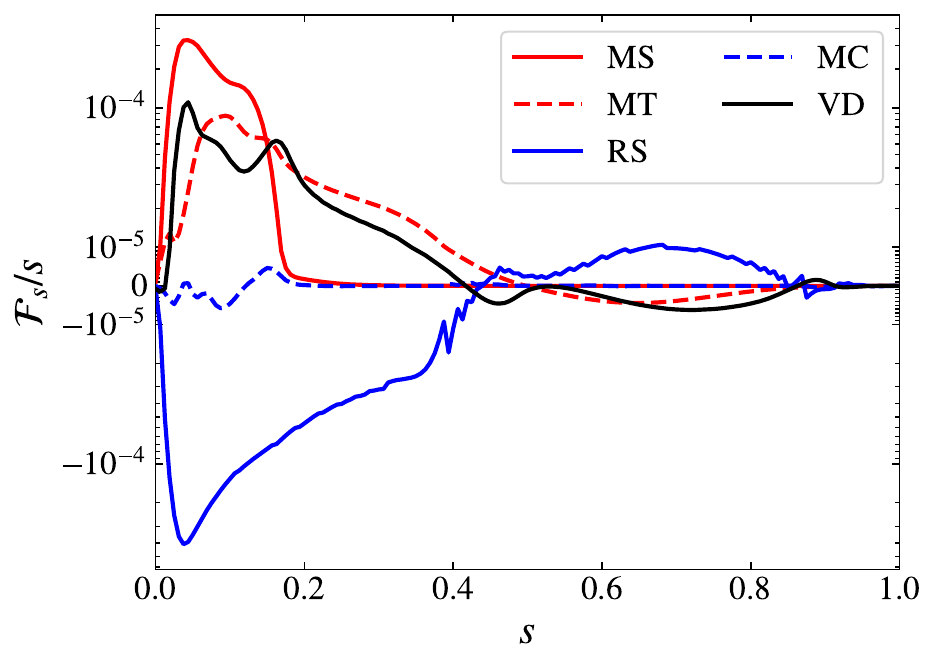}
    \caption{Flux terms within the radial (cylindrical) $\mathcal{F}_s/s$ angular momentum flux integrated along $z$, for $\Le=4\cdot10^{-4}$ and $\Pm=5$ before and close to the peak of $\Mp$ (see Fig. \ref{fig:compE_pm5}) at $t=4200$ (\textit{left}) and $t=4850$ (\textit{right}), respectively. They are Maxwell stresses (MS), magnetic torques (MT), Reynolds stresses (RS), meridional circulations (MC), and viscous diffusion contributions (VD).}
    \label{fig:maxStress}
\end{figure*}
For $\Le=4\cdot10^{-4}$, the surge (drop) of $\Mp$ ($\edr$ and $\Mt$), seems to be concomitant with a strong increase of the $m=1$ toroidal component, and to a lesser extent, to the rise of the $m=2$ components, as shown in Fig. \ref{fig:compE_pm5}. These modes (along with the $m=3$ mode to a lesser degree, not shown here) are excited at the forcing frequency $\omega=1.1$, which precludes possible triadic resonances and parametric instabilities as found in AB22. For $\Le=10^{-3}$, the $m=1$ mode does not seem to trigger the initial instability associated with the bursty behaviour of the magnetic field, but may rather be a consequence of it, since it arises latter in the simulation. 
Before the $m=1$ mode kicks in, a second sharper slope is visible in $\Mp(m=2)$ and $\Mt(m=2)$ peaks, but it is not present in the axisymmetric poloidal magnetic energy $\Mp(m=0)$, so both $m=0$ and $m\ne 0$ (first linked to $m=2$ and then to $m=1$ modes) instabilities could be in operation.
Thus, it is difficult to determine from these time series alone which magnetic components (axisymmetric or non-axisymmetric) induce the other -- i.e., whether the instability is axisymmetric or non-axisymmetric (or both), and which component could be a consequence of the other. 
It should be noted than higher growth rates are estimated for non-axisymmetric $m=2$ (and $m=1$) magnetic energy components in both simulations. Lastly, an important contribution of the (non-axisymmetric) radial Maxwell stresses stands out close to the pole 
when the instability is triggered for $\Le=4\cdot10^{-4}$ (see Fig. \ref{fig:maxStress}). These strong Maxwell stresses, that are also present for $\Le=10^{-3}$, have been observed, to a lesser extent, in some simulations for $\Pm=2$ (see Fig. \ref{fig:flux_osc_Pm2}). They seem to partially balance the Reynolds stresses generating the differential rotation, thereby requiring somewhat weaker viscous diffusion to maintain the zonal flow.
\begin{table}
    \centering
    \begin{tabular}{c||c|c|c|c}
    $\Le$ & $\gamma(\Mp)$ & $\gamma\left(\Mp^{m=0}\right)$ & $\gamma\left(\Mp^{m\neq0}\right)$ &  $\gamma\left(\Mt^{m\neq0}\right)$\\ 
    \hline
    $4\cdot10^{-4}$ & $6.5\cdot10^{-3}$ & $4.7\cdot10^{-3}$ & $7.1\cdot10^{-3}$ & $8.3\cdot10^{-3}$ \\
    $10^{-3}$ (1) & $7.1\cdot10^{-4}$ & $5.1\cdot10^{-4}$ & $1.7\cdot10^{-3}$ & $2.5\cdot10^{-3}$\\
    $10^{-3}$ (2) & $3.4\cdot10^{-3}$ & $\diagup$ & $1.2\cdot10^{-2}$ & $10^{-2}$ \\
    \end{tabular}
    \caption{Measured growth rates of the (non-)axisymmetric poloidal and toroidal magnetic energies assuming that $M_\mathrm{p,t}\propto \exp(2\gamma t)$ at the burst and performing a linear interpolation. (1) and (2) refers to the first and second slopes, before and after $t=2800$, respectively.}
    \label{tab:gamma}
\end{table}

Without any magnetic fields, the tidally-driven zonal flow for $\omega=1.1$ is hydrodynamically stable according to the Rayleigh criterion\footnote{This states that rotating flows in which angular momentum does not decrease outwards from the axis are hydrodynamically stable to axisymmetric instabilities.} (see AB22). However, since the angular velocity decreases outwards radially from the pole to the equator in our simulations, the cylindrical zonal flow could host magneto-rotational instabilities (MRI) in the presence of magnetic fields \citep[see e.g.][for an introduction]{BH1998,B2009}. These instabilities have been widely studied, for example in the context of accretion disks, for their capacity to transport angular momentum outwards through turbulence \citep[e.g.][]{BH1991,OP1996}, or in the context of the geodynamo \citep[e.g.][]{A1983,PD2013}.
Local analytical models have been developed to study both axisymmetric and non-axisymmetric MRI with different topologies of the magnetic field \citep[e.g.][and many others]{A1978,BH1991,KS2010}. 
The instability of an axial (vertical) magnetic field in a cylindrical Taylor-Couette flow was initiated by \citet{V1959}, but its importance was realised when it was rediscovered by \citet{BH1991}, and it is now known as the standard MRI (SMRI). The axisymmetric instability occurring with a combination of an axial and azimuthal magnetic fields is now sometimes called the helical MRI \citep[HMRI e.g.][where the latter points out its relation to SMRI]{HR2005,KS2010}, and the non-axisymmetric instability of a purely azimuthal magnetic field is called the azimuthal MRI \citep[e.g.][]{A1978,RH2007,HT2010,G2017}.

In Appendix \ref{sec:HMRI}, we have explored the possible roles of axisymmetric SMRI or HMRI to explain the instability observed close to the poles that varies along the rotation axis, based on applying the local linear stability analysis of \citet{KS2010}. The growth rate of the most unstable mode has been calculated by solving the resulting quartic dispersion relation. Predicted growth rates match relatively well those measured in our simulations (see Fig.~\ref{fig:azi_axi_mri} and Tables \ref{tab:gamma} and \ref{tab:params}), but the predicted vertical wavenumbers are typically lower than the measured ones.

Nevertheless, there are difficulties in applying the local analysis, partly because we must adopt appropriate values of the parameters used in the model to compare with simulations, such as the vertical and azimuthal magnetic field strengths, and the shear rate, which we have estimated from the simulations. These vary substantially as a function of position, even within the region where the instability is observed, and with time. In addition, the short-wavelength (WKBJ) approximations used to derive the dispersion relation may not be fully satisfied, as we discuss further in the appendix. Finally, the basic state is evolving in time (i.e., specifically the differential rotation and magnetic field profiles), whereas the local WKBJ theory assumes a static or slowly varying (compared to the growth time of the mode) basic state.
\begin{figure*}
    \centering
    \includegraphics[width=0.247\linewidth]{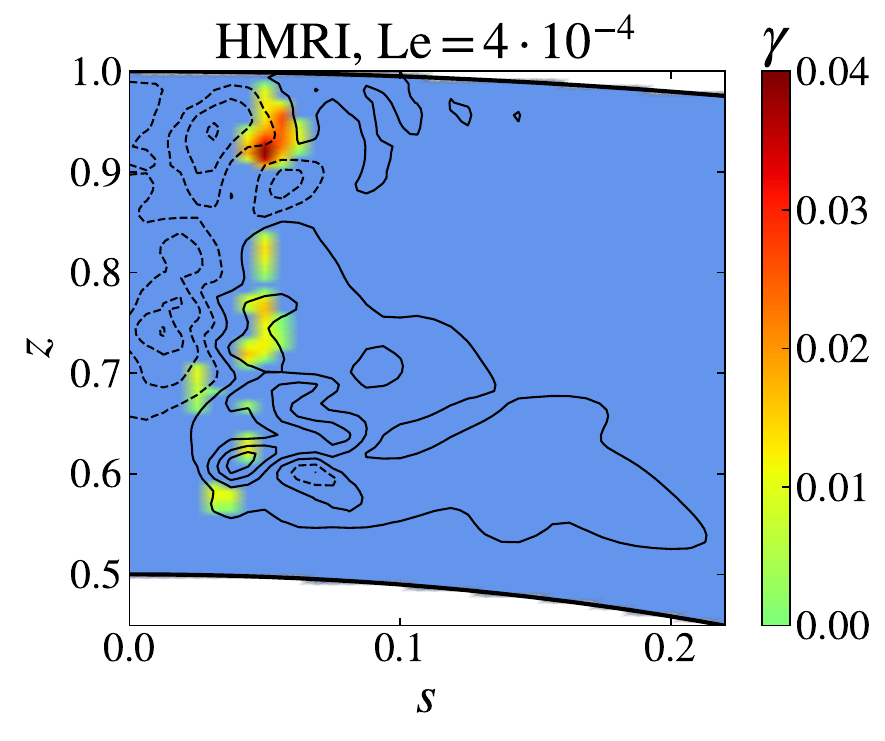}
    \includegraphics[width=0.247\linewidth]{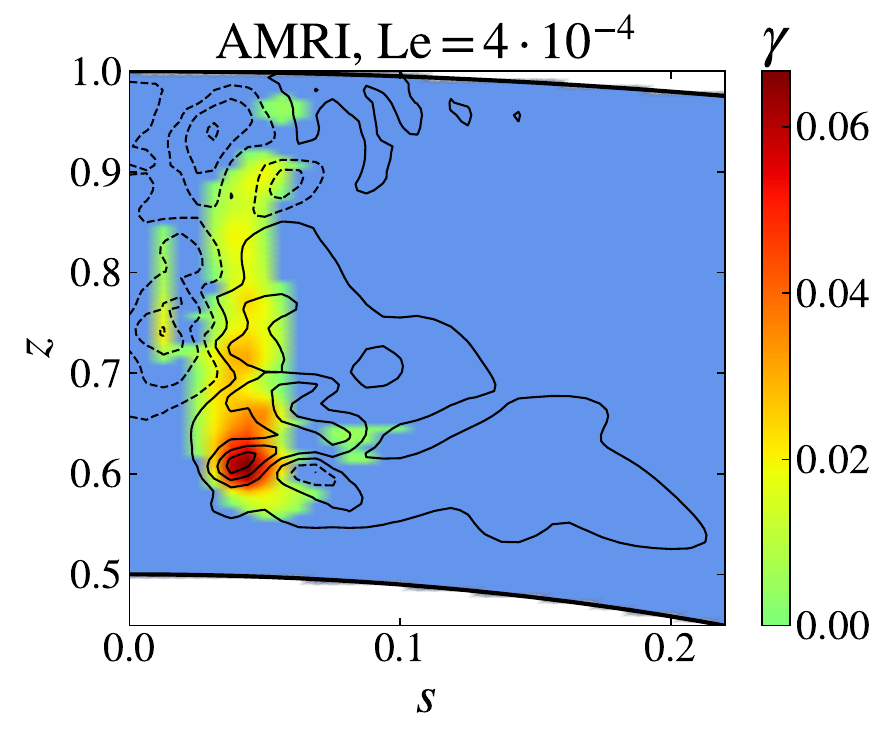}
    \includegraphics[width=0.247\linewidth]{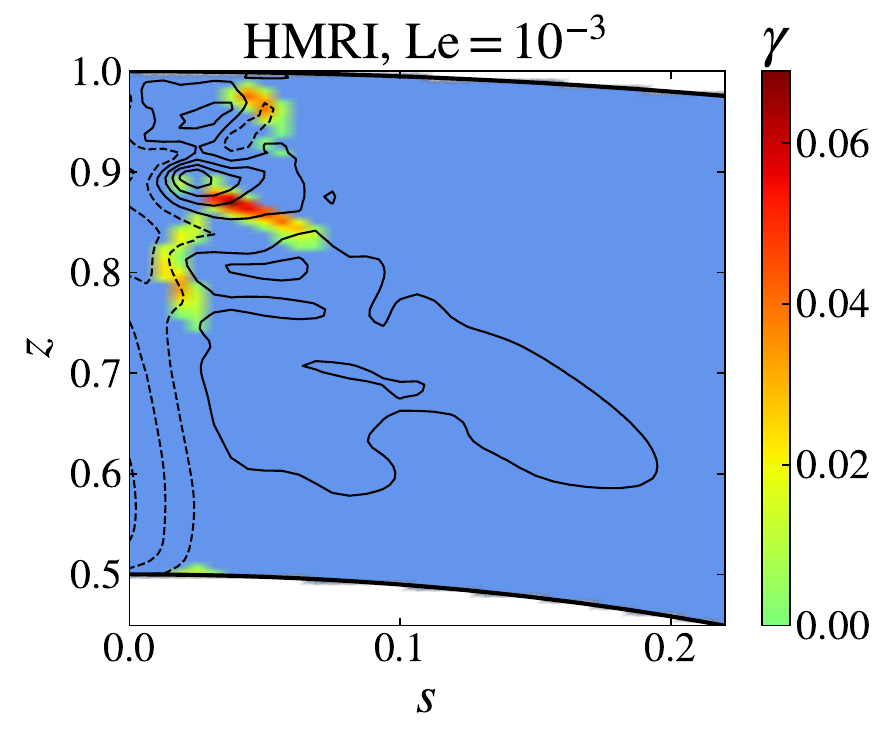}
    \includegraphics[width=0.247\linewidth]{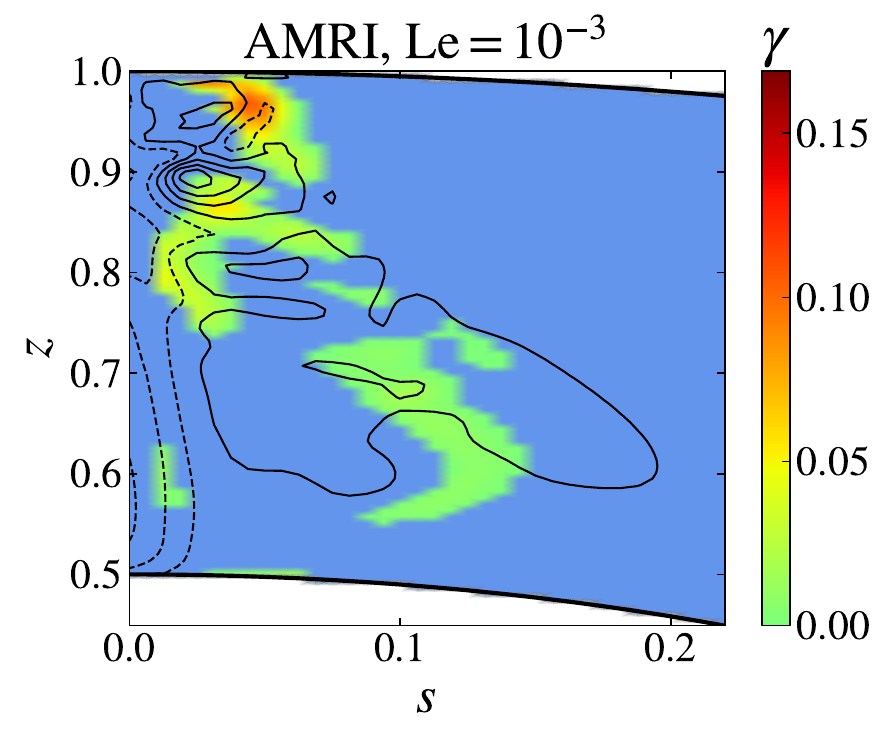}
    \caption{Two-dimensional colourmaps of the growth rate $\gamma$ in the polar region for the two simulations with $\Pm=5$ and $\Le=4\cdot10^{-4}$ (the two leftmost panels) and $\Le=10^{-3}$ (the two rightmost panels), from local stability analyses of the axisymmetric helical (HMRI) and non-axisymmetric azimuthal (AMRI) with $m=1$ (2$^\mathrm{nd}$ panel) and $m=2$ (4$^\mathrm{th}$ panel) magneto-rotational instabilities. Blue regions (also found everywhere for $s>0.22$) indicate decaying modes (i.e. with $\gamma<0$). Black contours represent the radial magnetic field $B_r(\varphi=0)$.}
    \label{fig:mapMRI}
\end{figure*}

In the two simulations examined in this section the axisymmetric azimuthal magnetic field is strongly dominant (see Fig.~\ref{fig:compE_pm5}). Therefore, and as discussed earlier, non-axisymmetric perturbations driven by this field may also be excited \citep[see e.g.][]{KS2010,HT2010}. We have thus also explored the possibility of non-axisymmetric AMRI in Appendix \ref{sec:AMRI} following \citet{A1978} and \citet{ML2019}. Although the same limitations apply to this model as for SMRI/HMRI, the analytically predicted growth rates could also be consistent with the measured ones (see Fig.~\ref{fig:AMRI}), both for $m=1$ or $m=2$ non-axisymmetric perturbations.

In conclusion, both HMRI and non-axisymmetric  AMRI could predict magneto-hydrodynamical instabilities observed in the polar regions of these simulations, as summarized in Fig.~\ref{fig:mapMRI}, where the growth rate $\gamma$ is higher in some regions (especially for AMRI) but overall compatible with the measured values in Table \ref{tab:gamma}, and with (radial) magnetic perturbations shown in black contours \citep[similarly as Fig. 10 of][]{JL2020}. To apply these local stability analyses and compute $\gamma$, azimuthal averages of the rotation and shear rates, vertical and azimuthal magnetic fields, and their gradients, have been computed at each point of the polar regions (Eqs.~(\ref{eq:params}) and (\ref{eq:par_AMRI})), and we choose $k_z=126$ for $\Le=4\cdot10^{-4}$ and $k_z=94$ for $\Le=10^{-3}$, along with $k_x=63$ (see Table \ref{tab:params}) based on by-eye fitting.

A more comprehensive local analysis involving non-axisymmetric perturbations with both axisymmetric poloidal and toroidal magnetic fields may be useful to have better predictions of the unstable modes involved in the instabilities seen in the simulations for $\Le=4\cdot10^{-4}$ and $\Le=10^{-3}$. But any such local analysis is unlikely to substantially improve agreement with simulations since the basic state is evolving in time, and the modes are not truly local since the basic state varies on a scale not much larger than the wavelengths of the modes (particularly with cylindrical radius). Finally, the non-axisymmetric analysis assumes exponentially growing normal modes, which we show in the appendix is unlikely to be strictly valid due to the effects of the differential rotation. Nevertheless, our analysis has identified the likely origin of the instabilities observed to be (axisymmetric and/or non-axisymmetric) MRI driven by the differential rotation. Broadly similar wavelengths and growth rates can be obtained in these analyses with suitable parameter choices, keeping in mind all of the above caveats. The role of the MRI in controlling tidally-driven differential rotation -- and its impact on tidal dissipation rates -- is therefore worth studying further.

\section{Conclusions}
\label{sec:con}
We have explored the interactions of magnetic fields and tidally excited inertial waves in magnetohydrodynamic nonlinear models of the convective envelopes of stars and gaseous planets. Our goals were to determine the influence of the field on tidal dissipation rates, and hence for the spin-orbit evolution of many astrophysical systems, but also to study the effects of the flow on the magnetic field. We imposed an initial dipolar field (thought to be generated by a convective dynamo) aligned with the rotation axis, with a strength determined by its Lehnert number $\Le$, and explored the nonlinear evolution of both the field and the tidal flow using magnetohydrodynamical simulations. Our simulations restricted the tidal frequency to one value in the inertial wave range \citep[$\omega=1.1\Omega_0$, since this case has been well explored hydrodynamically in][AB22, AB23]{O2009,FB2014}, considered a deep convective shell with fractional radius $0.5$ (relevant for some low-mass stars or giant planets with large stably stratified dilute cores), and fixed the Ekman number to a value approximately consistent with mixing-length expectations for a turbulent viscosity in solar-like stars. The nonlinear interactions of tidal waves and magnetic fields have been found to be complex, with several different regimes depending on $\Le$ and the magnetic Prandtl number $\Pm$. 

For small $\Le\lesssim \Lec\ll 1$, below a critical threshold $\Lec$, the tidal waves behave similarly to in hydrodynamical simulations, nonlinearly generating differential rotation in the form of zonal flows (strongest near the poles in our simulations). These strong zonal flows effectively stretch the initial dipolar magnetic field to produce an axisymmetric toroidal field via the $\Omega-$effect. Moreover, non-axisymmetric ($m=2$) wavelike tidal flows also interacts with these axisymmetric magnetic components to create quadrupolar poloidal and toroidal magnetic components. However, these magnetic fields are not observed to noticeably modify tidal dissipation rates over hydrodynamical nonlinear simulations, where zonal flows remain strong and produce observable differences with linear hydrodynamical predictions (as in AB22 and AB23).

For stronger fields with $\Le\gtrsim \Lec$, tidally-generated zonal flows are substantially inhibited, primarily by magnetic torques (from the contributions of axisymmetric components of the field to the total Maxwell stresses\footnote{(Total) Maxwell stresses often refers to both axisymmetric (torque) and non-axisymmetric (stress) magnetic contributions that we distinguish in the present study.}) arising from the large-scale poloidal field. For $\Le\sim \Lec$, we observe complex interactions involving the excitation of torsional Alfv\'{e}n waves (outwardly propagating from their launching sites near the poles) leading to oscillations in both the zonal flows and the field with a frequency proportional to the (cylindrical) radial field strength.
For smaller $\Le$, their frequency decreases, with magnetic torques becoming weaker, differential rotation becoming stronger, and the Alfv\'{e}n timescale becoming longer than the Ohmic diffusion timescale, leading to a predominantly hydrodynamical regime. For $\Le\gtrsim \Lec$, (viscous) tidal dissipation rates transition to a regime where they attain values close to the hydrodynamic linear theoretical predictions without zonal flows (despite the initial field being strong enough to modify linear predictions by itself).

For some simulations with $\Le\sim \Lec$, we have identified the magnetorotational instability (MRI) to operate, which strongly modifies the differential rotation when it occurs. To verify this interpretation, we have performed local analyses of the axisymmetric MRI in the presence of a poloidal and toroidal field, as well as the (weakly) non-axisymmetric MRI with a toroidal field. Our results suggest that the MRI is likely to be in operation where the differential rotation is strongest in our simulations, with growth rates and axial wavelengths broadly comparable to those observed numerically.
In the simulation with $\Le=10^{-3}$ and $\Pm=5$, the MRI produces cyclic, bursty behaviour in the differential rotation and magnetic field, somewhat reminiscent of predator-prey dynamics, while in the simulation with $\Le=4\cdot10^{-4}$, it has the sole effect of strongly mitigating differential rotation. 

The transitional $\Lec$ found in our simulations is similar to some estimates of $\Le$ near the surfaces, and in pre-main sequence phases, of low-mass stars (see Table \ref{tab:Le}) and in hot Jupiters. This suggests that we might expect a complex interplay between magnetic fields and tidal flows in stars and planets. This is particularly the case if we interpret our diffusivities as turbulent ones, so that we might expect $\Pm=O(1)$ in stars and planets, but less clear if the relevant diffusivities are microscopic ones. 

There are many ways in which our study should be extended. Firstly, we should explore a wider range of tidal frequencies, amplitudes, and convective shell thicknesses. In our simulations, the magnetic field led to tidal dissipation rates approaching hydrodynamic linear theoretical predictions, and it is important to explore whether this result is robust and is also found in other cases (including those adopting an imposed ``background magnetic field" instead of an initial imposed field). Introducing convection and performing self-consistent simulations of convective dynamos interacting with tidal flows would be particularly interesting, as well as exploring possible tidally driven dynamos \citep[as done for the elliptical instability in][in intermediate mass stars with radiative envelopes]{CH2014,VC2018}. The influence of density stratification within the anelastic approximation, and the roles of an interior radiation zone, are also important to explore in future work.

\appendix
\section*{Acknowledgements}
Funded by STFC grants ST/S000275/1 and ST/W000873/1, and by a Leverhulme Trust Early Career Fellowship (ECF-2022-362) to AA. Simulations were performed using the DiRAC Data Intensive service in Leicester, operated by the University of Leicester IT Services, part of the STFC DiRAC HPC Facility (\href{www.dirac.ac.uk}{www.dirac.ac.uk}). The equipment was funded by BEIS capital funding via STFC capital grants ST/K000373/1 and ST/R002363/1 and STFC DiRAC Operations grant ST/R001014/1. DiRAC is part of the National e-Infrastructure. AB is grateful for the support and hospitality at the Isaac Newton Institute for Mathematical Sciences, Cambridge, during the programme “Anti-diffusive dynamics: from sub-cellular to astrophysical scales”. AA thanks K. Hori for fruitful discussions and advice about torsional oscillations. We also thank the referee, L. Jouve and Y. Lazovik for their helpful comments that helped us to improve the paper. This research has made use of the NASA’s Astrophysics Data System Bibliographic Services. 
\section*{Data Availability}
The data underlying this article will be shared on reasonable request to the corresponding author.

\bibliographystyle{mnras}
\bibliography{biblio}


\appendix

\section{Decay of free modes}
\label{sec:appfm}
When injecting the poloidal $\bn\wedge\bn\wedge[g(r, \theta, \varphi,t){\bm e}_r]$ or the toroidal $\bn\wedge[h(r, \theta, \varphi,t){\bm e}_r]$ magnetic field component into the magnetic diffusion equation $\partial_t\bm B = \Em\, \Delta \bm B$, each potential satisfies a partial differential equation of the form \citep{M1978}:
\begin{equation}
\frac{\partial f}{\partial t} = \frac{\partial^2 f}{\partial r^2} -\frac{l(l+1)}{r^2}f,
\label{eq:pde}
\end{equation}
where $f$ stands for either $g$ or $h$ when projecting onto a spherical harmonic of degree $l$ \citep[see also][]{WG2015}. 
Setting $f=f_l\exp(p_\alpha t)$ with decay rate $-p_\alpha$ and changing variables such that $x^2=-(p_\alpha/\Em)\, r^2$ as in \citet[Section 2.7]{M1978}, we find that for free decay modes, $f_l$ satisfies Bessel's equation 
\begin{equation}
x^2\frac{\partial^2 f_l}{\partial x^2}+2x\frac{\partial f_l}{\partial x} +\left[x^2-l(l+1)\right] f_l=0,
\end{equation}
whose solutions are a linear combination of spherical Bessel functions of the $1^\mathrm{st}$ and $2^\mathrm{nd}$ kinds, of fractional order, respectively $J_{l+1/2}$ and $Y_{l+1/2}$ \citep{AS1972}:
\begin{equation}
    f_l=\frac{A}{\sqrt{r}}J_{l+1/2}(k_\alpha r)+\frac{B}{\sqrt{r}}Y_{l+1/2}(k_\alpha r).
    \label{eq:bessel}
\end{equation}
Here $A$ and $B$ are complex constants, and $k_\alpha^2=-p_\alpha/\Em$. In addition, the poloidal $g_l$ and toroidal $h_l$ potentials are constrained by the insulating boundary conditions at the inner:
\begin{equation}
    \partial_r g_{l}-\frac{l+1}{r}g_{l}=0, \text{ and }~ \partial_r h_{l}=0, \quad \text{at}\quad r=\alpha,
    \label{eq:bc_in}
\end{equation}
and outer spherical surfaces:
\begin{equation}
    \partial_r g_{l}+\frac{l}{r}g_{l}=0, \text{ and }~ \partial_r h_{l}=0 \quad \text{at}\quad r=1.
    \label{eq:bc_outi}
\end{equation}
When combining Eqs. (\ref{eq:bessel}), (\ref{eq:bc_in}), and (\ref{eq:bc_outi}), we can obtain the decay rate $2\ p_\alpha=-2\ \Em\, k_\alpha^2$ of the poloidal or toroidal magnetic energies ($\propto g^2$ or $h^2$) for a given spherical harmonic degree $l$ at fixed $\Em$. This provides the appropriate application of the derivation of free decay modes in, e.g.~\citet{M1978}, to our boundary conditions.
\section{Induced poloidal magnetic field during the kinematic phase}
\label{sec:ind}
In the simulations for low enough Lehnert numbers, we observe an increase in the poloidal magnetic energy at early times (like in Fig.~\ref{fig:M_t_Pm1}, left panel) possibly due to interactions between non-axisymmetric tidal flows and the toroidal magnetic field. If this interpretation is correct, the generation of a poloidal magnetic field comes from the particular induction term in the induction equation Eq. (\ref{eq:ind}):
\begin{equation}
    \bn\wedge(\bm u_\mathrm{w}\wedge\bm B_\mathrm{t})=(\bm B_\mathrm{t}\cdot\bn)\bm u_\mathrm{w}-(\bm u_\mathrm{w}\cdot\bn)\bm B_\mathrm{t},
    \label{eq:indw}
\end{equation}
where the non-axisymmetric tidal flow $\bm u_\mathrm{w}$ (mostly the $m=2$ component) couples with the axisymmetric ($m=0$) toroidal component of the magnetic field $\bm B_\mathrm{t}= \langle B_\varphi\rangle_\varphi\bm e_\varphi$. Indeed, we observe that while the induced toroidal magnetic component is mostly axisymmetric, the axisymmetric component of the poloidal magnetic field just decays with time, so the bump in $M_\mathrm{p}$ is purely non-axisymmetric (see Fig.~\ref{fig:M_t_m}).  
Note that the second term on the right hand side of Eq. (\ref{eq:indw}) does not contribute to the energy balance after volume-averaging (only the first term contributes), which makes sense since it locally cancels with part of the first term. Thus, the wavelike poloidal magnetic field in the meridional plane can be written as 
\begin{equation}
\bm B_\mathrm{pw}=\int_{t_1}^{t_2}\frac{\langle B_\varphi\rangle_\varphi}{s}\,\left(\bm e _r\partial_\varphi u^\mathrm{w}_{r} + \bm e_\theta\partial_\varphi u^\mathrm{w}_\theta\right) \,\mathrm{d} t,
\label{eq:Bpw}
\end{equation}
when integrating Eq. (\ref{eq:indw}) with time. To compute the non-axisymmetric meridional tidal flow, we set $u^\mathrm{w}_{r,\theta}=u_{r,\theta}-\langle u_{r,\theta} \rangle_\varphi$ to remove the background axisymmetric meridional flow. 
\begin{figure*}
    \centering
    \includegraphics[width=0.48\linewidth]{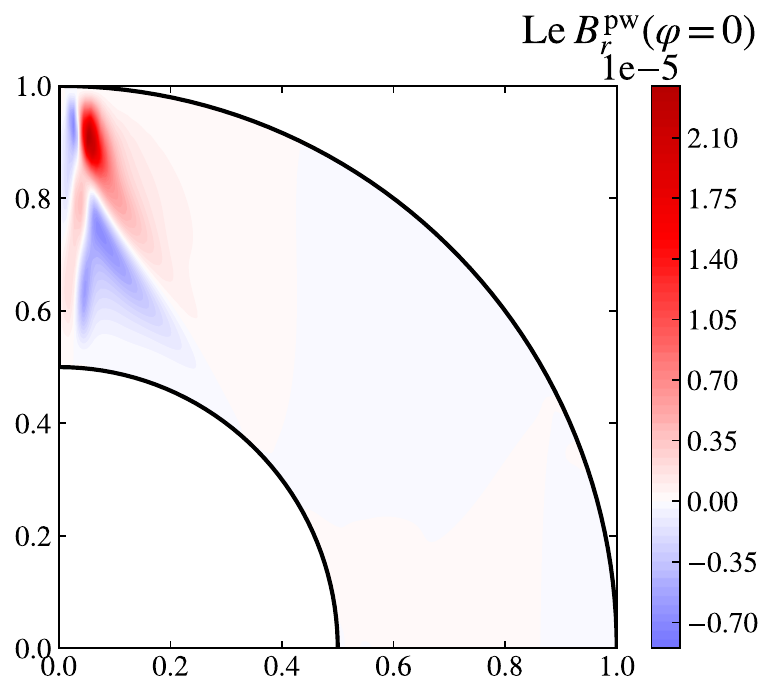}    
    \includegraphics[width=0.48\linewidth]{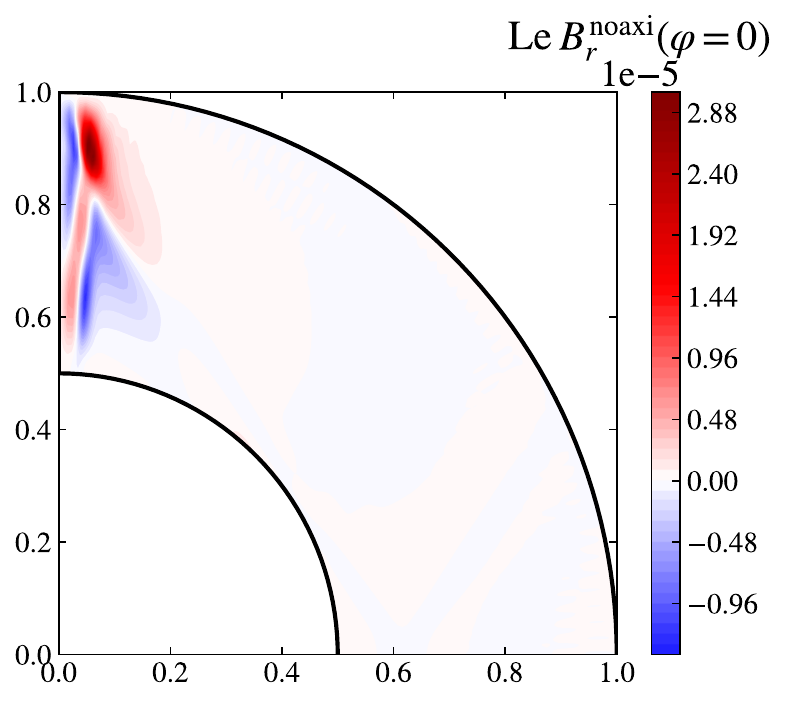}
    \includegraphics[width=0.48\linewidth]{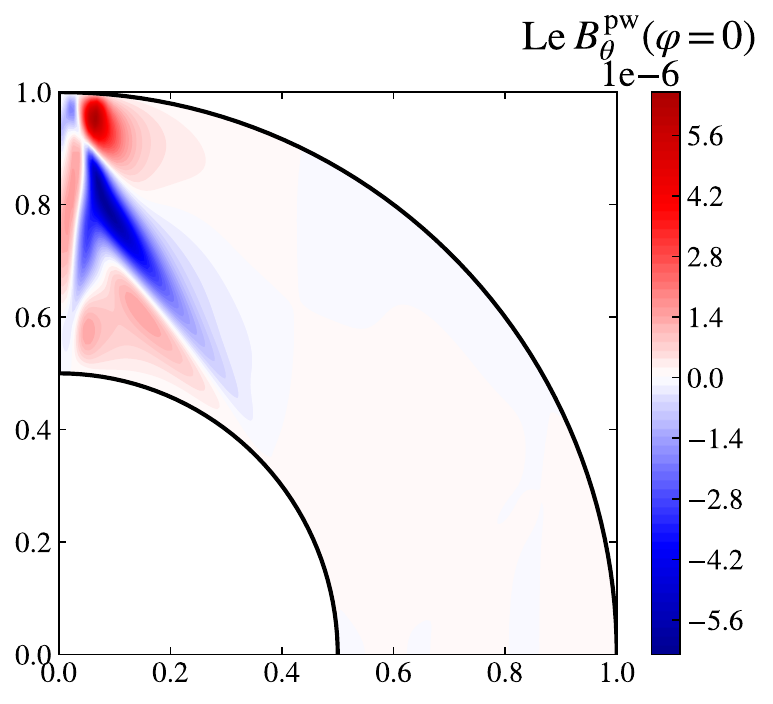}    
    \includegraphics[width=0.48\linewidth]{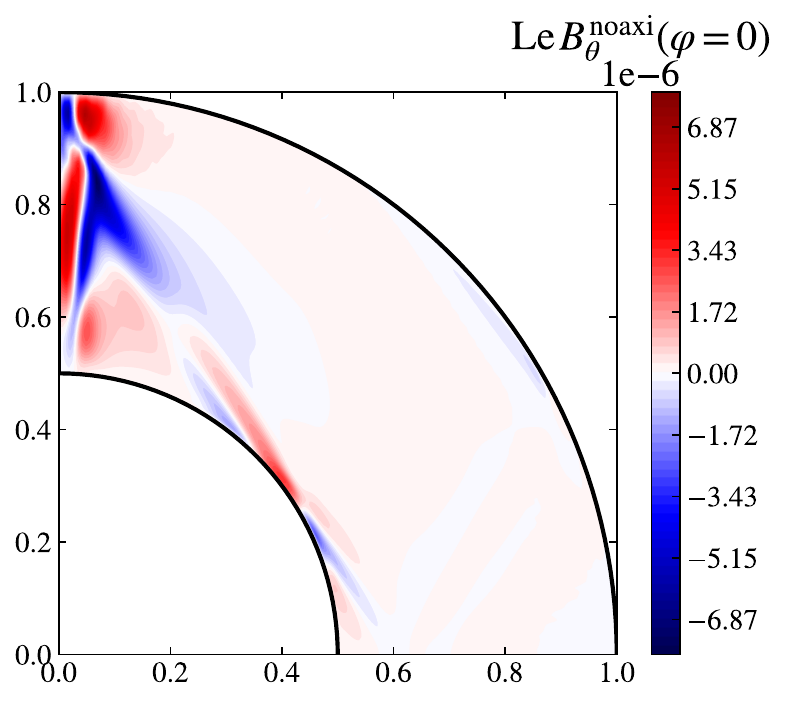}        
    \caption{Meridional snapshots in one quadrant at $\varphi=0$ of the induced wavelike magnetic field Eq. (\ref{eq:Bpw}) (\textit{left panels}) and the non-axisymmetric magnetic field Eq. (\ref{eq:Bnoaxi}) (\textit{right panels}) between $t_1=600$ and $t_2=650$ from a simulation with $\Le=10^{-5}$ and $\Pm=1$. For each row, the intensity of the colour scales with the same extrema (which are symmetric).
    \textit{Top:} $r$-components.  \textit{Bottom:} $\theta$-components.
    }
   \label{fig:ind}
\end{figure*}
We display the $r$ and $\theta$ components of $\bm B_\mathrm{pw}$ in Fig.~\ref{fig:ind} (left panels) choosing $\varphi=0$. The time integration has been performed between $t_1=600$ and $t_2=650$ with a timestep of one rotational unit, namely when the poloidal energy is rising and for a short period of time such that Ohmic diffusion doesn't have time to act. For comparison, the non-axisymmetric components 
\begin{equation}
B^\mathrm{noaxi}_{r,\theta}=B_{r,\theta}(t_2)-\langle B_{r,\theta}(t_2)\rangle_\varphi - \left[B_{r,\theta}(t_1)-\langle B_{r,\theta}(t_1)\rangle_\varphi\right],
\label{eq:Bnoaxi}
\end{equation}
are shown at $\varphi=0$ between $t_2$ and $t_1$ in the right panels. Fig.~\ref{fig:ind} verifies that our interpretation regarding the leading cause of the generation of poloidal magnetic energy at early times is correct.
\begin{figure*}
    \centering
    \includegraphics[width=0.49\linewidth]{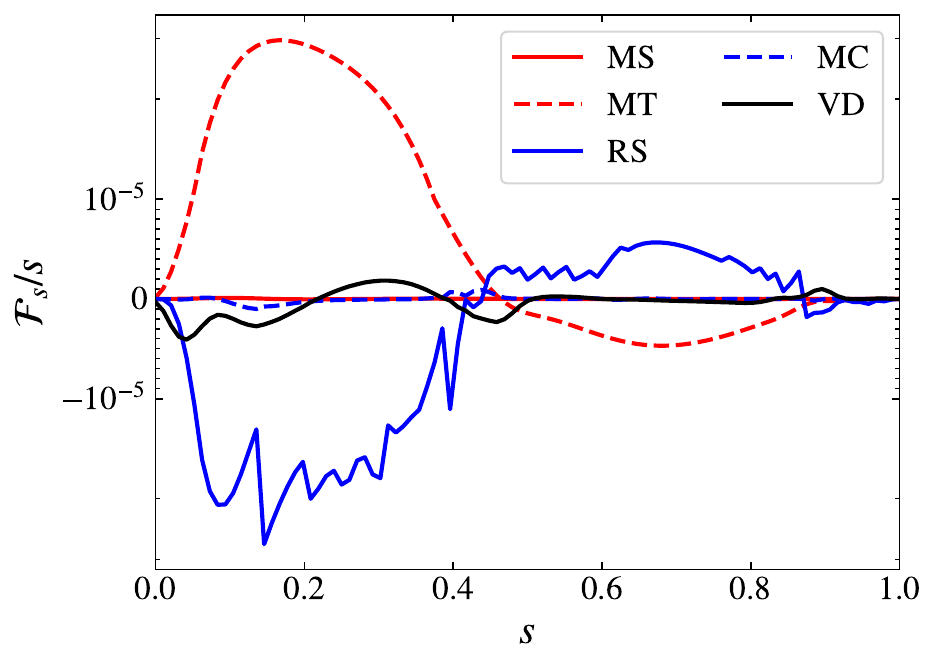}
    \includegraphics[width=0.49\linewidth]{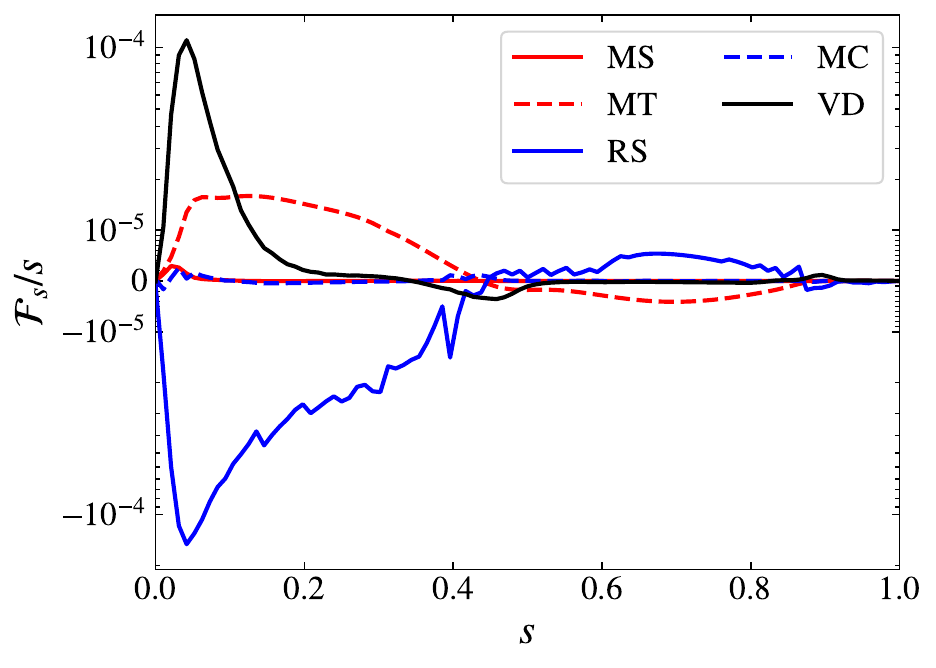}
    \includegraphics[width=0.49\linewidth]{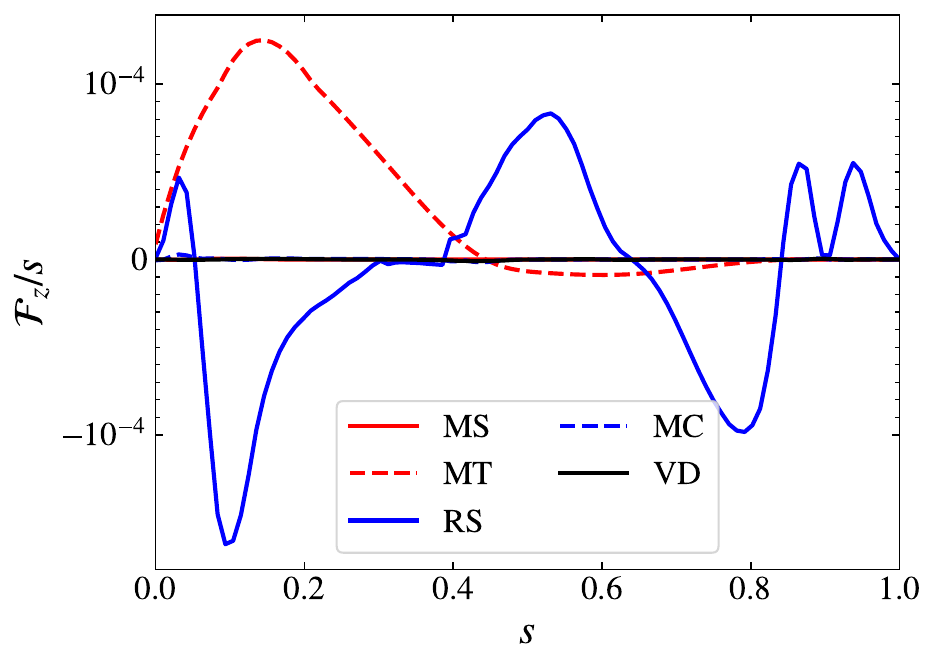}
    \includegraphics[width=0.49\linewidth]{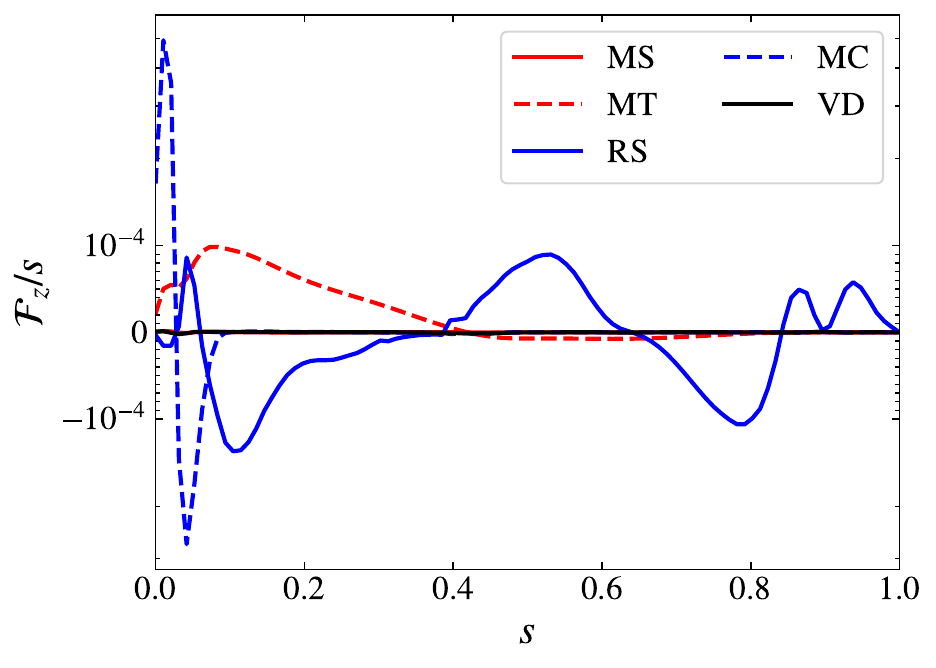}
    \caption{Flux terms within the cylindrical $\mathcal{F}_s/s$ (\textit{top}) and vertical $\mathcal{F}_z/s$ (\textit{bottom}) angular momentum fluxes, integrated along $z$ over the north hemisphere only, for $\Le=10^{-2}$ and $\Pm=1$. They include Maxwell stresses (MS), magnetic torques (MT), Reynolds stresses (RS), meridional circulations (MC), viscous diffusion contributions (VD). \textit{Left:} $t\approx6500$ close to a local minimum for the zonal flow amplitude. \textit{Right:} $t\approx7150$ close to a local maximum for the  zonal flow amplitude.}
    \label{fig:flux_osc}
\end{figure*}
\begin{figure*}
    \centering
    \includegraphics[width=0.49\textwidth]{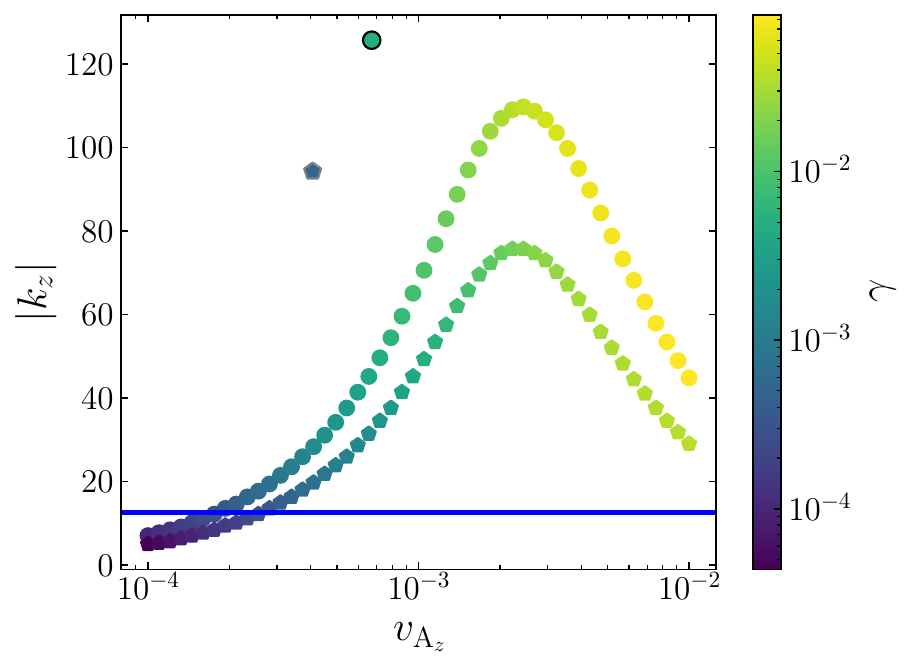}
    \includegraphics[width=0.49\textwidth]{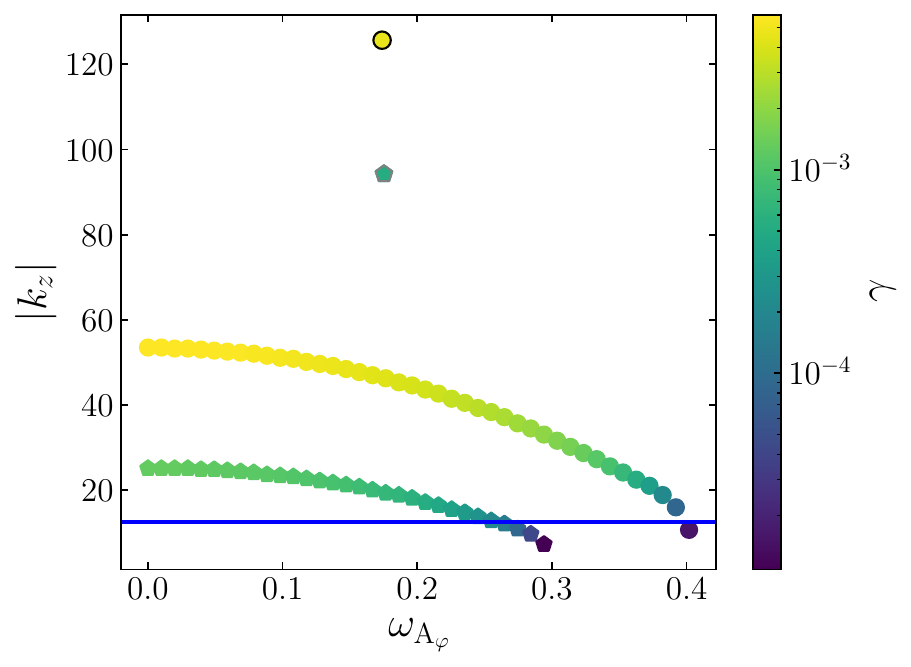}
    \caption{Fastest growing modes with growth rates $\gamma$ and vertical wavenumbers $k_z$ from SMRI/HMRI local stability analysis with axisymmetric vertical and azimuthal magnetic fields whose strengths are given by $\vaz$ (\textit{left}) and $\Lp$ (\textit{right}). Bullets and pentagons correspond respectively to predictions for $\Le=4\cdot10^{-4}$ and $\Le=10^{-3}$ using the specific sets of parameters $(\Som,\Omega,\Lp)$ (\textit{left panel}) and $(\Som,\Omega,\vaz)$ (\textit{right}) given in Table \ref{tab:params} for the 2 simulations. Black/grey outlined outlined symbols correspond to the associated estimations using the values in Table \ref{tab:params} and the axisymmetric growth rates in Table \ref{tab:gamma}. 
    The lower limit $k_z=2\pi/(1-\alpha)$ is indicated by a blue line.} 
    \label{fig:azi_axi_mri}
\end{figure*}
\section{Local linear stability analyses}
\subsection{Axisymmetric (helical and standard) magnetorotational instabilities}
\label{sec:HMRI}
We adopt the approach of \citet{KS2010} to analyse the growth of axisymmetric perturbations in a differentially rotating incompressible fluid in the presence of both a poloidal and toroidal magnetic field. This employs a local WKBJ approximation around a fiducial point $(s,z)$. 
Linearised axisymmetric ($m=0$) perturbations can be sought for each variable proportional to $\mathrm{exp}(\gamma t+\mathrm{i}k_s s+\mathrm{i}k_z z)$, where $\gamma$ is the complex growth rate (at least when $\mathrm{Re}[\gamma]>0$), and $k_s$ and $k_z$ are the cylindrical radial and vertical wavenumbers, respectively (assumed to be real). The unstable modes can be determined by solving the resulting dispersion relation, which is the following degree 4 polynomial \citep[see][for details of the derivation]{KS2010}:
\begin{equation}
    \gamma^4 + a_1 \gamma^3 + a_2 \gamma^2 + (a_3 + \mathrm{i}b_3 )\gamma + a_4 + \mathrm{i}b_4 = 0,
    \label{eq:pol4}
\end{equation}
where the coefficients are:
\begin{equation}
\begin{aligned}
a_1 &= 2(\Ekk + \Emk ),\\
a_2 &= (\Ekk + \Emk )^2 + 2(\Lz^2+ \Ekk \Emk) + \alpha^2 \kappa^2 + 4\alpha^2 \Lp^2 ,\\
a_3 &= 2(\Emk + \Ekk )(\Lz^2+ \Emk \Ekk) + 2\alpha^2 \kappa^2 \Emk + 4\alpha^2 (\Emk + \Ekk )\Lp^2 ,\\
a_4 &= (\Lz^2+ \Ekk \Emk)^2 - 4\alpha^2 \Lz^2\Omega^2 + \alpha^2 \kappa^2(\Lz^2+ \Emk^2)
+ 4\alpha^2 \Ekk \Emk \Lp^2 ,\\
b_3 &= -8\alpha^2\Omega \Lz \Lp ,\\
b_4 &= -4\alpha^2\Omega \Lz \Lp (2\Emk + \Ekk ) - \kappa^2 \alpha^2\Omega^{-1} \Lz \Lp (\Emk - \Ekk ).
\end{aligned}    
\label{eq:ab}
\end{equation}
We have also defined
\begin{equation}
\begin{aligned}
    &k=\sqrt{k_z^2+k_s^2},~~ \alpha=\frac{k_z}{k},~~ \Omega=\Omega_0+\delta\Omega(s)\\
    &\Lz=\frac{k_z B_z}{\sqrt{\mu_0\rho}},~~ \Lp=\frac{B_\varphi}{\sqrt{\mu_0\rho}\,s},\\
    &\Ekk=\nu k^2,~~ \Emk=\eta k^2,~~ \kappa^2=2\Omega\left(2\Omega+\frac{\mathrm{d}\,\delta\Omega}{\mathrm{d}\ln s}\right)=2\Omega(2\Omega-\Som).
\end{aligned}
\label{eq:params}
\end{equation}
This dispersion relation for axisymmetric modes describes the standard MRI (SMRI) acting on a purely poloidal field if $B_\varphi=0$ \citep[e.g.][]{BH1991,BH1998} and the Helical MRI (HMRI) involving a combination of a poloidal and toroidal field otherwise \citep[e.g.][]{HR2005,KS2010}. The presence of toroidal magnetic fields can affect some axisymmetric modes through hoop stresses, which is why $B_\varphi$ appears in the dispersion relation, unlike in analyses that adopt a purely local Cartesian model \citep[e.g.][]{DB2024}.

Since it is not straightforward to choose a specific fiducial point $(s,z)$ where to apply the local stability analysis to our simulations (especially in the vertical direction), we arbitrarily choose to take $s=s_\mathrm{m}$ where the $\varphi$ and $z$-averaged shear parameter $\Som=-\mathrm{d}\delta\Omega/\mathrm{d}\ln s$ is maximised at a specific time $t$. We then take the $\varphi$ and $z$ average of the azimuthal Alfvén frequency $\Lp$, the vertical Alfvén velocity $\vaz=\omega_\mathrm{A}/k_z$ (defined to get rid of $k_z$ in $\Lz$), and the rotation rate $\Omega$, at this location $s_\mathrm{m}$. 
\begin{table}
    \centering
    \setlength{\tabcolsep}{5pt}
    \begin{tabular}{c||c|c|c|c|c|c|c}
        $\Le$ & $t$ & $s_\mathrm{m}$ & $\vaz$ & $\Lp$ & $\Omega$ & $ \Som$ & $k_z$ \\
        \hline
        $4\cdot10^{-4}$ & $4850$ & $0.044$ & $6.7\cdot10^{-4}$ & $0.17$ & $1.2$  & $0.23$& $126$ \\
        $10^{-3}$ & $2950$ & $0.044$ & $4.1\cdot10^{-4}$ & $0.18$ & $1.1$ & $0.13$ & $94$
    \end{tabular}
    \caption{Values chosen to apply the HMRI and AMRI local stability analyses. The $\varphi$ and $z$ averaged parameters $\vaz$, $\Lp$, $\Omega$, and $S_\Omega$ have been evaluated at a fiducial point $s_\mathrm{m}$ which maximises $\langle S_\Omega\rangle_z$ in the upper hemisphere at a time $t$. The vertical wavenumber $k_z$ is measured from the left panels of Fig. \ref{fig:mri_map}.}
    \label{tab:params}
\end{table}
The values of these parameters are displayed in Table \ref{tab:params} for the two simulations showing signs of magnetohydrodynamic instabilities at a time where $\Mp$ is maximised ($2^\mathrm{nd}$ peak for $\Le=10^{-3}$ shown in Fig. \ref{fig:compE_pm5}). The vertical wavenumber $k_z=2\pi/\lambda_z$ is estimated by computing the number of vertical wavelengths $\lambda_z$ in the domain in Fig.~\ref{fig:mri_map} (left panels). Lastly, the viscous and Ohmic decay rates $\omega_\nu$ and $\omega_\eta$ are computed using the global (magnetic) Ekman number $\Ek=10^{-5}$ and $\Em=\Ek/\Pm=2\cdot10^{-6}$ for $\Pm=5$.

In Fig.~\ref{fig:azi_axi_mri}, we present predictions for growth rates $\gamma$ and vertical wavenumbers $k_z$ from solving Eq.~(\ref{eq:pol4}) with the parameters given in Eqs. (\ref{eq:ab}) and (\ref{eq:params}). We do so by varying either $\Lp$ or $\vaz$, and fixing the other parameters to the values listed in Table \ref{tab:params}. Only the fastest growing mode is selected for each set of parameters, which is always found to have $k_s=0$ (``channel modes"). Increasing the strength of the azimuthal magnetic field (by increasing $\Lp$) reduces both $\gamma$ and $k_z$ compared to SMRI with $B_\varphi=0$. Conversely, increasing $\vaz$ boosts the growth rate, but not necessarily the value of $k_z$, which decays after a maximum around $\vaz\approx2.3\cdot10^{-3}$. 
This may be related to some known properties for SMRI. Without diffusion, this would predict $\lambda_z\propto B_z$, so $k_z\propto 1/B_z\propto 1/\vaz$. With diffusion, this would be modified for small $B_z$ where those smaller scale modes would be damped. Also, $B_\varphi$ would presumably play a role also in causing this maximum.
Monotonic growth of both the maximum $\gamma$ and the corresponding $k_z$ is observed with increases in the shear parameter $\Som$ (not shown here). The modes predicted below the threshold $k_z=2\pi/(1-\alpha)$ (blue line) should not be able to develop in the shell (with size $1-\alpha$) since they would be larger than the domain size.
Values of the growth rate measured in the simulations (from $\Mp(m=0)$ in Table \ref{tab:gamma}) are indicated in grey/black outlined symbols in both panels of Fig.~\ref{fig:azi_axi_mri} using the values of $k_z$, $\Lp$, and $\vaz$ listed in Table \ref{tab:params}. The measured growth rates match quite well the analytically predicted ones, which are around $4\cdot10^{-3}$ and $7\cdot10^{-4}$ for $\Le=4\cdot10^{-4}$ and $\Le=10^{-3}$, respectively. However, the predicted vertical wavenumbers are lower than the estimates in Table \ref{tab:params}. If the most unstable mode is not taken but $k_z$ is instead fixed to the last column of Table \ref{tab:params}, the predicted modes would be stable.
However, it must be stressed that the values for $\Som$, $\Lp$, $\Omega$, and $\vaz$ vary substantially from one fiducial point to another (and the $z$-average reduces values of $\Som$ and $\vaz$, and thus of $k_z$), while the growth rate has been measured from a global poloidal quantity, which does not make the comparison with the local analytical model straightforward. Moreover, the validity of this local dispersion relation relies on several assumptions that may not be fully satisfied in our global simulations. First, the cylindrical and vertical variations are assumed to be small compared with the characteristic lengthscales of variation of background flow and field quantities in the same directions. This is probably justified in the vertical direction, where the zonal flow and the vertical and azimuthal magnetic fields are mainly invariant of $z$. However, it is marginally satisfied in the radial direction, since the variation along $s$ in Fig.~\ref{fig:mri_map} is only slightly smaller than the typical lengthscale on which the shear varies $\ls\sim0.1$. In addition, the short-wavelength approximation relies on the fact that $k_s s\gg 1$, which is not satisfied here. Nevertheless, these results are suggestive that the MRI is in operation in our simulations. 
\begin{figure*}
    \centering
    \includegraphics[width=0.49\linewidth]{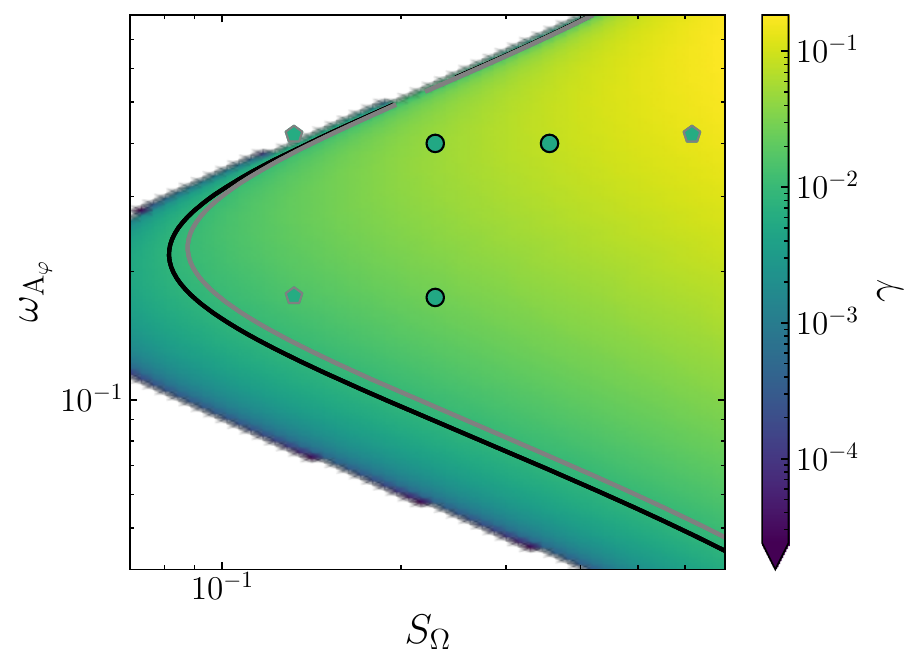}
    \includegraphics[width=0.49\linewidth]{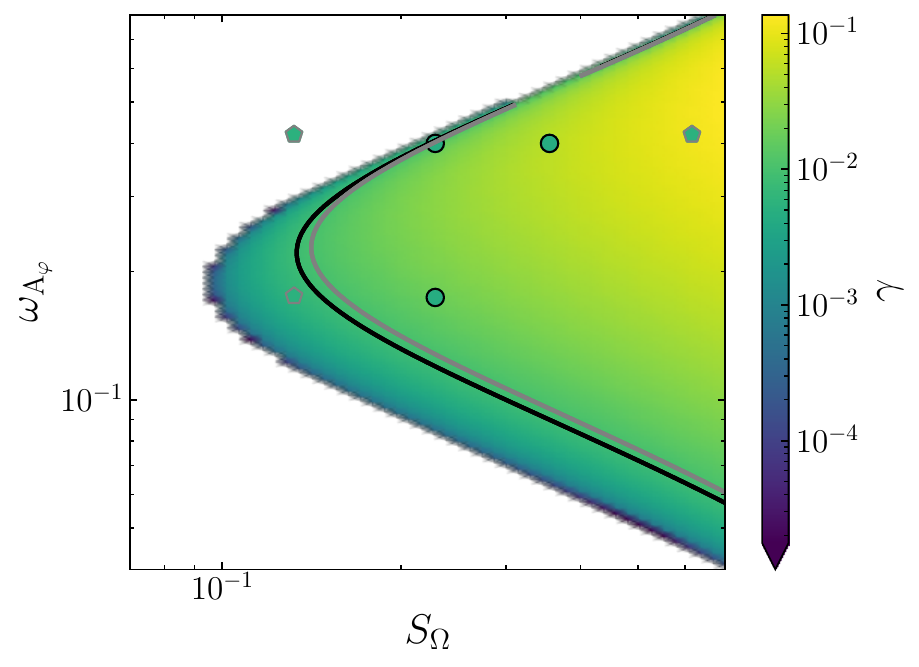}    
    \includegraphics[width=0.49\linewidth]{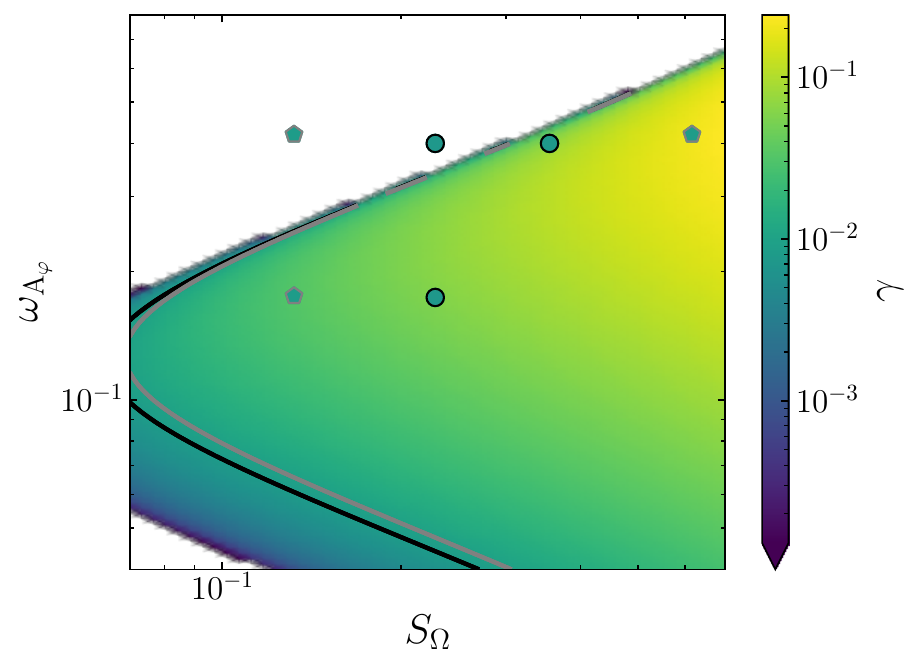}
    \includegraphics[width=0.49\linewidth]{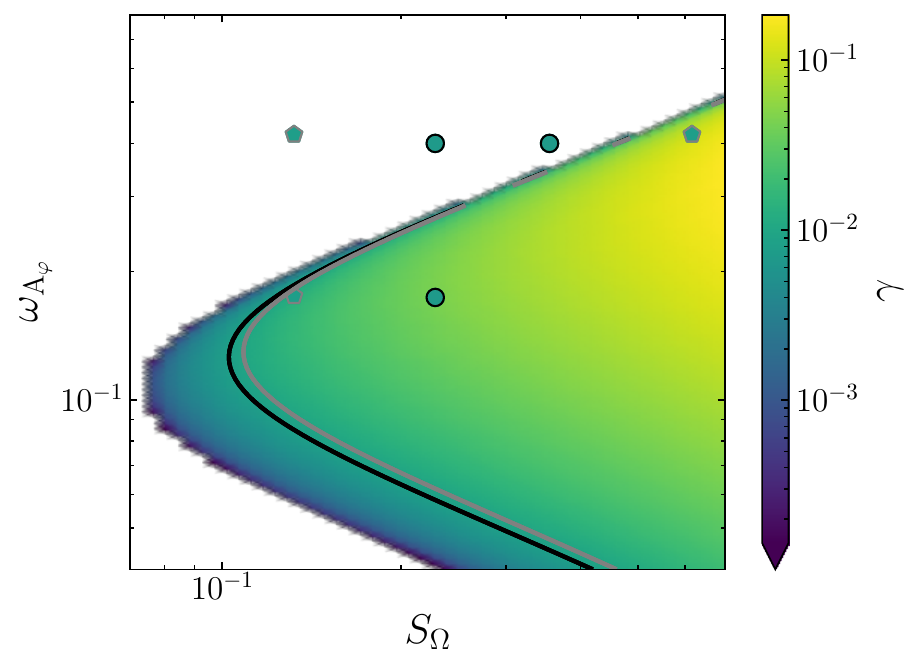}    
    \caption{Predictions for the growth rate $\gamma$ due to non-axisymmetric AMRI for a fixed vertical wavenumber $k_z=94$ as a function of the shear parameter $\Som$ and the azimuthal Alfvén frequency $\Lp$. As in Fig.~\ref{fig:azi_axi_mri}, black outlined bullets are for $\Le=4\cdot10^{-4}$ and grey pentagons are for $\Le=10^{-3}$. These refer to growth rate estimates using either $z$-averaged $(\Som,\Lp)$ from Table \ref{tab:params} (lowest pair of symbols), the maximum of $(\Som,\Lp)$ over $z$ at $s_\mathrm{m}$ (upper right pair symbols), or a mix of the two (upper left pair of symbols), and the highest non-axisymmetric growth rates of the toroidal magnetic energy in Table \ref{tab:gamma}. Associated contour lines are also indicated in black (for $\gamma=8\cdot10^{-3}$) and grey (for $\gamma=10^{-2}$). The white region indicates stable modes (with zero growth). \textit{Left:} $k_x=0$. \textit{Right:} $k_x=63$. \textit{Up:} $m=1$. \textit{Down:} $m=2$.
    }
    \label{fig:AMRI}
\end{figure*}
\subsection{Non-axisymmetric azimuthal magnetic instabilities}
\label{sec:AMRI}
In this section, we apply the local analytical model derived in \cite{ML2019} to investigate azimuthal MRI (AMRI) involving the growth of non-axisymmetric perturbations in an incompressible and differentially rotating fluid with a dominant azimuthal magnetic field \citep[first derived by][but here we omit buoyancy and thermal diffusion]{A1978}. The overall approach is similar to that in Sec.~\ref{sec:HMRI} to derive a dispersion relation, but here we consider the presence of only an axisymmetric azimuthal magnetic field. By injecting non-axisymmetric small-amplitude perturbations of the form $\exp\{\mathrm{i}(k_ss+k_zz+m\varphi-\sigma t)\}$ into the governing equations and linearising (around the background axisymmetric state)
we can obtain the dispersion relation below. The growth rate is $\gamma=\Im{\sigma}$ when the fluid is unstable ($\Im{\sigma}>0$) and the vertical and (cylindrical) radial wavenumbers are again $k_z$ and $k_s$. The azimuthal wavenumber $m$ is fixed to $1$ or $2$ in the following analysis, motivated by our simulations. In this framework, the dispersion relation is again a $4$-th order polynomial:
\begin{equation}
a_4\tilde\omega^4+a_3\tilde\omega^3+a_2\tilde\omega^2+a_1\tilde\omega+a_0=0,
\end{equation}
with $\tilde\omega=\sigma/\Omega-m$ the dimensionless Doppler-shifted frequency, and with coefficients:
\begin{equation}
\begin{aligned}
a_4 =& 1+\beta^2,\\
a_3 =&  2\mathrm{i} \left(1+\beta^2\right) \left(\Re^{-1}+\Rm^{-1}\right),\\
a_2 =& -2 \left(q+2\right)+2 \Lo^2 \left[b-1-m^2\left(1+\beta^2\right) \right]\\
&-\left(1+\beta^2\right) \left(\Re^{-2}+\Rm^{-2}-4 \Re^{-1} \Rm^{-1}\right),\\
a_1 =& -8 m \Lo^2+\mathrm{i} \left\{2 \Lo^2 \left[b-1-\left(1+\beta^2\right) m^2\right] \left(\Re^{-1}+\Rm^{-1}\right)\right.\\
&\left.-4 \left(2+q\right) \Rm^{-1}-2 \left(1+\beta^2\right) \left(\Re^{-2} \Rm^{-1}+\Re^{-1} \Rm^{-2}\right)\right\},\\
a_0 =& m^2 \Lo^2 \left\{2 q-\Lo^2 \left[2 \left(b+1\right)-\left(1+\beta^2\right) m^2\right]\right\}\\
&-2 \Lo^2 \left[b-1-\left(1+\beta^2\right) m^2\right] \Re^{-1} \Rm^{-1}+\left(1+\beta^2\right) \Re^{-2} \Rm^{-2}\\
&+2 \left(2+q\right) \Rm^{-2} +\mathrm{i} \left\{2 m \Lo^2 \left[q \Re^{-1}-(4+q) \Rm^{-1}\right]\right\}.
\end{aligned}
\label{eq:par_AMRI}
\end{equation}
In Eq. \ref{eq:par_AMRI}, six additional dimensionless parameters have been introduced (along with $k=\sqrt{k_s^2+k_z^2}$):
\begin{equation}
\begin{aligned}
    &\beta=\frac{k_s}{k_z},~~ \Re=\frac{\Omega}{\nu k^2}, ~~ \Rm=\frac{\Omega}{\eta k^2},~~q=\frac{\partial\ln\Omega}{\partial\ln s}-\beta\frac{s}{z}\frac{\partial\ln \Omega}{\partial\ln z}, \\
    &\Lo=\frac{\Lp}{\Omega}=\frac{B_\varphi}{\sqrt{\mu\rho}s\Omega},~~ b=\frac{1}{2}\left(\frac{\partial\ln B_\varphi^2}{\partial\ln s}-\beta\frac{s}{z}\frac{\partial\ln B_\varphi^2}{\partial\ln z}\right),
\end{aligned}
\end{equation}
quantifying, in order, the poloidal wavenumber ratio, the hydrodynamical and magnetic Reynolds numbers, the shear rate, the azimuthal magnetic field amplitude and its gradient. This dispersion relation describes non-axisymmetric magnetorotational (differential rotation-driven) instabilities and the Tayler (current-driven) instability \citep[for 
instance in MHD simulations]{T1973,JF2023}.

In our simulations, the zonal flow depends mainly on $s$, such that $\Omega=1+\delta\Omega(s)$, so its vertical gradient can be neglected and the shear parameter can be approximated by $q\approx \partial\ln\Omega/\partial\ln s\approx-S_\Omega/\Omega$. The vertical gradient in $b$ is also quite weak in the region where instability is observed, with $(s/z)\,\partial\ln B_\varphi^2/\partial\ln z\lesssim10^{-3}$, and the radial gradient is of the order of unity with $\partial\ln B_\varphi^2/\partial\ln s\approx1$. 
We choose the same fiducial point $s_\mathrm{m}$ as for SMRI/HMRI in Table \ref{tab:params}, as well as the same values for $\Lp$, $\Omega$, and $\Som$. To illustrate the variability of the parameters in the simulations, the growth rate computed using the maximum (along $z$) shear rate and azimuthal Alfvén frequency are also displayed when computing the estimates (symbols in Fig.~\ref{fig:AMRI}). The Reynolds numbers $\Re$ and $\Rm$ are computed using the global (magnetic) Ekman number $\Ek=10^{-5}$ and $\Em=\Ek/\Pm=2\cdot10^{-6}$ together with the appropriate value of $k$. 

Contrary to our axisymmetric SMRI/HMRI stability analysis, we choose to fix the wavenumbers\footnote{We adopt this approach, as in \cite{ML2019}, because the fastest growing mode typically has very small $k_s,k_z\ll 1$.}. Here $k_z$ and $k_s$ are fixed and we vary $\Lp$ and $\Som$, as is shown in Fig.~\ref{fig:AMRI}, setting the local rotation rate to the maximum of the two simulations $\Omega=1.2$. The vertical wavenumber has been set to $k_z=94$ as estimated for $\Le=4\cdot10^{-4}$, and $k_s=63$ in the right panel is a rough estimation of the radial wavenumber $k_s=2\pi/\lambda_s$ in Fig.~\ref{fig:mri_map}, with $\lambda_s$ the cylindrical radial wavelength. We pick $m=1$ since it is expected to be the azimuthal wavenumber of the most unstable non-axisymmetric mode \citep[see e.g.][and Fourier transform of the toroidal velocity reveals that this component is non-negligible after the instability is triggered for $\Le=4\cdot10^{-4}$ and $\Le=10^{-3}$]{HT2010}, along with the $m=2$ forced mode in our simulations. The main effect of increasing $k_z$ (not shown here) is to reduce the extent of the unstable region for small $\Lp$ and small $\Som$. Increasing $k_s$ has the same effect, and more specifically shifts the region of unstable modes to higher shear rates $\Som$, as seen when comparing the two columns of Fig.~\ref{fig:AMRI}. Higher azimuthal wavenumbers shift the region of unstable modes to lower azimuthal Alfvén frequencies for which the restoring effects of magnetic tension are weaker.  
Unlike for SMRI/HMRI, an increase of the azimuthal Alfvén frequency implies a larger growth rate. 
Larger growth rates are also found for higher shear $\Som$. The term encoding radial and vertical gradient of the azimuthal magnetic field in Eq. \ref{eq:par_AMRI} may not play an important in our model, since putting $b=0$ instead does not alter much the growth rate, which probably rules out the possibility of a Tayler-type instability.

Depending on the value used for the shear (either $\mathrm{max}\langle \Som\rangle_z$ or $\max[S(s_\mathrm{m},z)]$) and for the azimuthal Alfvén wave frequency (either $\langle\Lp(s_m)\rangle_z$ or $\max[\Lp(s_\mathrm{m},z)]$), analytical predictions for the growth rate can be found quite close to the measured values for both $m=1$ and $m=2$, though higher growth rates (when values are in the unstable region) are often predicted theoretically. Hence, non-axisymmetric MRI is a plausible explanation for the instability observed in these simulations, potentially in addition to axisymmetric MRI. 

We caution once again regarding the difficulty of choosing the fiducial point, which has a major impact on the parameter values and the uncertainty surrounding our predictions of the growth rates and wavenumbers. Moreover, the short-wavelength approximation assumes the meridional wavelength perturbation 
\begin{equation}
    \lambda_\mathrm{m}=\sqrt{\lambda_s^2+\lambda_z^2}\ll r, l_{B_\varphi}, \ls ,
    \label{eq:cond_AMRI}
\end{equation}
with $r$ the spherical radius, $l_{B_\varphi}=|\bm\nabla\ln B_\varphi|^{-1}$ and $\ls=|\bm\nabla\ln\Omega|^{-1}$ the scale heights of the azimuthal magnetic field and rotation. Since vertical variations of $B_\varphi$ and $\Omega$ are negligible compared to radial variations, the scale heights can be written as $l_{B_\varphi}\approx2s/|\partial\ln B_\varphi^2/\partial\ln s|$ and $\ls\approx|\partial_s\ln\Omega|^{-1}$, both taking values of order 0.1, like $r$. Depending on the value of $\lambda_s$, the condition Eq. (\ref{eq:cond_AMRI}) is either not satisfied (if $k_s=63$, $\lambda_\mathrm{m}\sim0.1\sim r,l_{B_\varphi},\ls$) or is only marginally satisfied (if $k_s=0$, $\lambda_\mathrm{m}\sim5\cdot10^{-2}\lesssim r,l_{B_\varphi},\ls$). 
Moreover, the azimuthal wavelength of the perturbations must also satisfy $\lambda_\mathrm{m}\ll\lambda_\varphi$, which is not well verified with $\lambda_\varphi=2\pi s_\mathrm{m}/m\sim10^{-1}\sim\lambda_\mathrm{m}$ if $m=1,2$. Finally, and perhaps most importantly, we also require the growth time for the instability to be (much) shorter than the timescale to shear out non-axisymmetric perturbations for a (exponentially growing in time) normal mode analysis to be valid. This requires 
\begin{equation}
m \ll k_s \frac{\ls}{\Omega} \gamma_\mathrm{max},
\label{eq:cond3}
\end{equation}
with $\gamma_\mathrm{max}$ the maximum growth rate of the most unstable mode for a specific set of parameters $(\Lp,\Som)$
\citep[see Eq. (26) of][and the associated arguments]{ML2019}. The right hand side of Eq. (\ref{eq:cond3}) is evaluated here to be, at most,  $63\times 0.1\times 10^{-1}\lesssim 1\lesssim m$, implying that non-axisymmetric perturbations with $m=1$ and 2 will be substantially modified by the shear over the predictions of this normal mode theory. In this case, a transient amplification of nonaxisymmetric modes is predicted instead, and the consequent increase in $k_s$ with time due to the action of the shear will enhance the effects of diffusion and ultimately stabilise the modes. In addition, the larger magnetic tension acting on these larger $k_s$ modes may also help to stabilise them. Hence, we expect the observed growth rates to be smaller than the theoretical predictions, just as we have observed numerically.

\bsp	
\label{lastpage}
\end{document}